\documentclass[longauth]{aa}

\usepackage{graphicx}
\usepackage{natbib}
\usepackage{scalerel}
\usepackage[dvipsnames]{xcolor}

\usepackage{txfonts}
\usepackage[pdfencoding=auto,psdextra]{hyperref}
\hypersetup{
    colorlinks=true,
    linkcolor=blue,
    filecolor=magenta,      
    urlcolor=blue,
    citecolor=blue
}
\makeatletter
\renewcommand*\aa@pageof{, page \thepage{} of \pageref*{LastPage}}
\makeatother

\usepackage[utf8]{inputenc}

\usepackage[switch, modulo]{lineno}

\usepackage{orcidlink} 
\newcommand{\orcid}[1]{\orcidlink{#1}}

\usepackage{lastpage}

\usepackage{euclid}

\usepackage{placeins} 

\begin{document}

\title{\Euclid: Early Release Observations -- Interplay between dwarf galaxies and their globular clusters in the Perseus galaxy cluster\thanks{This paper is published on
       behalf of the Euclid Consortium}}

\author{T.~Saifollahi\orcid{0000-0002-9554-7660}\thanks{\email{Teymoor.saifollahi@astro.unistra.fr}}\inst{\ref{aff1}}
\and A.~Lan\c{c}on\orcid{0000-0002-7214-8296}\inst{\ref{aff1}}
\and M.~Cantiello\orcid{0000-0003-2072-384X}\inst{\ref{aff2}}
\and J.-C.~Cuillandre\orcid{0000-0002-3263-8645}\inst{\ref{aff3}}
\and M.~Bethermin\orcid{0000-0002-3915-2015}\inst{\ref{aff1}}
\and D.~Carollo\orcid{0000-0002-0005-5787}\inst{\ref{aff4}}
\and P.-A.~Duc\orcid{0000-0003-3343-6284}\inst{\ref{aff1}}
\and A.~Ferr\'e-Mateu\orcid{0000-0002-6411-220X}\inst{\ref{aff5},\ref{aff6}}
\and N.~A.~Hatch\orcid{0000-0001-5600-0534}\inst{\ref{aff7}}
\and M.~Hilker\orcid{0000-0002-2363-5522}\inst{\ref{aff8}}
\and L.~K.~Hunt\orcid{0000-0001-9162-2371}\inst{\ref{aff9}}
\and F.~R.~Marleau\orcid{0000-0002-1442-2947}\inst{\ref{aff10}}
\and J.~Rom\'an\orcid{0000-0002-3849-3467}\inst{\ref{aff11}}
\and R.~S\'anchez-Janssen\orcid{0000-0003-4945-0056}\inst{\ref{aff12}}
\and C.~Tortora\orcid{0000-0001-7958-6531}\inst{\ref{aff13}}
\and M.~Urbano\orcid{0000-0001-5640-0650}\inst{\ref{aff1}}
\and K.~Voggel\orcid{0000-0001-6215-0950}\inst{\ref{aff1}}
\and M.~Bolzonella\orcid{0000-0003-3278-4607}\inst{\ref{aff14}}
\and H.~Bouy\orcid{0000-0002-7084-487X}\inst{\ref{aff15},\ref{aff16}}
\and M.~Kluge\orcid{0000-0002-9618-2552}\inst{\ref{aff17}}
\and M.~Schirmer\orcid{0000-0003-2568-9994}\inst{\ref{aff18}}
\and C.~Stone\orcid{0000-0002-9086-6398}\inst{\ref{aff19},\ref{aff20},\ref{aff21}}
\and C.~Giocoli\orcid{0000-0002-9590-7961}\inst{\ref{aff14},\ref{aff22}}
\and J.~H.~Knapen\orcid{0000-0003-1643-0024}\inst{\ref{aff5},\ref{aff6}}
\and M.~N.~Le\orcid{0009-0003-0674-9813}\inst{\ref{aff5},\ref{aff6}}
\and M.~Mondelin\orcid{0009-0004-5954-0930}\inst{\ref{aff3}}
\and M.~Poulain\orcid{0000-0002-7664-4510}\inst{\ref{aff23}}
\and N.~Aghanim\orcid{0000-0002-6688-8992}\inst{\ref{aff24}}
\and B.~Altieri\orcid{0000-0003-3936-0284}\inst{\ref{aff25}}
\and S.~Andreon\orcid{0000-0002-2041-8784}\inst{\ref{aff26}}
\and N.~Auricchio\orcid{0000-0003-4444-8651}\inst{\ref{aff14}}
\and C.~Baccigalupi\orcid{0000-0002-8211-1630}\inst{\ref{aff27},\ref{aff4},\ref{aff28},\ref{aff29}}
\and D.~Bagot\inst{\ref{aff30}}
\and M.~Baldi\orcid{0000-0003-4145-1943}\inst{\ref{aff31},\ref{aff14},\ref{aff22}}
\and A.~Balestra\orcid{0000-0002-6967-261X}\inst{\ref{aff32}}
\and S.~Bardelli\orcid{0000-0002-8900-0298}\inst{\ref{aff14}}
\and A.~Basset\inst{\ref{aff30}}
\and P.~Battaglia\orcid{0000-0002-7337-5909}\inst{\ref{aff14}}
\and A.~Biviano\orcid{0000-0002-0857-0732}\inst{\ref{aff4},\ref{aff27}}
\and A.~Bonchi\orcid{0000-0002-2667-5482}\inst{\ref{aff33}}
\and D.~Bonino\orcid{0000-0002-3336-9977}\inst{\ref{aff34}}
\and W.~Bon\inst{\ref{aff35}}
\and E.~Branchini\orcid{0000-0002-0808-6908}\inst{\ref{aff36},\ref{aff37},\ref{aff26}}
\and M.~Brescia\orcid{0000-0001-9506-5680}\inst{\ref{aff38},\ref{aff13}}
\and J.~Brinchmann\orcid{0000-0003-4359-8797}\inst{\ref{aff39},\ref{aff40}}
\and S.~Camera\orcid{0000-0003-3399-3574}\inst{\ref{aff41},\ref{aff42},\ref{aff34}}
\and V.~Capobianco\orcid{0000-0002-3309-7692}\inst{\ref{aff34}}
\and C.~Carbone\orcid{0000-0003-0125-3563}\inst{\ref{aff43}}
\and J.~Carretero\orcid{0000-0002-3130-0204}\inst{\ref{aff44},\ref{aff45}}
\and S.~Casas\orcid{0000-0002-4751-5138}\inst{\ref{aff46}}
\and M.~Castellano\orcid{0000-0001-9875-8263}\inst{\ref{aff47}}
\and G.~Castignani\orcid{0000-0001-6831-0687}\inst{\ref{aff14}}
\and S.~Cavuoti\orcid{0000-0002-3787-4196}\inst{\ref{aff13},\ref{aff48}}
\and K.~C.~Chambers\orcid{0000-0001-6965-7789}\inst{\ref{aff49}}
\and A.~Cimatti\inst{\ref{aff50}}
\and C.~Colodro-Conde\inst{\ref{aff5}}
\and G.~Congedo\orcid{0000-0003-2508-0046}\inst{\ref{aff51}}
\and C.~J.~Conselice\orcid{0000-0003-1949-7638}\inst{\ref{aff52}}
\and L.~Conversi\orcid{0000-0002-6710-8476}\inst{\ref{aff53},\ref{aff25}}
\and Y.~Copin\orcid{0000-0002-5317-7518}\inst{\ref{aff54}}
\and F.~Courbin\orcid{0000-0003-0758-6510}\inst{\ref{aff55},\ref{aff56}}
\and H.~M.~Courtois\orcid{0000-0003-0509-1776}\inst{\ref{aff57}}
\and M.~Cropper\orcid{0000-0003-4571-9468}\inst{\ref{aff58}}
\and A.~Da~Silva\orcid{0000-0002-6385-1609}\inst{\ref{aff59},\ref{aff60}}
\and H.~Degaudenzi\orcid{0000-0002-5887-6799}\inst{\ref{aff61}}
\and G.~De~Lucia\orcid{0000-0002-6220-9104}\inst{\ref{aff4}}
\and H.~Dole\orcid{0000-0002-9767-3839}\inst{\ref{aff24}}
\and M.~Douspis\orcid{0000-0003-4203-3954}\inst{\ref{aff24}}
\and F.~Dubath\orcid{0000-0002-6533-2810}\inst{\ref{aff61}}
\and C.~A.~J.~Duncan\orcid{0009-0003-3573-0791}\inst{\ref{aff51},\ref{aff52}}
\and X.~Dupac\inst{\ref{aff25}}
\and S.~Dusini\orcid{0000-0002-1128-0664}\inst{\ref{aff62}}
\and S.~Escoffier\orcid{0000-0002-2847-7498}\inst{\ref{aff63}}
\and M.~Farina\orcid{0000-0002-3089-7846}\inst{\ref{aff64}}
\and R.~Farinelli\inst{\ref{aff14}}
\and F.~Faustini\orcid{0000-0001-6274-5145}\inst{\ref{aff47},\ref{aff33}}
\and S.~Ferriol\inst{\ref{aff54}}
\and S.~Fotopoulou\orcid{0000-0002-9686-254X}\inst{\ref{aff65}}
\and M.~Frailis\orcid{0000-0002-7400-2135}\inst{\ref{aff4}}
\and E.~Franceschi\orcid{0000-0002-0585-6591}\inst{\ref{aff14}}
\and M.~Fumana\orcid{0000-0001-6787-5950}\inst{\ref{aff43}}
\and S.~Galeotta\orcid{0000-0002-3748-5115}\inst{\ref{aff4}}
\and K.~George\orcid{0000-0002-1734-8455}\inst{\ref{aff66}}
\and B.~Gillis\orcid{0000-0002-4478-1270}\inst{\ref{aff51}}
\and J.~Gracia-Carpio\inst{\ref{aff17}}
\and A.~Grazian\orcid{0000-0002-5688-0663}\inst{\ref{aff32}}
\and F.~Grupp\inst{\ref{aff17},\ref{aff66}}
\and S.~V.~H.~Haugan\orcid{0000-0001-9648-7260}\inst{\ref{aff67}}
\and J.~Hoar\inst{\ref{aff25}}
\and H.~Hoekstra\orcid{0000-0002-0641-3231}\inst{\ref{aff68}}
\and W.~Holmes\inst{\ref{aff69}}
\and I.~M.~Hook\orcid{0000-0002-2960-978X}\inst{\ref{aff70}}
\and F.~Hormuth\inst{\ref{aff71}}
\and A.~Hornstrup\orcid{0000-0002-3363-0936}\inst{\ref{aff72},\ref{aff73}}
\and K.~Jahnke\orcid{0000-0003-3804-2137}\inst{\ref{aff18}}
\and M.~Jhabvala\inst{\ref{aff74}}
\and E.~Keih\"anen\orcid{0000-0003-1804-7715}\inst{\ref{aff75}}
\and S.~Kermiche\orcid{0000-0002-0302-5735}\inst{\ref{aff63}}
\and A.~Kiessling\orcid{0000-0002-2590-1273}\inst{\ref{aff69}}
\and B.~Kubik\orcid{0009-0006-5823-4880}\inst{\ref{aff54}}
\and M.~K\"ummel\orcid{0000-0003-2791-2117}\inst{\ref{aff66}}
\and M.~Kunz\orcid{0000-0002-3052-7394}\inst{\ref{aff76}}
\and H.~Kurki-Suonio\orcid{0000-0002-4618-3063}\inst{\ref{aff77},\ref{aff78}}
\and O.~Lahav\orcid{0000-0002-1134-9035}\inst{\ref{aff79}}
\and R.~Laureijs\inst{\ref{aff80}}
\and A.~M.~C.~Le~Brun\orcid{0000-0002-0936-4594}\inst{\ref{aff81}}
\and D.~Le~Mignant\orcid{0000-0002-5339-5515}\inst{\ref{aff35}}
\and S.~Ligori\orcid{0000-0003-4172-4606}\inst{\ref{aff34}}
\and P.~B.~Lilje\orcid{0000-0003-4324-7794}\inst{\ref{aff67}}
\and V.~Lindholm\orcid{0000-0003-2317-5471}\inst{\ref{aff77},\ref{aff78}}
\and I.~Lloro\orcid{0000-0001-5966-1434}\inst{\ref{aff82}}
\and D.~Maino\inst{\ref{aff83},\ref{aff43},\ref{aff84}}
\and E.~Maiorano\orcid{0000-0003-2593-4355}\inst{\ref{aff14}}
\and O.~Mansutti\orcid{0000-0001-5758-4658}\inst{\ref{aff4}}
\and O.~Marggraf\orcid{0000-0001-7242-3852}\inst{\ref{aff85}}
\and M.~Martinelli\orcid{0000-0002-6943-7732}\inst{\ref{aff47},\ref{aff86}}
\and N.~Martinet\orcid{0000-0003-2786-7790}\inst{\ref{aff35}}
\and F.~Marulli\orcid{0000-0002-8850-0303}\inst{\ref{aff87},\ref{aff14},\ref{aff22}}
\and R.~Massey\orcid{0000-0002-6085-3780}\inst{\ref{aff88}}
\and S.~Maurogordato\inst{\ref{aff89}}
\and E.~Medinaceli\orcid{0000-0002-4040-7783}\inst{\ref{aff14}}
\and S.~Mei\orcid{0000-0002-2849-559X}\inst{\ref{aff90},\ref{aff91}}
\and Y.~Mellier\inst{\ref{aff92},\ref{aff93}}
\and M.~Meneghetti\orcid{0000-0003-1225-7084}\inst{\ref{aff14},\ref{aff22}}
\and E.~Merlin\orcid{0000-0001-6870-8900}\inst{\ref{aff47}}
\and G.~Meylan\inst{\ref{aff94}}
\and A.~Mora\orcid{0000-0002-1922-8529}\inst{\ref{aff95}}
\and M.~Moresco\orcid{0000-0002-7616-7136}\inst{\ref{aff87},\ref{aff14}}
\and L.~Moscardini\orcid{0000-0002-3473-6716}\inst{\ref{aff87},\ref{aff14},\ref{aff22}}
\and R.~Nakajima\orcid{0009-0009-1213-7040}\inst{\ref{aff85}}
\and C.~Neissner\orcid{0000-0001-8524-4968}\inst{\ref{aff96},\ref{aff45}}
\and S.-M.~Niemi\orcid{0009-0005-0247-0086}\inst{\ref{aff97}}
\and C.~Padilla\orcid{0000-0001-7951-0166}\inst{\ref{aff96}}
\and S.~Paltani\orcid{0000-0002-8108-9179}\inst{\ref{aff61}}
\and F.~Pasian\orcid{0000-0002-4869-3227}\inst{\ref{aff4}}
\and K.~Pedersen\inst{\ref{aff98}}
\and W.~J.~Percival\orcid{0000-0002-0644-5727}\inst{\ref{aff99},\ref{aff100},\ref{aff101}}
\and V.~Pettorino\inst{\ref{aff97}}
\and S.~Pires\orcid{0000-0002-0249-2104}\inst{\ref{aff3}}
\and G.~Polenta\orcid{0000-0003-4067-9196}\inst{\ref{aff33}}
\and M.~Poncet\inst{\ref{aff30}}
\and L.~A.~Popa\inst{\ref{aff102}}
\and L.~Pozzetti\orcid{0000-0001-7085-0412}\inst{\ref{aff14}}
\and F.~Raison\orcid{0000-0002-7819-6918}\inst{\ref{aff17}}
\and R.~Rebolo\orcid{0000-0003-3767-7085}\inst{\ref{aff5},\ref{aff103},\ref{aff6}}
\and A.~Renzi\orcid{0000-0001-9856-1970}\inst{\ref{aff104},\ref{aff62}}
\and J.~Rhodes\orcid{0000-0002-4485-8549}\inst{\ref{aff69}}
\and G.~Riccio\inst{\ref{aff13}}
\and E.~Romelli\orcid{0000-0003-3069-9222}\inst{\ref{aff4}}
\and M.~Roncarelli\orcid{0000-0001-9587-7822}\inst{\ref{aff14}}
\and R.~Saglia\orcid{0000-0003-0378-7032}\inst{\ref{aff66},\ref{aff17}}
\and Z.~Sakr\orcid{0000-0002-4823-3757}\inst{\ref{aff105},\ref{aff106},\ref{aff107}}
\and A.~G.~S\'anchez\orcid{0000-0003-1198-831X}\inst{\ref{aff17}}
\and D.~Sapone\orcid{0000-0001-7089-4503}\inst{\ref{aff108}}
\and B.~Sartoris\orcid{0000-0003-1337-5269}\inst{\ref{aff66},\ref{aff4}}
\and J.~A.~Schewtschenko\orcid{0000-0002-4913-6393}\inst{\ref{aff51}}
\and P.~Schneider\orcid{0000-0001-8561-2679}\inst{\ref{aff85}}
\and T.~Schrabback\orcid{0000-0002-6987-7834}\inst{\ref{aff10}}
\and G.~Seidel\orcid{0000-0003-2907-353X}\inst{\ref{aff18}}
\and M.~Seiffert\orcid{0000-0002-7536-9393}\inst{\ref{aff69}}
\and S.~Serrano\orcid{0000-0002-0211-2861}\inst{\ref{aff109},\ref{aff110},\ref{aff111}}
\and C.~Sirignano\orcid{0000-0002-0995-7146}\inst{\ref{aff104},\ref{aff62}}
\and G.~Sirri\orcid{0000-0003-2626-2853}\inst{\ref{aff22}}
\and L.~Stanco\orcid{0000-0002-9706-5104}\inst{\ref{aff62}}
\and J.~Steinwagner\orcid{0000-0001-7443-1047}\inst{\ref{aff17}}
\and P.~Tallada-Cresp\'{i}\orcid{0000-0002-1336-8328}\inst{\ref{aff44},\ref{aff45}}
\and A.~N.~Taylor\inst{\ref{aff51}}
\and I.~Tereno\orcid{0000-0002-4537-6218}\inst{\ref{aff59},\ref{aff112}}
\and S.~Toft\orcid{0000-0003-3631-7176}\inst{\ref{aff113},\ref{aff114}}
\and R.~Toledo-Moreo\orcid{0000-0002-2997-4859}\inst{\ref{aff115}}
\and F.~Torradeflot\orcid{0000-0003-1160-1517}\inst{\ref{aff45},\ref{aff44}}
\and A.~Tsyganov\inst{\ref{aff116}}
\and I.~Tutusaus\orcid{0000-0002-3199-0399}\inst{\ref{aff106}}
\and E.~A.~Valentijn\inst{\ref{aff80}}
\and L.~Valenziano\orcid{0000-0002-1170-0104}\inst{\ref{aff14},\ref{aff117}}
\and J.~Valiviita\orcid{0000-0001-6225-3693}\inst{\ref{aff77},\ref{aff78}}
\and T.~Vassallo\orcid{0000-0001-6512-6358}\inst{\ref{aff66},\ref{aff4}}
\and G.~Verdoes~Kleijn\orcid{0000-0001-5803-2580}\inst{\ref{aff80}}
\and A.~Veropalumbo\orcid{0000-0003-2387-1194}\inst{\ref{aff26},\ref{aff37},\ref{aff36}}
\and Y.~Wang\orcid{0000-0002-4749-2984}\inst{\ref{aff118}}
\and J.~Weller\orcid{0000-0002-8282-2010}\inst{\ref{aff66},\ref{aff17}}
\and G.~Zamorani\orcid{0000-0002-2318-301X}\inst{\ref{aff14}}
\and F.~M.~Zerbi\inst{\ref{aff26}}
\and E.~Zucca\orcid{0000-0002-5845-8132}\inst{\ref{aff14}}
\and C.~Burigana\orcid{0000-0002-3005-5796}\inst{\ref{aff119},\ref{aff117}}
\and J.~Mart\'{i}n-Fleitas\orcid{0000-0002-8594-569X}\inst{\ref{aff95}}
\and V.~Scottez\inst{\ref{aff92},\ref{aff120}}}

\institute{Universit\'e de Strasbourg, CNRS, Observatoire astronomique de Strasbourg, UMR 7550, 67000 Strasbourg, France\label{aff1}
\and
INAF - Osservatorio Astronomico d'Abruzzo, Via Maggini, 64100, Teramo, Italy\label{aff2}
\and
Universit\'e Paris-Saclay, Universit\'e Paris Cit\'e, CEA, CNRS, AIM, 91191, Gif-sur-Yvette, France\label{aff3}
\and
INAF-Osservatorio Astronomico di Trieste, Via G. B. Tiepolo 11, 34143 Trieste, Italy\label{aff4}
\and
Instituto de Astrof\'{\i}sica de Canarias, V\'{\i}a L\'actea, 38205 La Laguna, Tenerife, Spain\label{aff5}
\and
Universidad de La Laguna, Departamento de Astrof\'{\i}sica, 38206 La Laguna, Tenerife, Spain\label{aff6}
\and
School of Physics and Astronomy, University of Nottingham, University Park, Nottingham NG7 2RD, UK\label{aff7}
\and
European Southern Observatory, Karl-Schwarzschild-Str.~2, 85748 Garching, Germany\label{aff8}
\and
INAF-Osservatorio Astrofisico di Arcetri, Largo E. Fermi 5, 50125, Firenze, Italy\label{aff9}
\and
Universit\"at Innsbruck, Institut f\"ur Astro- und Teilchenphysik, Technikerstr. 25/8, 6020 Innsbruck, Austria\label{aff10}
\and
Departamento de F{\'\i}sica de la Tierra y Astrof{\'\i}sica, Universidad Complutense de Madrid, Plaza de las Ciencias 2, E-28040 Madrid, Spain\label{aff11}
\and
UK Astronomy Technology Centre, Royal Observatory, Blackford Hill, Edinburgh EH9 3HJ, UK\label{aff12}
\and
INAF-Osservatorio Astronomico di Capodimonte, Via Moiariello 16, 80131 Napoli, Italy\label{aff13}
\and
INAF-Osservatorio di Astrofisica e Scienza dello Spazio di Bologna, Via Piero Gobetti 93/3, 40129 Bologna, Italy\label{aff14}
\and
Institut universitaire de France (IUF), 1 rue Descartes, 75231 PARIS CEDEX 05, France\label{aff15}
\and
Laboratoire d'Astrophysique de Bordeaux, CNRS and Universit\'e de Bordeaux, All\'ee Geoffroy St. Hilaire, 33165 Pessac, France\label{aff16}
\and
Max Planck Institute for Extraterrestrial Physics, Giessenbachstr. 1, 85748 Garching, Germany\label{aff17}
\and
Max-Planck-Institut f\"ur Astronomie, K\"onigstuhl 17, 69117 Heidelberg, Germany\label{aff18}
\and
Department of Physics, Universit\'{e} de Montr\'{e}al, 2900 Edouard Montpetit Blvd, Montr\'{e}al, Qu\'{e}bec H3T 1J4, Canada\label{aff19}
\and
Ciela Institute - Montr{\'e}al Institute for Astrophysical Data Analysis and Machine Learning, Montr{\'e}al, Qu{\'e}bec, Canada\label{aff20}
\and
Mila - Qu{\'e}bec Artificial Intelligence Institute, Montr{\'e}al, Qu{\'e}bec, Canada\label{aff21}
\and
INFN-Sezione di Bologna, Viale Berti Pichat 6/2, 40127 Bologna, Italy\label{aff22}
\and
Space physics and astronomy research unit, University of Oulu, Pentti Kaiteran katu 1, FI-90014 Oulu, Finland\label{aff23}
\and
Universit\'e Paris-Saclay, CNRS, Institut d'astrophysique spatiale, 91405, Orsay, France\label{aff24}
\and
ESAC/ESA, Camino Bajo del Castillo, s/n., Urb. Villafranca del Castillo, 28692 Villanueva de la Ca\~nada, Madrid, Spain\label{aff25}
\and
INAF-Osservatorio Astronomico di Brera, Via Brera 28, 20122 Milano, Italy\label{aff26}
\and
IFPU, Institute for Fundamental Physics of the Universe, via Beirut 2, 34151 Trieste, Italy\label{aff27}
\and
INFN, Sezione di Trieste, Via Valerio 2, 34127 Trieste TS, Italy\label{aff28}
\and
SISSA, International School for Advanced Studies, Via Bonomea 265, 34136 Trieste TS, Italy\label{aff29}
\and
Centre National d'Etudes Spatiales -- Centre spatial de Toulouse, 18 avenue Edouard Belin, 31401 Toulouse Cedex 9, France\label{aff30}
\and
Dipartimento di Fisica e Astronomia, Universit\`a di Bologna, Via Gobetti 93/2, 40129 Bologna, Italy\label{aff31}
\and
INAF-Osservatorio Astronomico di Padova, Via dell'Osservatorio 5, 35122 Padova, Italy\label{aff32}
\and
Space Science Data Center, Italian Space Agency, via del Politecnico snc, 00133 Roma, Italy\label{aff33}
\and
INAF-Osservatorio Astrofisico di Torino, Via Osservatorio 20, 10025 Pino Torinese (TO), Italy\label{aff34}
\and
Aix-Marseille Universit\'e, CNRS, CNES, LAM, Marseille, France\label{aff35}
\and
Dipartimento di Fisica, Universit\`a di Genova, Via Dodecaneso 33, 16146, Genova, Italy\label{aff36}
\and
INFN-Sezione di Genova, Via Dodecaneso 33, 16146, Genova, Italy\label{aff37}
\and
Department of Physics "E. Pancini", University Federico II, Via Cinthia 6, 80126, Napoli, Italy\label{aff38}
\and
Instituto de Astrof\'isica e Ci\^encias do Espa\c{c}o, Universidade do Porto, CAUP, Rua das Estrelas, PT4150-762 Porto, Portugal\label{aff39}
\and
Faculdade de Ci\^encias da Universidade do Porto, Rua do Campo de Alegre, 4150-007 Porto, Portugal\label{aff40}
\and
Dipartimento di Fisica, Universit\`a degli Studi di Torino, Via P. Giuria 1, 10125 Torino, Italy\label{aff41}
\and
INFN-Sezione di Torino, Via P. Giuria 1, 10125 Torino, Italy\label{aff42}
\and
INAF-IASF Milano, Via Alfonso Corti 12, 20133 Milano, Italy\label{aff43}
\and
Centro de Investigaciones Energ\'eticas, Medioambientales y Tecnol\'ogicas (CIEMAT), Avenida Complutense 40, 28040 Madrid, Spain\label{aff44}
\and
Port d'Informaci\'{o} Cient\'{i}fica, Campus UAB, C. Albareda s/n, 08193 Bellaterra (Barcelona), Spain\label{aff45}
\and
Institute for Theoretical Particle Physics and Cosmology (TTK), RWTH Aachen University, 52056 Aachen, Germany\label{aff46}
\and
INAF-Osservatorio Astronomico di Roma, Via Frascati 33, 00078 Monteporzio Catone, Italy\label{aff47}
\and
INFN section of Naples, Via Cinthia 6, 80126, Napoli, Italy\label{aff48}
\and
Institute for Astronomy, University of Hawaii, 2680 Woodlawn Drive, Honolulu, HI 96822, USA\label{aff49}
\and
Dipartimento di Fisica e Astronomia "Augusto Righi" - Alma Mater Studiorum Universit\`a di Bologna, Viale Berti Pichat 6/2, 40127 Bologna, Italy\label{aff50}
\and
Institute for Astronomy, University of Edinburgh, Royal Observatory, Blackford Hill, Edinburgh EH9 3HJ, UK\label{aff51}
\and
Jodrell Bank Centre for Astrophysics, Department of Physics and Astronomy, University of Manchester, Oxford Road, Manchester M13 9PL, UK\label{aff52}
\and
European Space Agency/ESRIN, Largo Galileo Galilei 1, 00044 Frascati, Roma, Italy\label{aff53}
\and
Universit\'e Claude Bernard Lyon 1, CNRS/IN2P3, IP2I Lyon, UMR 5822, Villeurbanne, F-69100, France\label{aff54}
\and
Institut de Ci\`{e}ncies del Cosmos (ICCUB), Universitat de Barcelona (IEEC-UB), Mart\'{i} i Franqu\`{e}s 1, 08028 Barcelona, Spain\label{aff55}
\and
Instituci\'o Catalana de Recerca i Estudis Avan\c{c}ats (ICREA), Passeig de Llu\'{\i}s Companys 23, 08010 Barcelona, Spain\label{aff56}
\and
UCB Lyon 1, CNRS/IN2P3, IUF, IP2I Lyon, 4 rue Enrico Fermi, 69622 Villeurbanne, France\label{aff57}
\and
Mullard Space Science Laboratory, University College London, Holmbury St Mary, Dorking, Surrey RH5 6NT, UK\label{aff58}
\and
Departamento de F\'isica, Faculdade de Ci\^encias, Universidade de Lisboa, Edif\'icio C8, Campo Grande, PT1749-016 Lisboa, Portugal\label{aff59}
\and
Instituto de Astrof\'isica e Ci\^encias do Espa\c{c}o, Faculdade de Ci\^encias, Universidade de Lisboa, Campo Grande, 1749-016 Lisboa, Portugal\label{aff60}
\and
Department of Astronomy, University of Geneva, ch. d'Ecogia 16, 1290 Versoix, Switzerland\label{aff61}
\and
INFN-Padova, Via Marzolo 8, 35131 Padova, Italy\label{aff62}
\and
Aix-Marseille Universit\'e, CNRS/IN2P3, CPPM, Marseille, France\label{aff63}
\and
INAF-Istituto di Astrofisica e Planetologia Spaziali, via del Fosso del Cavaliere, 100, 00100 Roma, Italy\label{aff64}
\and
School of Physics, HH Wills Physics Laboratory, University of Bristol, Tyndall Avenue, Bristol, BS8 1TL, UK\label{aff65}
\and
Universit\"ats-Sternwarte M\"unchen, Fakult\"at f\"ur Physik, Ludwig-Maximilians-Universit\"at M\"unchen, Scheinerstrasse 1, 81679 M\"unchen, Germany\label{aff66}
\and
Institute of Theoretical Astrophysics, University of Oslo, P.O. Box 1029 Blindern, 0315 Oslo, Norway\label{aff67}
\and
Leiden Observatory, Leiden University, Einsteinweg 55, 2333 CC Leiden, The Netherlands\label{aff68}
\and
Jet Propulsion Laboratory, California Institute of Technology, 4800 Oak Grove Drive, Pasadena, CA, 91109, USA\label{aff69}
\and
Department of Physics, Lancaster University, Lancaster, LA1 4YB, UK\label{aff70}
\and
Felix Hormuth Engineering, Goethestr. 17, 69181 Leimen, Germany\label{aff71}
\and
Technical University of Denmark, Elektrovej 327, 2800 Kgs. Lyngby, Denmark\label{aff72}
\and
Cosmic Dawn Center (DAWN), Denmark\label{aff73}
\and
NASA Goddard Space Flight Center, Greenbelt, MD 20771, USA\label{aff74}
\and
Department of Physics and Helsinki Institute of Physics, Gustaf H\"allstr\"omin katu 2, 00014 University of Helsinki, Finland\label{aff75}
\and
Universit\'e de Gen\`eve, D\'epartement de Physique Th\'eorique and Centre for Astroparticle Physics, 24 quai Ernest-Ansermet, CH-1211 Gen\`eve 4, Switzerland\label{aff76}
\and
Department of Physics, P.O. Box 64, 00014 University of Helsinki, Finland\label{aff77}
\and
Helsinki Institute of Physics, Gustaf H{\"a}llstr{\"o}min katu 2, University of Helsinki, Helsinki, Finland\label{aff78}
\and
Department of Physics and Astronomy, University College London, Gower Street, London WC1E 6BT, UK\label{aff79}
\and
Kapteyn Astronomical Institute, University of Groningen, PO Box 800, 9700 AV Groningen, The Netherlands\label{aff80}
\and
Laboratoire d'etude de l'Univers et des phenomenes eXtremes, Observatoire de Paris, Universit\'e PSL, Sorbonne Universit\'e, CNRS, 92190 Meudon, France\label{aff81}
\and
SKA Observatory, Jodrell Bank, Lower Withington, Macclesfield, Cheshire SK11 9FT, UK\label{aff82}
\and
Dipartimento di Fisica "Aldo Pontremoli", Universit\`a degli Studi di Milano, Via Celoria 16, 20133 Milano, Italy\label{aff83}
\and
INFN-Sezione di Milano, Via Celoria 16, 20133 Milano, Italy\label{aff84}
\and
Universit\"at Bonn, Argelander-Institut f\"ur Astronomie, Auf dem H\"ugel 71, 53121 Bonn, Germany\label{aff85}
\and
INFN-Sezione di Roma, Piazzale Aldo Moro, 2 - c/o Dipartimento di Fisica, Edificio G. Marconi, 00185 Roma, Italy\label{aff86}
\and
Dipartimento di Fisica e Astronomia "Augusto Righi" - Alma Mater Studiorum Universit\`a di Bologna, via Piero Gobetti 93/2, 40129 Bologna, Italy\label{aff87}
\and
Department of Physics, Institute for Computational Cosmology, Durham University, South Road, Durham, DH1 3LE, UK\label{aff88}
\and
Universit\'e C\^{o}te d'Azur, Observatoire de la C\^{o}te d'Azur, CNRS, Laboratoire Lagrange, Bd de l'Observatoire, CS 34229, 06304 Nice cedex 4, France\label{aff89}
\and
Universit\'e Paris Cit\'e, CNRS, Astroparticule et Cosmologie, 75013 Paris, France\label{aff90}
\and
CNRS-UCB International Research Laboratory, Centre Pierre Bin\'etruy, IRL2007, CPB-IN2P3, Berkeley, USA\label{aff91}
\and
Institut d'Astrophysique de Paris, 98bis Boulevard Arago, 75014, Paris, France\label{aff92}
\and
Institut d'Astrophysique de Paris, UMR 7095, CNRS, and Sorbonne Universit\'e, 98 bis boulevard Arago, 75014 Paris, France\label{aff93}
\and
Institute of Physics, Laboratory of Astrophysics, Ecole Polytechnique F\'ed\'erale de Lausanne (EPFL), Observatoire de Sauverny, 1290 Versoix, Switzerland\label{aff94}
\and
Aurora Technology for European Space Agency (ESA), Camino bajo del Castillo, s/n, Urbanizacion Villafranca del Castillo, Villanueva de la Ca\~nada, 28692 Madrid, Spain\label{aff95}
\and
Institut de F\'{i}sica d'Altes Energies (IFAE), The Barcelona Institute of Science and Technology, Campus UAB, 08193 Bellaterra (Barcelona), Spain\label{aff96}
\and
European Space Agency/ESTEC, Keplerlaan 1, 2201 AZ Noordwijk, The Netherlands\label{aff97}
\and
DARK, Niels Bohr Institute, University of Copenhagen, Jagtvej 155, 2200 Copenhagen, Denmark\label{aff98}
\and
Waterloo Centre for Astrophysics, University of Waterloo, Waterloo, Ontario N2L 3G1, Canada\label{aff99}
\and
Department of Physics and Astronomy, University of Waterloo, Waterloo, Ontario N2L 3G1, Canada\label{aff100}
\and
Perimeter Institute for Theoretical Physics, Waterloo, Ontario N2L 2Y5, Canada\label{aff101}
\and
Institute of Space Science, Str. Atomistilor, nr. 409 M\u{a}gurele, Ilfov, 077125, Romania\label{aff102}
\and
Consejo Superior de Investigaciones Cientificas, Calle Serrano 117, 28006 Madrid, Spain\label{aff103}
\and
Dipartimento di Fisica e Astronomia "G. Galilei", Universit\`a di Padova, Via Marzolo 8, 35131 Padova, Italy\label{aff104}
\and
Institut f\"ur Theoretische Physik, University of Heidelberg, Philosophenweg 16, 69120 Heidelberg, Germany\label{aff105}
\and
Institut de Recherche en Astrophysique et Plan\'etologie (IRAP), Universit\'e de Toulouse, CNRS, UPS, CNES, 14 Av. Edouard Belin, 31400 Toulouse, France\label{aff106}
\and
Universit\'e St Joseph; Faculty of Sciences, Beirut, Lebanon\label{aff107}
\and
Departamento de F\'isica, FCFM, Universidad de Chile, Blanco Encalada 2008, Santiago, Chile\label{aff108}
\and
Institut d'Estudis Espacials de Catalunya (IEEC),  Edifici RDIT, Campus UPC, 08860 Castelldefels, Barcelona, Spain\label{aff109}
\and
Satlantis, University Science Park, Sede Bld 48940, Leioa-Bilbao, Spain\label{aff110}
\and
Institute of Space Sciences (ICE, CSIC), Campus UAB, Carrer de Can Magrans, s/n, 08193 Barcelona, Spain\label{aff111}
\and
Instituto de Astrof\'isica e Ci\^encias do Espa\c{c}o, Faculdade de Ci\^encias, Universidade de Lisboa, Tapada da Ajuda, 1349-018 Lisboa, Portugal\label{aff112}
\and
Cosmic Dawn Center (DAWN)\label{aff113}
\and
Niels Bohr Institute, University of Copenhagen, Jagtvej 128, 2200 Copenhagen, Denmark\label{aff114}
\and
Universidad Polit\'ecnica de Cartagena, Departamento de Electr\'onica y Tecnolog\'ia de Computadoras,  Plaza del Hospital 1, 30202 Cartagena, Spain\label{aff115}
\and
Centre for Information Technology, University of Groningen, P.O. Box 11044, 9700 CA Groningen, The Netherlands\label{aff116}
\and
INFN-Bologna, Via Irnerio 46, 40126 Bologna, Italy\label{aff117}
\and
Infrared Processing and Analysis Center, California Institute of Technology, Pasadena, CA 91125, USA\label{aff118}
\and
INAF, Istituto di Radioastronomia, Via Piero Gobetti 101, 40129 Bologna, Italy\label{aff119}
\and
ICL, Junia, Universit\'e Catholique de Lille, LITL, 59000 Lille, France\label{aff120}}    

 \abstract{
  We present an analysis of globular clusters (GCs) of dwarf galaxies in the Perseus galaxy cluster to explore the relationship between dwarf galaxy properties and their GCs. Our focus is on GC numbers ($N_{\rm GC}$) and GC half-number radii ($R_{\rm GC}$) around dwarf galaxies, and their relations with host galaxy stellar masses ($M_{\ast}$), central surface brightnesses ($\mu_0$), and effective radii ($R_{\rm e}$). {This work is unique due to its large sample size and the absence of pre-selection based on $\mu_0$ and $R_{\rm GC}$ for dwarf galaxies.} Interestingly, we find that at a given stellar mass, $R_{\rm GC}$ is almost independent of the host galaxy $\mu_0$ and $R_{\rm e}$, while $R_{\rm GC}/R_{\rm e}$ depends on $\mu_0$ and $R_{\rm e}$; lower surface brightness and diffuse dwarf galaxies show $R_{\rm GC}/R_{\rm e}\approx 1$ while higher surface brightness and compact dwarf galaxies show $R_{\rm GC}/R_{\rm e} \approx $\,$1.5$--$2$. This means that for dwarf galaxies of similar stellar mass, the GCs have a similar median extent; however, their distribution is different from the field stars of their host. Additionally, low surface brightness and diffuse dwarf galaxies on average have a higher $N_{\rm GC}$ than high surface brightness and compact dwarf galaxies at any given stellar mass. We also find that UDGs (ultra-diffuse galaxies) and non-UDGs in the sample have similar $R_{\rm GC}$, while UDGs have smaller $R_{\rm GC}/R_{\rm e}$ (typically less than 1) and 3--4 times higher $N_{\rm GC}$ than non-UDGs. Furthermore, examining nucleated versus not-nucleated dwarf galaxies, we find that for $M_{\ast} > 10^{8} \si{M_{\odot}}$, nucleated dwarf galaxies seem to have smaller $R_{\rm GC}$ and $R_{\rm GC}/R_{\rm e}$, with no significant differences seen between their $N_{\rm GC}$, except at $M_{\ast} < 10^{8} \si{M_{\odot}}$ where the nucleated dwarf galaxies tend to have a higher $N_{\rm GC}$. Lastly, we explore the stellar-to-halo mass ratio (SHMR) of dwarf galaxies (halo mass based on $N_{\rm GC}$) and conclude that the Perseus cluster dwarf galaxies follow the expected SHMR at $z=0$ extrapolated down to $M_{\ast} = 10^{6} \si{M_{\odot}}$.}

\titlerunning{GCs of dwarf galaxies in the Perseus cluster}
\authorrunning{Saifollahi et al.}
   
\maketitle

\section{\label{sc:Intro}Introduction}

Current galaxy formation and evolution models do not fully reproduce the properties of dwarf galaxies as observed today (e.g., \citealp{sales}). Dwarf galaxies are low-mass dark matter-dominated objects (\citealp{mcc,eftekhari}), and their evolution and morphological appearance are driven by various internal and external (environmental) processes, along with the physics of dark matter, all of which act simultaneously on these systems. Depending on their initial conditions at the time of formation, dwarf galaxies can follow very diverse evolutionary paths, resulting in the diverse population of dwarf galaxies observed today. This diversity is particularly pronounced in high-density environments, such as those in galaxy clusters. Such environments appear to play a significant role in the formation of extreme dwarf galaxies, namely ultra-compact dwarf galaxies (UCDs, \citealp{Hilker-1999,Drinkwater-1999,Mieske-2013,Liu-2020}) and ultra-diffuse dwarf galaxies (UDGs, \citealp{vd15,francine,iodice23}), with effective radii varying by a factor of 100, spanning the range of a few tens of parsec to several kiloparsec. The latter are dwarf galaxies with low central surface ($\mu_0$) brightness and large effective radius ($R_{\rm e}$), defined as galaxies with $\mu_{0,\,g} > 24$\,mag\,arcsec$^{-2}$ and $R_{\rm e}>1.5$\,kpc (\citealp{vd15}). {UDGs are not a stand-alone category but a subset of low surface brightness dwarf (elliptical/spheroidal) galaxies (\citealp{Conselice2018,Canossa-Gosteinski-2024})}.

In addition to the observed diversity in galaxy size and light distribution, the globular cluster (GC) populations of dwarf galaxies also exhibit considerable diversity. Previous studies of GCs in dwarf galaxies did not reveal significant abnormalities in GC properties (\citealp{Georgiev2009,Liu_etal19_ACSFCS,prole2019,jones2023}). However, some studies indicate that a small fraction of the UDG population host more GCs than dwarf galaxies of comparable luminosity and stellar mass (\citealp{lim2018,lim2020,gannon2021}). The significance of this observed excess of GCs remains uncertain due to inconsistencies in GC number counts ($N_{\rm GC}$) reported in the literature, particularly for DF44 and DFX1 in the Coma cluster (\citealp{vd17,saifollahi2022}) and MATLAS-2019, also known as NGC5648-UDG1 (\citealp{muller21,matlas-danieli}).
In addition to the debate surrounding $N_{\rm GC}$, some studies have aimed to quantify the radial profiles of GCs around dwarf galaxies, including UDGs, by estimating the ratio between the half-number radius of the GC population ($R_{\rm GC}$, the distance from the host galaxy within which half of the total number of GCs are found) and the host galaxy's effective radius ($R_{\rm e}$), expressed as $R_{\rm GC}/R_{\rm e}$. Compared with those focused on GC number counts, fewer studies have examined the constraints on the GC radial distribution. Overall, $R_{\rm GC}/R_{\rm e}$ appears to span a range of 0.7 to 1.5 (\citealp{lim2018,carleton20,saifollahi2022,roman2023,matlas-2024,janssens2024,li2025})\footnote{\citet{lim2018} reports $R_{\rm GC}/R_{\rm e} = 1.5$ for dwarf galaxies based on an analysis of samples from ACS Virgo Cluster Surveys (ACSVCS, \citealp{acsvcs}), though the paper does not provide further details on this analysis.}. 
These ranges of $N_{\rm GC}$ and $R_{\rm GC}/R_{\rm e}$ suggest diversity among the GC properties of dwarf galaxies. However, since most recent studies have concentrated on UDGs, a comprehensive examination of GC properties as a function of host galaxy characteristics and environmental factors remains warranted. Given that dwarf galaxies typically host fewer GCs than massive ones, substantial samples of such objects are essential to improve the statistics for an adequate analysis, highlighting the need for extensive observational efforts.

At distances closer than 5\,Mpc, GCs are bright and large enough in angular size to be spatially resolved from the ground, {using medium class telescopes (2-4\,m)}; otherwise, most are barely or not resolved and tend to appear similar to the foreground stars and background galaxies. Combining deep optical data in the $u$ band with near-infrared data in the $K$ band helps to distinguish GCs from non-GCs (foreground stars and background galaxies), due to their distinct spectral energy distribution (SED) in this wide colour range. This method has proven effective in detecting GCs out to 20\,Mpc distance, such as in the Virgo and Fornax galaxy clusters, respectively at 16\,Mpc and 20\,Mpc (\citealp{Munoz-2014,Powalka2016,cantiello2018,saifollahi2021b}). Observations at high spatial resolution using the \textit{Hubble} Space Telescope (HST) have been an invaluable alternative for studying GCs, both in nearby systems within 10\,Mpc (e.g. \citealp{sharina2005,Georgiev2009}) and in the more remote Virgo and Fornax clusters (\citealp{jordan2004,jordan2015}). At the distances of the latter, GCs with typical half-light radii of 2--3\,pc can appear slightly resolved (larger than point sources), making them marginally distinguishable from unresolved foreground stars and compact background galaxies. HST has also played a crucial role in investigating GCs in more distant galaxy clusters, such as the Perseus cluster at 72\,Mpc (\citealp{harris2020}) and the Coma cluster at 100\,Mpc (\citealp{amorisco2018}). Although GCs appear as point sources in HST images at these distances, HST's high spatial resolution enables the rejection of background galaxies. The Coma cluster currently represents the distance limit of our observational capabilities for studying GCs in dwarf galaxies. Although it is possible to extend the studies further with longer exposures, this approach is inefficient in terms of telescope time. Another caveat is that the small field of view of HST cameras makes it impractical to collect data on large samples of dwarf galaxies. {These limitations also exist for the \textit{James Webb} Space Telescope (JWST), even-though the deep JWST/NIRCAM images would detect GCs at larger distances, up to $z=0.3$ (\citealp{Harris2023,Berkheimer2024})}.

To address these observational limitations, a wide-field of a significant portion of the extragalactic sky survey at space-based spatial resolution would be beneficial. The recently launched ESA \Euclid mission offers such capabilities. \Euclid is particularly powerful for studying dwarf galaxies (\citealp{EuclidSkyOverview}) and is expected to identify over 10\,000 nearby dwarf galaxies during its six-year survey. The deep, high-resolution images provided by \Euclid in optical (VIS, \citealp{EuclidSkyVIS}) combined with near-infrared (NIR/NISP, \citealp{EuclidSkyNISP}) and auxiliary ground-based data in $u$, $g$, $r$, $i$, and $z$, will enable the identification of GCs in dwarf galaxies within the Local Universe. According to \citet{EP-Voggel}, for galaxies at distances smaller than 50\,Mpc, Euclid Wide Survey (EWS) \IE images will detect GCs brighter than the turn-over magnitude (TOM) of the GC luminosity function (GCLF). The GCLF TOM is the peak magnitude of the (nearly) symmetric Gaussian distribution of magnitudes of GCs around galaxies (\citealp{secker,rejkuba2012}) and therefore, detecting GCs brighter than this magnitude corresponds to detecting about half of the total GCs around galaxies. At 20\,Mpc, GC detection in \IE with the Euclid Wide Survey is near to complete. This is also shown in the \Euclid Early Release Observations (ERO) of the Fornax galaxy cluster (\citealp{EROFornaxGCs}). The completeness of GC detection drops when including detections in the NISP bands and, therefore, the overall completeness of GC selection depends on the adopted methodology for a colour-based GC selection.

Given the forthcoming data from the \Euclid mission, it is timely to start investigating the GC properties of dwarf galaxies in detail, as a function of their host galaxies and the environment on a larger scale. The uniform character of \Euclid imaging data with a stable point spread function (PSF) {facilitates} a homogeneous study of GCs across the observed sky. This is particularly important because the current literature often adopts different methodologies and assumptions when studying GCs. Moreover, previous studies have largely focused on UDGs, resulting in a lack of comprehensive analyses of unbiased samples of dwarf galaxies. This uncertainty in GC properties should be addressed in the coming years using data from \Euclid and other wide-field deep extragalactic surveys such as the Legacy Survey of Space and Time (LSST) with the Rubin Observatory, the \textit{Nancy Grace Roman} Space Telescope's High Latitude Wide Area Survey (\citealp{usher-lsst,dage-roman}), {and the Chinese Space Station Telescope (CSST, \citealp{csst})}.

The \Euclid ERO programme (\citealp{EROData}) included a pointing towards the Perseus galaxy cluster (\citealp{EROPerseusOverview,EROPerseusICL}). This galaxy cluster, with a mass of $10^{15} \si{M_{\odot}}$ (\citealp{simionescu2012,aguerri2020}), is similar in mass but closer than the Coma cluster (72\,Mpc versus 100\,Mpc) and represents one of the densest environments in the local cosmic web. The ERO data for the Perseus cluster are deeper than the standard EWS image stacks and reach the GCLF TOM. With more than 1000 dwarf galaxies identified within this data set (\citealp{EROPerseusDGs}), this field is a unique and valuable resource for studies of the GCs of dwarf galaxies. \citet{EROPerseusDGs} conducted a first analysis of the GC counts in that data set. In this paper, using a different technique and source catalogue, we present a detailed analysis and study of GCs in dwarf galaxies within the Perseus cluster, and examine the differences between the GC number and the GC radial distribution in different categories of dwarf galaxies in terms of surface brightness and effective radius. Section \ref{sc:Data} describes the \Euclid data, while Sect.\,\ref{sc:Methods} outlines the methodology used to analyse the data and search for GCs. Section \ref{sc:Results} examines the properties of the GCs and explores trends between dwarf galaxies and their GCs. Our findings are discussed in Sect.\,\ref{sc:discussion}, and Sect.\,\ref{sc:Summary} summarises our results.

\section{\label{sc:Data}Data} 

\begin{figure*}[htbp!]
\begin{center}
\includegraphics[trim={0 0 0 0},angle=0,width=\linewidth]{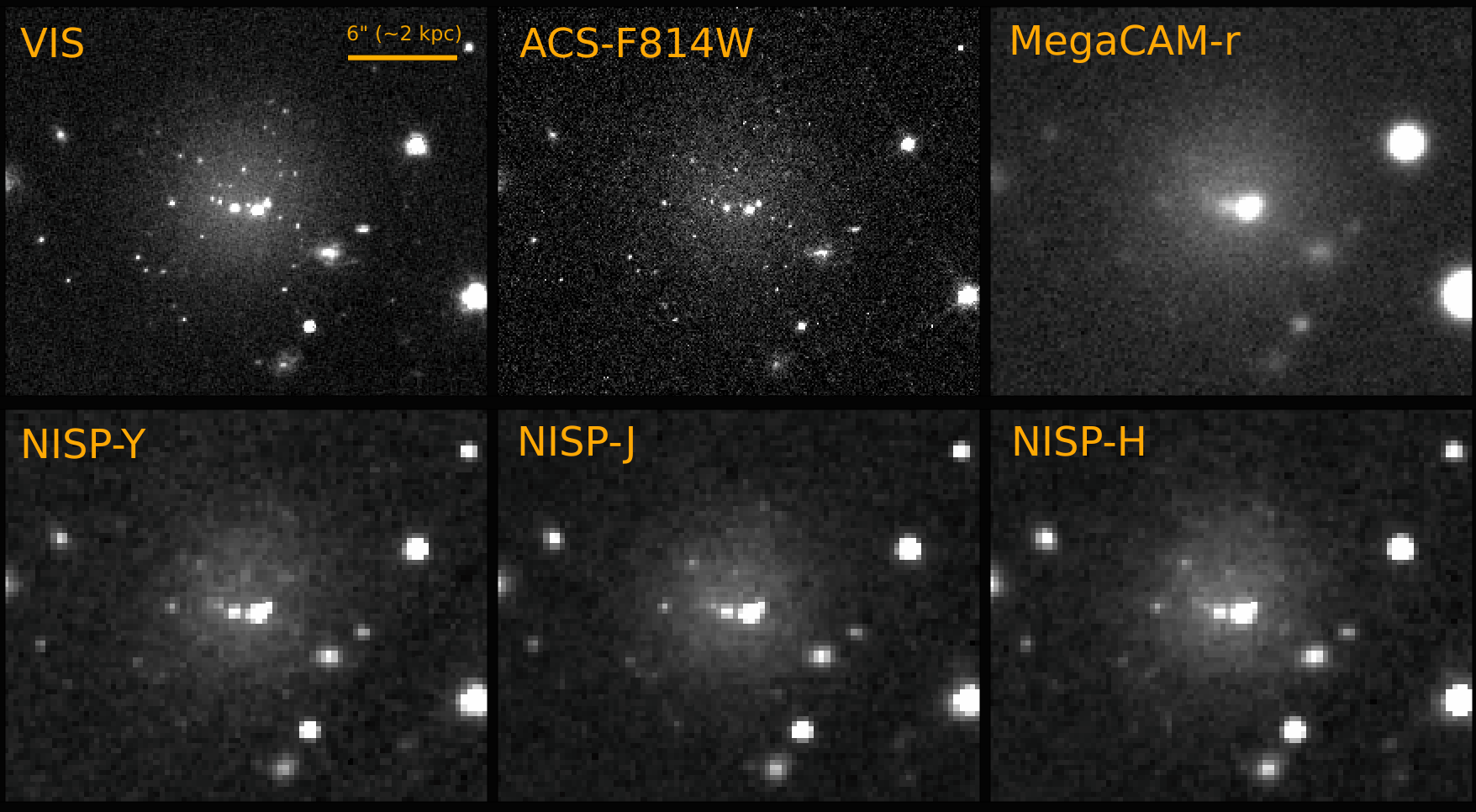}
\end{center}
\caption{A view of a Perseus cluster dwarf galaxy (EDwC-0120 in \citealp{EROPerseusDGs}, and R84 in \citealp{janssens2024}) as seen by the VIS and NISP instruments onboard \Euclid (in \IE, \YE, \JE, and \HE), by HST/ACS (in F814W) and by CFHT/MegaCAM (in $r$). The galaxy seems to host several GCs, which appear as point sources on top of the galaxy, as well as a nuclear star cluster (NSC) in the centre of the diffuse component. The brightest object projected near the centre of the galaxy could be a foreground star, or a massive GC destined to merge with the already existing NSC.}
\label{vis-vs-hst}
\end{figure*}

\begin{figure*}[htbp!]
\begin{center}
\includegraphics[trim={0 0 0 0},angle=0,width=\linewidth]{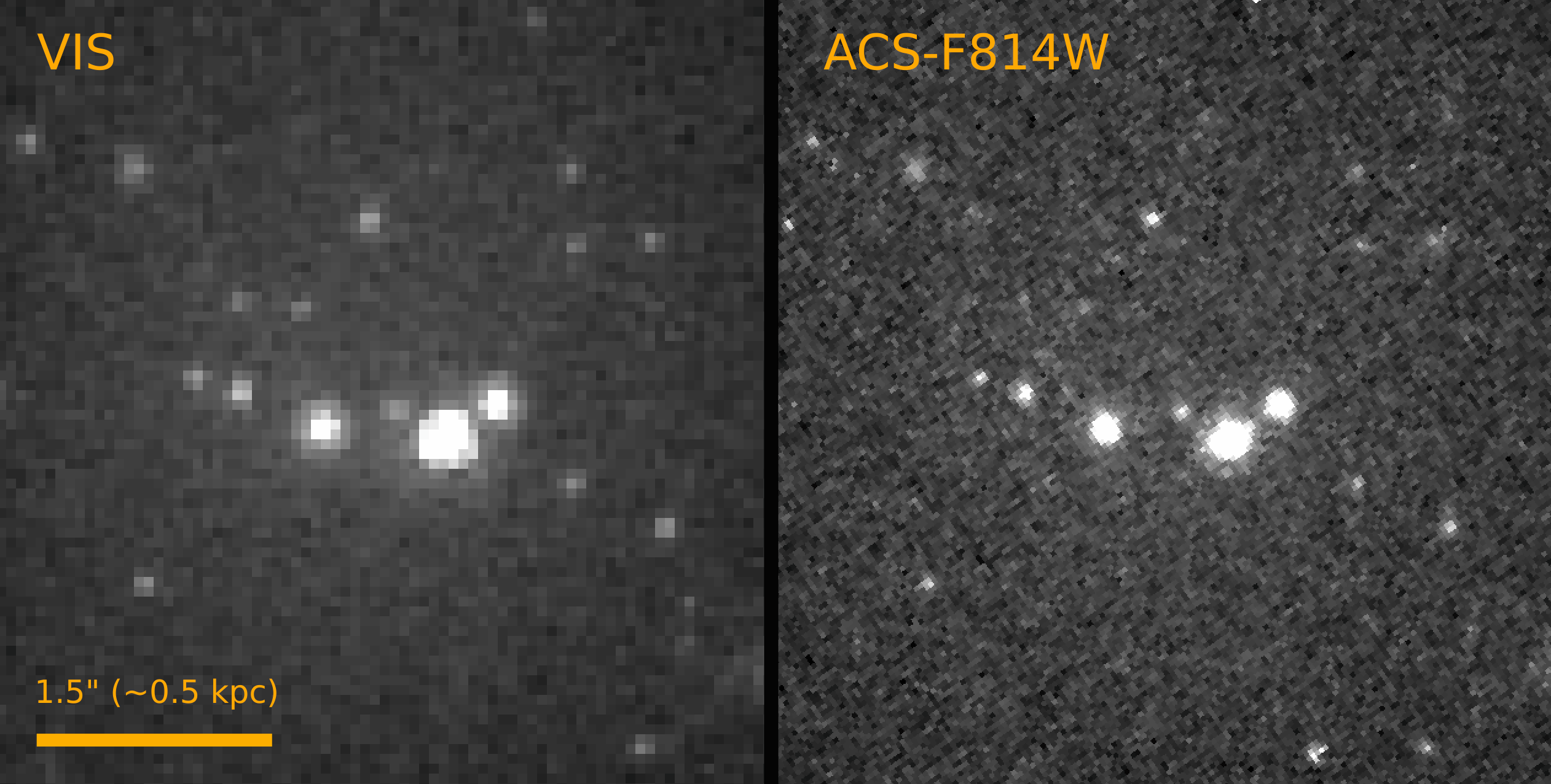}
\end{center}
\caption{A zoomed view of the dwarf galaxy in Fig.\,\ref{vis-vs-hst} as seen by {\Euclid}/VIS in \IE and HST/ACS in F814W (PIPER survey, \citealp{harris2020}). The total integration times for these images are 8960\,s and 2100\,s respectively. The depths of both data sets are comparable while HST/ACS provides a higher spatial resolution with a better sampled PSF than \textit{Euclid}/VIS.}
\label{vis-vs-hst+}
\end{figure*}

\subsection{\label{sc:Data-image}Imaging data}
The \Euclid ERO images (\citealp{EROcite}) of the Perseus galaxy cluster (\citealp{EROPerseusOverview}) cover $0.6\,\deg^2$ of the Perseus cluster in \IE, \YE, \JE, and \HE, with pixel sizes \mbox{\ang{;;0.1}} for \IE and \mbox{\ang{;;0.3}} otherwise. The data and their processing are described in detail in \citet{EROData}. The ERO data set for Perseus consists of images from 16 exposures in each filter, which amounts to four times the Reference Observation Sequence (ROS) that will be executed on each field of the EWS. As a result, the stacked VIS images allow us to reach slightly deeper than the TOM of the GCLF at the 72\,Mpc distance of the Perseus cluster. The canonical TOM is expected at $\IE = 26.3$ (\citealp{EROPerseusDGs}). In Sect.\, \ref{sc:Methods-art} we assess the depth and completeness of this data set in \IE. The VIS data were already used to study the GC numbers of dwarf galaxies (\citealp{EROPerseusDGs}) as well as intracluster GCs (\citealp{EROPerseusICL}). Here we take advantage of the near-infrared (NIR) images, and adopt a slightly different approach in the selection and analysis of GCs.

The Perseus cluster has been observed in various HST programmes, including the PIPER survey (\citealp{harris2020}) which covers about 40 {low surface brightness dwarf galaxies (of which some are also categorized as UDGs by definition).} Figure\, \ref{vis-vs-hst} compares the \IE, \YE, \JE, and \HE images of a dwarf galaxy with HST/ACS observations of that survey in filter F814W, as well as with ground-based CFHT/MegaCAM data in the $r$ band. Figure\,\ref{vis-vs-hst+} shows a zoomed view of the central part of the dwarf galaxy taken by {\Euclid}/VIS in \IE and HST/ACS in F814W. \Euclid ERO and HST images are comparable in depth, while HST images have while HST images have a tighter PSF and better PSF sampling. The \Euclid ERO data have such quality for more than 1000 dwarf galaxies in the Perseus cluster. 

\subsection{\label{sc:Data-dwarf-cat}Dwarf galaxy catalogue}
Using the ERO data for the Perseus cluster, \citet{EROPerseusDGs} have identified more than 1000 dwarf galaxies {(with a majority to be early-type)} in a stellar mass range of $10^5$--$10^{9.5} \si{M_{\odot}}$, with surface brightnesses between 18 and $26\, {\rm mag \, arcsec^{-2}}$ and effective radii between 0.5 and 6\,kpc in \IE. Among many structural parameters, the Perseus cluster dwarf catalogue (\citealp{EROPerseusDGs}) contains the effective radius ($R_{\rm e}$), the central surface brightness ($\mu_0$), and the stellar mass ($M_{\ast}$) of Perseus cluster dwarf galaxies. These parameters are used frequently for the analysis in this paper. We examined all the dwarf galaxies and their parameters, and excluded from the analysis the dwarf galaxies whose parameters seem unreliable (e.g. affected by a bright star). The final sample contains about 870 dwarf galaxies, {all dwarf early-types, of which 86 are also categorized as UDGs} (defined as dwarf galaxies with central surface brightness $\mu_{0,\,\IE} > 23$\,mag\,arcsec$^{-2}$ and effective radius $R_{\rm e}>1.5$\,kpc). 

Overall, this data set contains a diverse population of {early-type dwarf galaxies} in terms of stellar mass, surface brightness, and effective radius. The absence of strong bias in size or surface brightness, when compared for instance to samples of UDGs, makes it ideal for studying the GC properties of {early-type dwarf galaxies} in a global way. The results presented in this work represent GCs of dwarf galaxies in dense cluster-like environments considering that all the dwarf galaxies are highly likely members of the Perseus cluster.

\section{\label{sc:Methods}Methods}

This section outlines the methodology for source detection, photometry, and identification of GC candidates around dwarf galaxies. Our approach is similar to that of \citet{EROFornaxGCs}, with some differences in GC selection criteria and enhancements to address flux-dependent uncertainties in colour and the source's ellipticity for GC selection. The pipeline used for this analysis, \texttt{GCEx}, is publicly available\footnote{ \url{https://github.com/teymursaif/GCEx}}. \texttt{GCEx} uses several astronomical packages and \texttt{Python} libraries and is developed specifically to analyse images and identify GCs in wide-field multi-wavelength data for large samples of galaxies. The methodology described below refers to the methodology adopted in \texttt{GCEx}.

\subsection{\label{sc:Methods-psf}Point-spread function (PSF)}

Empirical PSF models are necessary in several instances throughout this analysis. Therefore, in the first step, we produced PSF models in each filter. The \Euclid PSF is known to vary with the location of the image in the focal plane and the colour of the sources; however, such variations are minimal (\citealp{EROData,Q1-TP002}). For this analysis, we ignore these variations and use a single PSF model per filter. Here, PSF models are created by stacking 1000 cutouts of bright, non-saturated stars with magnitudes between 19 and 21 for $\IE$ and 17 and 19 for $\YE$, $\JE$, and $\HE$. Cutout sizes are 10\,arcsec in all the bands. The cutouts are oversampled by a factor of 10 using \texttt{SWarp} (\citealp{swarp}) with the \texttt{LANCZOS3} interpolation kernel, resulting in a PSF model with a pixel size 10 times smaller than the instrumental pixel size. Once the cutouts are created, the centroid is estimated, and the cutouts are aligned and stacked using \texttt{SWarp}. The final PSF models have a full-width half-maximum (FWHM) of \mbox{\ang{;;0.18}}, \mbox{\ang{;;0.52}}, \mbox{\ang{;;0.54}}, and \mbox{\ang{;;0.58}} in \IE, \YE, \JE, and \HE, respectively. These FWHM values are larger than the FWHM of single exposures before stacking (\mbox{\ang{;;0.16}} for \IE) mainly due to interpolation effects during resampling to create the final stacked frames (\citealp{EROData}). Our analysis of the PSF indicates that resampling primarily affects the very central region of the PSF model (within 0.1\,arcsec).

\subsection{\label{sc:Methods-photom}Source detection and photometry}

Source detection is carried out on $240\arcsec \times 240\arcsec$ ($85\,{\rm kpc} \times 85\,{\rm kpc}$) \IE cutouts of dwarf galaxies. For this, initially we perform unsharp-masking of the cutouts using a small background mesh size (\texttt{BACKPSIZE=8}) to subtract the extended light of galaxies, thus improving the detection of fainter sources and GC candidates around galaxies. Then, we use \texttt{SExtractor} for source detection. Once sources are detected, we apply an initial filtering based on an estimate of their magnitudes (\texttt{SExtractor} output parameter \texttt{MAG\_AUTO}), their FWHM (\texttt{FWHM\_IMAGE}), and their ellipticity (\texttt{ELLIPTICITY}) to remove bright and saturated sources (brighter than 19 in \IE), very faint sources (fainter than 28.5 in \IE), extended sources (\texttt{FWHM\_IMAGE} $> 20$\,pixels), very compact sources (\texttt{FWHM\_IMAGE} $< 1$\,pixel), and highly elongated sources (\texttt{ELLIPTICITY} $ > 0.9$). This filtering effectively cleans the catalogue of unwanted sources, such as large objects and artefacts, and reduces the size of the source catalogue for the next steps of the analysis. 

After the initial filtering, close to 500\,000 sources remain in the detection catalogue. For these detections, we then perform aperture photometry using the \texttt{photutils} library in \texttt{Python}, employing an aperture with a diameter of 1.5 times the average FWHM of the data in the corresponding band (single value across the data). This aperture diameter corresponds to \mbox{\ang{;;0.27}, \ang{;;0.78}, \ang{;;0.81}, and \ang{;;0.87}} in \IE, \YE, \JE, and \HE, respectively. The aperture photometry values are corrected for the fraction of the missing flux beyond the aperture using the previously created PSF models. This fraction is about 20\% and 10\%, respectively, in the VIS and NISP bands, estimated from the PSF models out to a radius of $5\arcsec$. The fraction of the light beyond $5\arcsec$ is less than 1\% (\citealp{EROData}). {Furthermore, the PSF FWHM varies between 1.5 and 1.9\,pixels in the field-of-view. Our analysis of the colours of bright point sources shows that the effect of this variation on the photometry and the measured colours is negligible.}

At the 72 Mpc distance of the Perseus cluster, and at the \Euclid spatial resolution, GCs appear as point sources. In this case, aperture photometry as described here properly measures the total flux of GCs, while it underestimates the total flux of more extended sources (more extended than the PSF). However, this is not a concern because these extended sources will be removed during the GC identification step. Furthermore, we measure the compactness index of the sources, $C_{2-4}$, defined as the difference between the magnitudes measured within aperture diameters of 2 and 4 pixels. This compactness index is used in subsequent steps to identify point sources in the data, which include the GCs as well as some faint foreground (halo) stars (see Fig.\,C.2 in \citealp{EROData}). 

\subsection{\label{sc:Methods-art}Artificial GCs and detection completeness}

We perform artificial GC injection into the data to assess the completeness of source detection and then GC selection. We generate 500 artificial GCs per dwarf galaxy, randomly (uniformly) distributed within their cutout in all the bands. The artificial GCs are created using a King profile (\citealp{king}) with a half-light radius between 1 and 5\,pc and an absolute \IE between $-12$ and $-5$, covering the full magnitude range of GCs, convolved with the PSF model of a given filter (over-sampled by a factor of 10). Before injection into the data, we re-bin the artificial GCs to the pixel scale of the images with a random sub-pixel shift in the position of their centre using \texttt{SWarp}. We assume an average colour of 0.45 for $\IE-\YE$, and 0 for $\YE-\JE$, and $\JE-\HE$. These average values are chosen considering the average GC colour in the ERO observations of the Fornax cluster (\citealp{EROFornaxGCs}) as well as synthetic photometry of single stellar population (SSP) models for old and metal-poor GCs (see Appendix\,A in \citealp{EROFornaxGCs}), {considering that the GCs of dwarf galaxies are mostly metal-poor (\citealp{forbes2024-gcs})}. Finally, we add noise to the artificial GCs before injecting them into the data. The added noise is three times larger than the pixel Poisson noise to take into account the correlated noise in the data, which we have estimated to be about 2--3 times the Poisson noise. 

\begin{figure}[htbp!]
\begin{center}
\includegraphics[trim={0 0 0 0},angle=0,width=\linewidth]{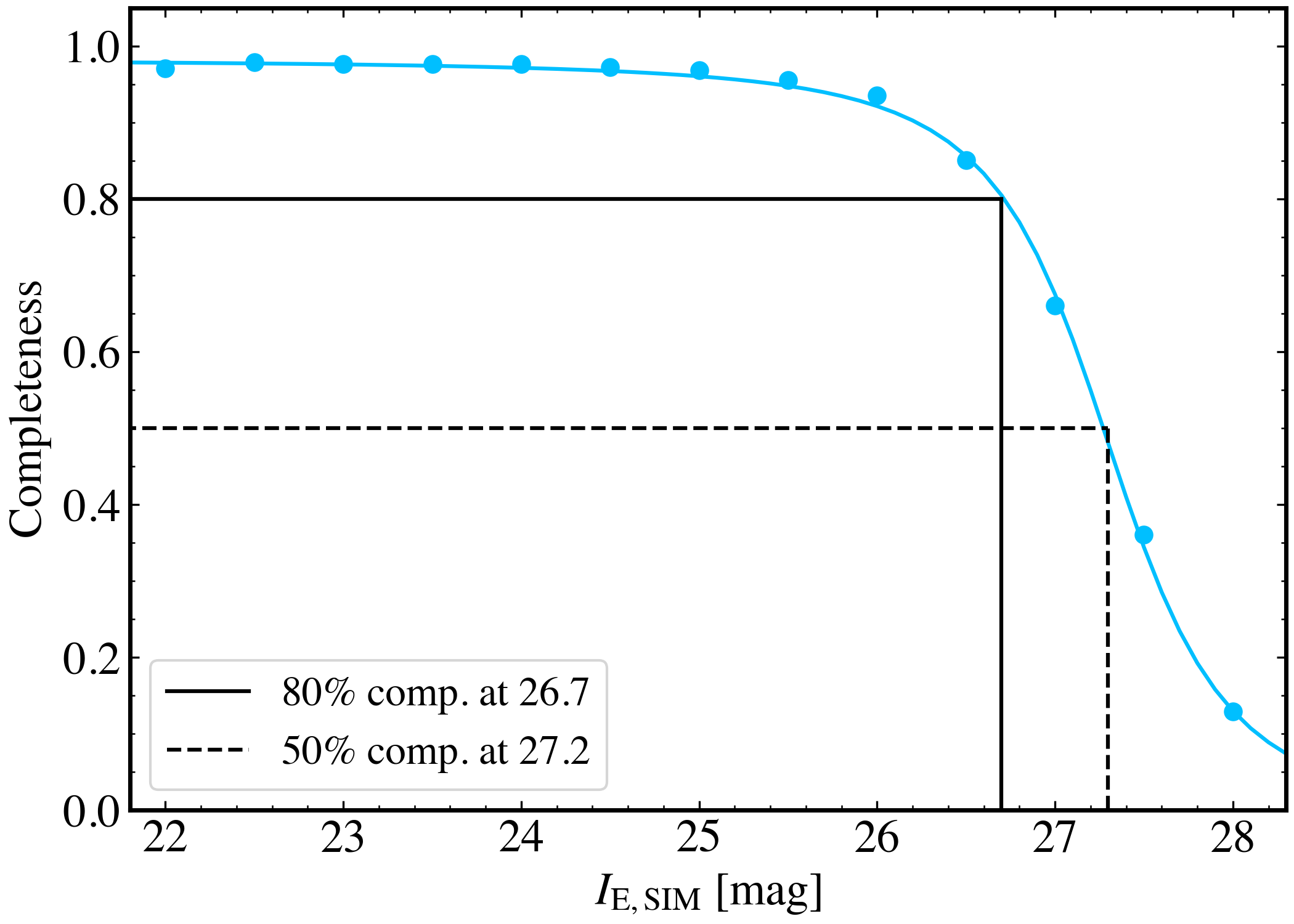}
\end{center}
\caption{Completeness of the performed source detection in \IE based on artificial GC tests. Our assessment shows that the GC detection is 80\% complete at $\IE=26.7$ (without considering the extinction). This magnitude is about the expected location of GCLF TOM for dwarf galaxies {at the distance of the Perseus cluster (72\,Mpc)} (taking into account the average Galactic foreground dust extinction $A_{\IE}=0.2$\,mag).}
\label{completeness}
\end{figure}

On the images thus modified, we carry out source detection and photometry as described above. Then, we examine the completeness of the source detection in \IE. The result is presented in Fig.\,\ref{completeness}. GC detection is almost complete to $\IE=26$, and starts to drop after this magnitude, with a completeness of 80\% and 50\% achieved respectively at $\IE = 26.7$ and $\IE = 27.2$. These values were estimated by fitting a modified version of the interpolation function of \citet{fleming}. Before accounting for the Galactic foreground extinction along the line of sight, the canonical TOM of the GCLF is expected at 26.3, while for dwarf galaxies it will be 0.3\,mag fainter, at 26.6 (\citealp{Liu_etal19_ACSFCS}). Considering that the extinction, $A_{\IE}$, across the field-of-view ranges from 0.1 to 0.3 mag in \IE (\citealp{EROPerseusDGs,EROPerseusICL}) with an average value of about 0.2 mag, we expect $\IE \approx 26.8$ for the GCLF TOM of dwarf galaxies {at the distance of the Perseus cluster (72\,Mpc)}, about the magnitude corresponding to 80\% detection completeness. This assessment is based on the average extinction correction, and in practice, this correction is different for every dwarf galaxy. We will take such differences into account when analysing GC properties of dwarf galaxies.

This assessment shows that overall, we detect most GCs brighter than the GCLF TOM. The fraction of missing GCs depends on the exact shape of the GCLF of dwarf galaxies. Assuming a Gaussian GCLF with TOM $\IE=26.6$ and width $\sigma=1$\,mag (\citealp{villegas2010}), and an average extinction of 0.2\,mag, we estimate that 90\% of the GCs are detected. This fraction changes to 85\% if we assume a narrower GCLF for dwarf galaxies with $\sigma=0.5$. 

Given the outcome of the completeness assessment, the data set allows us to study GC numbers (albeit with some assumptions on the shape of the GCLF, described later in Sect.\,\ref{sc:Results-numbers}) as well as radial profiles of GC populations. However, deeper data are required to study the shape of the GCLF of dwarf galaxies in detail, for example its behaviour as a function of the stellar mass of galaxies. Therefore, in this paper, we do not make an assessment of the GCLF. Such studies can be done with \Euclid with dwarf galaxies at distances below 20\,Mpc, where we measure the entire GCLF in \IE.

Considering the shallower depth (about 2\,mag) and the lower detection completeness of the near-infrared images (\YE, \JE, and \HE) compared to \IE (Fig.\,\ref{completeness-nisp} in Appendix\,\ref{app-comp-nisp}), in this work we do not limit our detection catalogue by requesting a measurement in any of those bands but instead consider every object detected in \IE. However, if an object is detected in one of the three near-infrared filters, we use the flux and estimated magnitude to measure its colour and apply flux-dependent colour cuts. This procedure is described in Sect.\,\ref{sc:Methods-gc-det}.

\begin{figure*}[htbp!]
\begin{center}
{\large \bf {GC selection criteria for compactness index}}
\\[6pt]
\includegraphics[trim={0 0 0 0},angle=0,width=\linewidth]{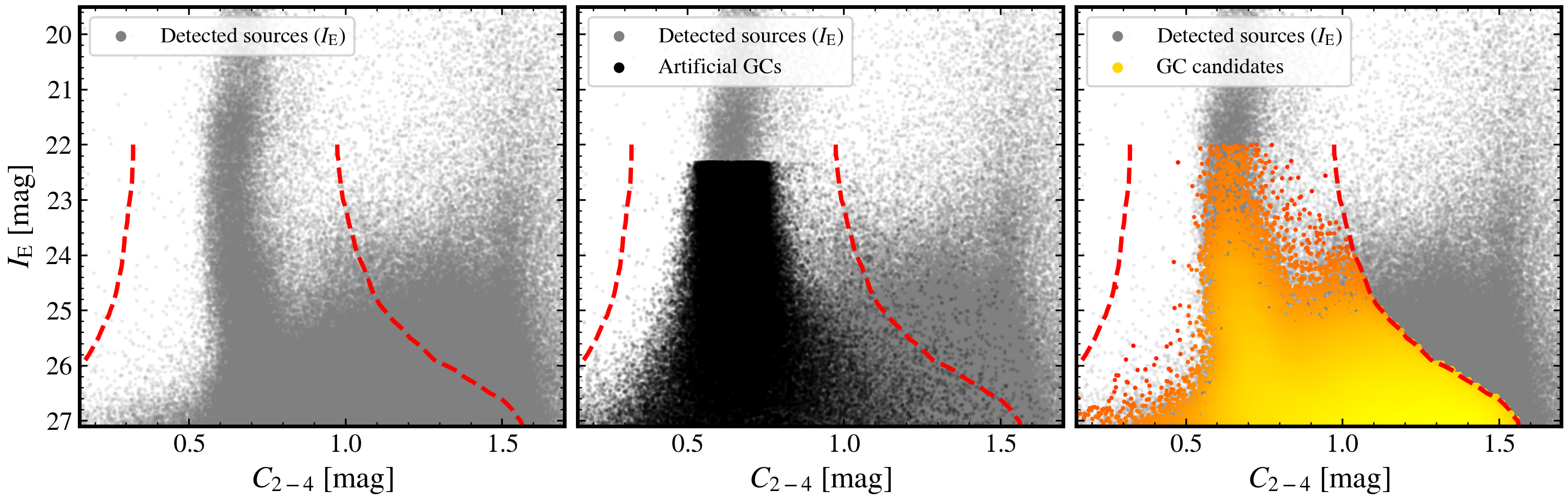}
\\[9pt]
{\large \bf {GC selection criteria for ellipticity}}
\\[6pt]
\includegraphics[trim={0 0 0 0},angle=0,width=\linewidth]{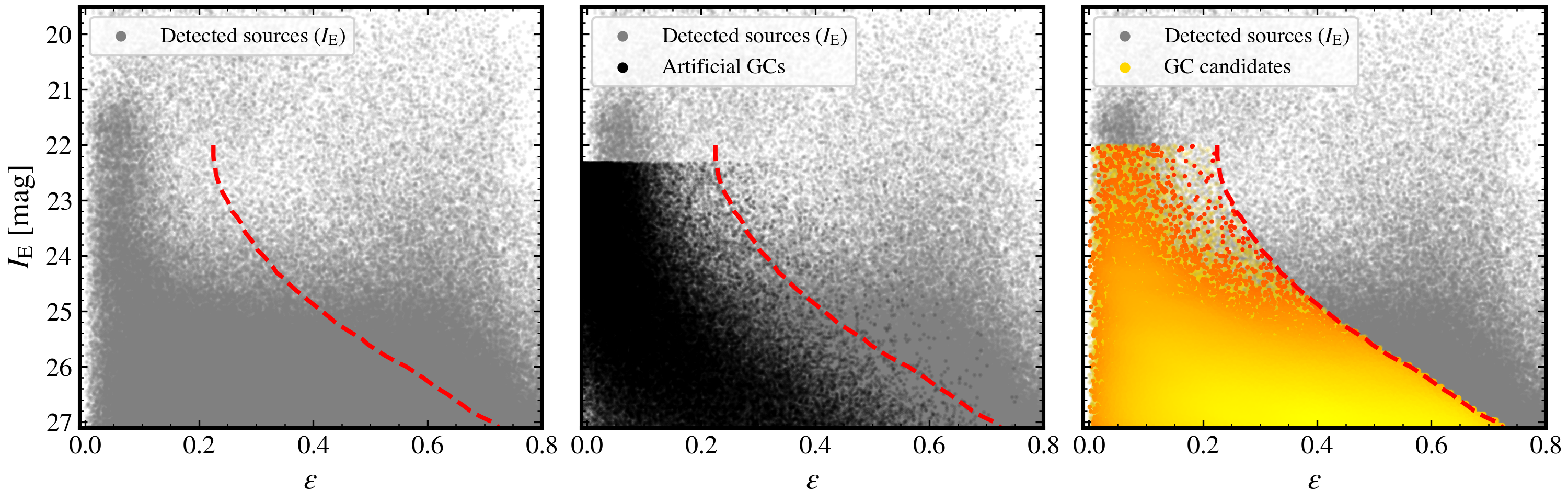}
\\[9pt]
{\large \bf {GC selection criteria for $\IE-\YE$ colour}}
\\[6pt]
\includegraphics[trim={0 0 0 0},angle=0,width=\linewidth]{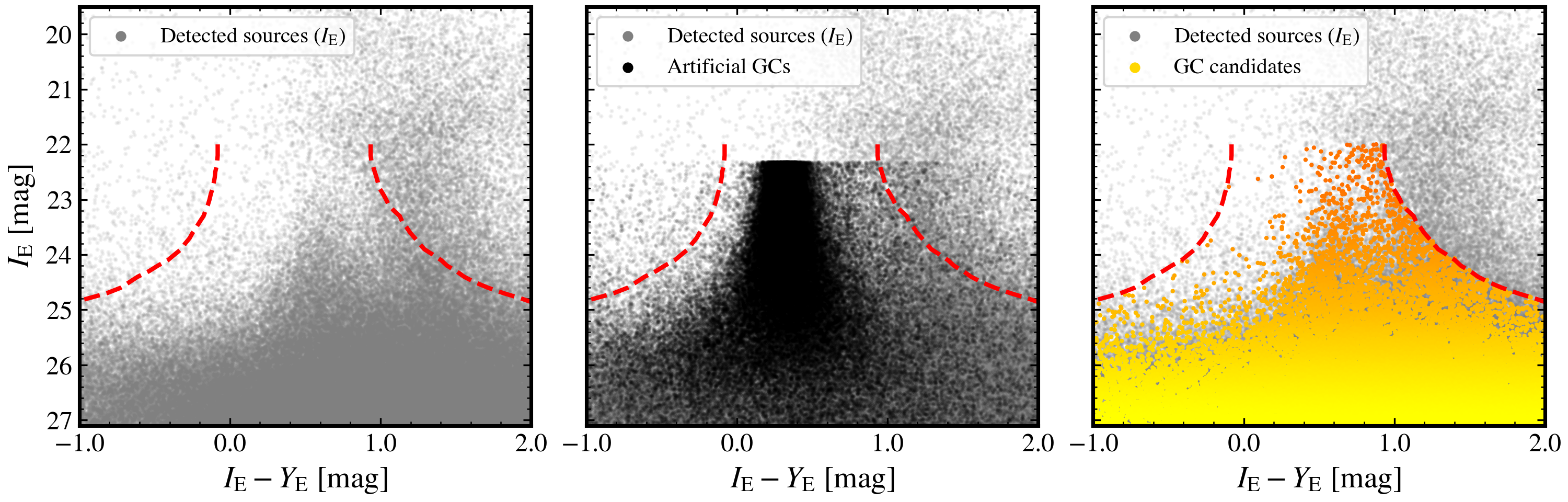}
\end{center}
\caption{Photometric criteria applied for GC identification, aiming for high completeness. The three rows demonstrate these criteria for compactness index ($C_{2-4}$), ellipticity ($\varepsilon$), and $\IE-\YE$ colour (as an example). In each row, we show the real-sky sources detected in \Euclid \IE images (grey points), the detected artificial GCs (black points), and the real-sky sources selected as GC candidates (yellow points). The dashed red curves represent the selection boundaries adopted based on the distribution of artificial GCs in each parameter space. In the top row, we see that the artificial GCs follow the vertical sequence of point sources (with an average $C_{2-4}=0.7$), and the scatter becomes larger with fainter magnitude due to an increase in photometric uncertainties. In the bottom row, we see that the selection boundaries reject the majority of objects with $\IE<24$ and $\IE-\YE>1$, which are foreground stars. However, a fraction of those stars end up in the final catalogue. This contamination is unavoidable considering the expected colour range of GCs. Tightening the selection criteria removes these contaminant stars, but leads to lower completeness.}
\label{compactness}
\end{figure*}

\begin{figure*}[htbp!]
\begin{center}
\textbf{High central surface brightness dwarf galaxies}
\\[5pt]
\includegraphics[trim={0 0 0 0},angle=0,width=0.24\linewidth]{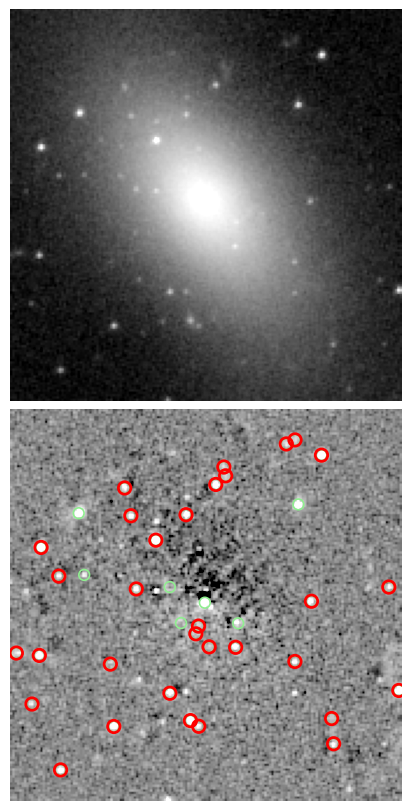}
\includegraphics[trim={0 0 0 0},angle=0,width=0.24\linewidth]{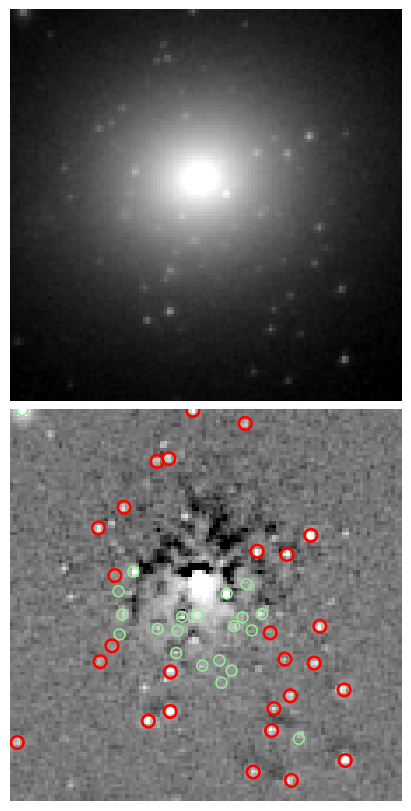}
\includegraphics[trim={0 0 0 0},angle=0,width=0.24\linewidth]{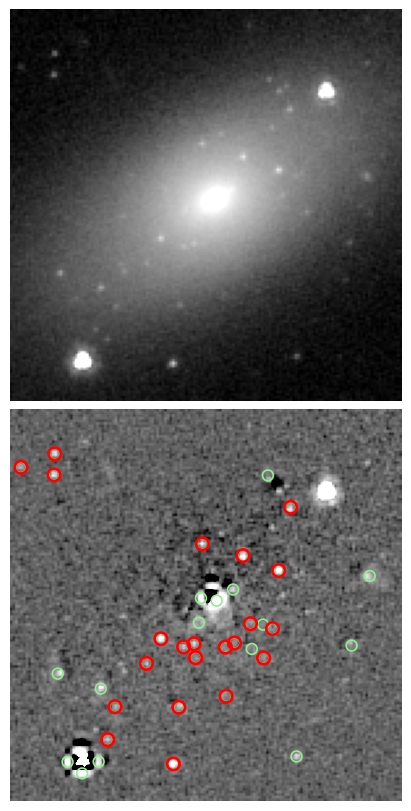}
\includegraphics[trim={0 0 0 0},angle=0,width=0.24\linewidth]{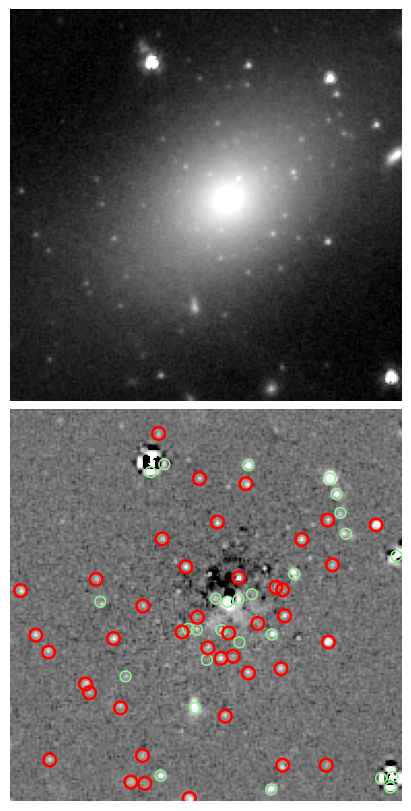}
\\[8pt]
\textbf{Low central surface brightness dwarf galaxies}
\\[5pt]
\includegraphics[trim={0 0 0 0},angle=0,width=0.24\linewidth]{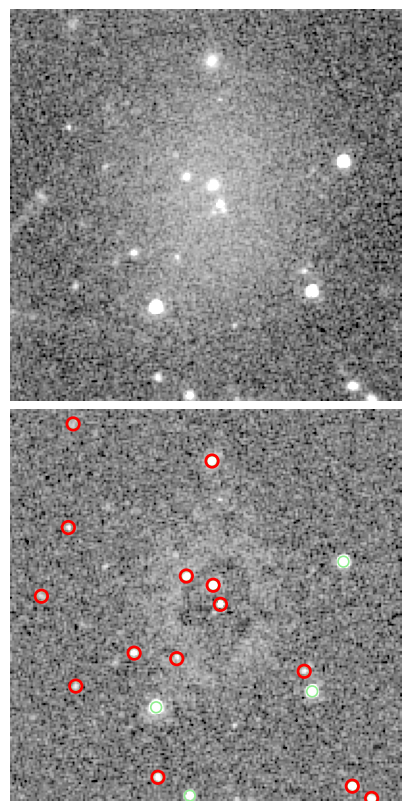}
\includegraphics[trim={0 0 0 0},angle=0,width=0.24\linewidth]{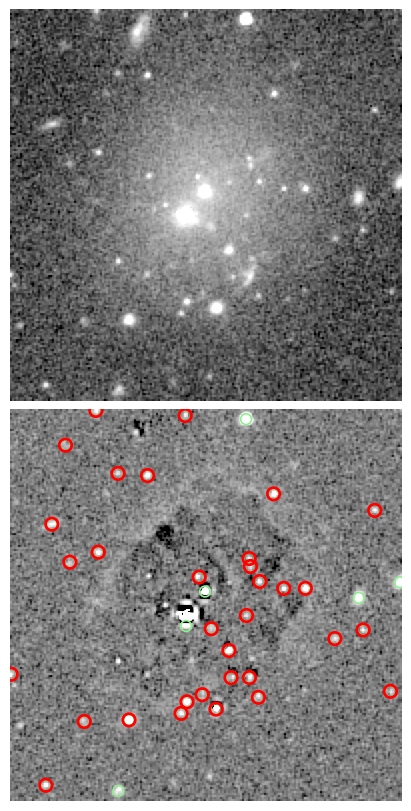}
\includegraphics[trim={0 0 0 0},angle=0,width=0.24\linewidth]{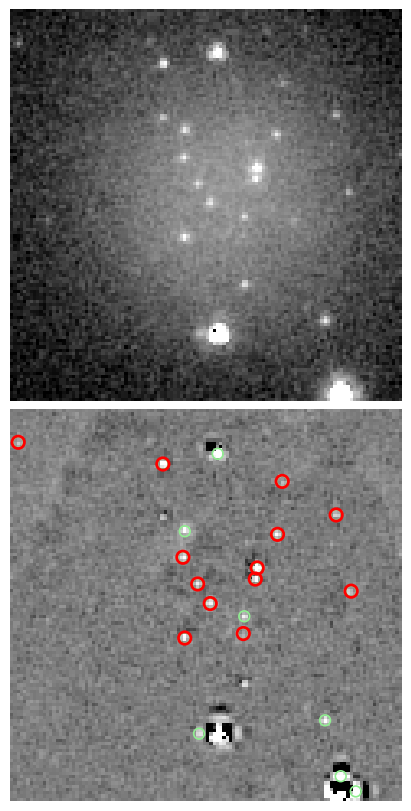}
\includegraphics[trim={0 0 0 0},angle=0,width=0.24\linewidth]{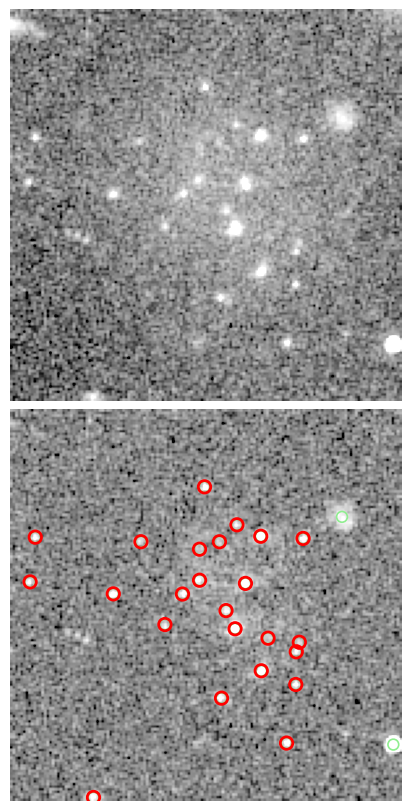}
\end{center}
\caption{Source detection and GC candidate selection for eight of the dwarf galaxies in this work. The figure shows galaxies with high and low central brightness respectively in the top and bottom half. For each dwarf galaxy, it includes a small \IE cutout with sides of $4R_{\rm e}$ and the galaxy-subtracted version of this cutout. {Galaxy-subtraction has been done by unsharp-masking of the cutouts using a small background mesh size (8 $\times$ 8\,pixels), using \textsc{SExtractor}.} Among the detected sources, GC candidates are highlighted in red; other sources (in green) did not match the selection criteria.}
\label{dwarfs-and-gcs-vis}
\end{figure*}

\subsection{\label{sc:Methods-gc-det}GC (candidate) selection around dwarf galaxies}

We used the artificial GCs injected into the data (all filters) to examine the expected range of compactness, colour, and apparent ellipticity of real GCs as a function of their magnitude and to define the GC selection criteria.
{Those sources that are selected through this procedure are considered as GC candidates and are used in this work to study the GC properties of dwarf galaxies. The GC (candidate) selection procedure} is demonstrated in Fig.\,\ref{compactness}, where the top, middle, and bottom panels show the selection based on the compactness index ($C_{2-4}$), ellipticity ($\varepsilon$), and colour (for example, $\IE-\YE$). Sources with \IE magnitude between 22 and 27 that meet all these criteria are selected as GC candidates. The cuts in \IE correspond to the brightest magnitude at which we expect to find typical GCs ($\IE=22$) and the magnitude at which the completeness of GC detection is about 50\% ($\IE=27.0$). For studying GC numbers and radial distributions, we only consider GCs brighter than the GCLF TOM, which, considering extinction, is always brighter than $\IE=27$. Therefore, our analysis does not rely on GCs fainter than this magnitude.

The boundaries of GC selection in each parameter space are shown with red curves in Fig.\,\ref{compactness}. These boundaries are chosen to include 98\% of the artificial GCs at a given magnitude. We initially find an offset of about 0.1\,mag between point sources and artificial GCs in $C_{2-4}$. This is due to the effects of resampling while using undersampled data to model the PSF. Therefore, we adjust the $C_{2-4}$ values of the artificial GCs by 0.1\,mag to the lower $C_{2-4}$ values. The point of the experiment with artificial GCs is to model the range of parameters within which measurement noise, and systematic offsets, such as the one for $C_{2-4}$, remain unaffected by a constant shift of the compactness index values. Additionally, to include other (unexpected and random) effects in $C_{2-4}$ of real GCs, we extend the selection boundaries (red curves) by 0.2\,mag on both sides. We also extend the boundaries of colour criteria for GC selection (red curves in the bottom row of Fig.\,\ref{compactness}) to cover the full range of expected GC colours. This is necessary considering that the artificial GCs are produced using one value (average colour) of GCs. We extend the range of colour selection by 0.3 for $\IE-\YE$ and 0.15 for $\YE-\JE$ and $\JE-\HE$. For more details on the selection of these values, we refer the reader to Appendix\,A in \citet{EROFornaxGCs}.

Figure\,\ref{dwarfs-and-gcs-vis} shows a few examples of dwarf galaxies and their identified GC candidates brighter than $\IE=27$ (in red), as well as all other detected sources brighter than $\IE=27$ that did not meet the GC selection criteria (in green). Figure\,\ref{dwarfs-and-gcs-vis} is divided into two parts: dwarf galaxies with high and low central surface brightness are shown at the top and bottom panels, respectively. Upon inspection, we see that source detection performs equally well across dwarf galaxies, GC identification can miss some GCs in central regions of high central surface brightness dwarf galaxies. This is due to residuals from unsharp masking (used to subtract the galaxy) in the centre, which can affect photometry and measurements regarding the most central GCs. As a consequence, these objects do not meet one or more GC selection criteria. This mostly limits GC detection in the central 500\,pc of the brightest (and most massive) dwarf galaxies. We consider this point when studying the radial distribution of GCs and show that it has a negligible overall effect. However, this could lead to lower GC numbers for the dwarf galaxies with highest central surface brightness. Considering that these objects are the ones with most GCs, we expect that this bias in GC numbers is not significant, but must be taken into account when interpreting the results. In contrast, GC identification in low central surface brightness dwarf galaxies does not have any limitations in the centre of their host galaxies.

Before finalising the GC catalogues, for cutouts of dwarf galaxies, we produced a mask frame where nearby galaxies (both dwarf and massive) and bright stars are masked so that detections corresponding to these objects are ignored. Within some of the cutouts, edges of massive galaxies have also been masked. The motivation for such masking is to clean the background from GC over-densities around other galaxies and any false detections around bright stars. Such objects, if not masked, will artificially increase the source count around galaxies and in the background (areas far from dwarf galaxies). Additionally, given extinction corrections ($A_{\IE}$) in \citet{EROPerseusDGs}, we correct all the magnitudes in four filters for extinction. In Sect.\,\ref{sc:Results} we use the masked and extinction-corrected GC catalogues to study the GC properties of dwarf galaxies.

\section{\label{sc:Results}Results}

The \Euclid ERO data of the Perseus cluster allow us to reach the GCLF TOM expected for dwarf galaxies (at $\IE=26.6$, without applying any extinction correction) with an overall estimated completeness of 85\% down to the TOM. With such a data set, we study the {GC total numbers} and their radial distribution after selecting GC candidates around dwarf galaxies. We begin by examining the GC radial distribution and the $R_{\rm GC}$ values around the Perseus sample. Using our results, we subsequently examine the total number of GCs, taking into account the differences in $R_{\rm GC}$ between galaxies. These results\footnote{The output of the analysis and associated catalogues will be available on the Strasbourg astronomical Data Center (CDS) database.} are presented in Sects.\,\ref{sc:Results-radial-profile} and \ref{sc:Results-numbers}. We study the average GC properties of the dwarf galaxies. Later, for the dwarf galaxies with more than 10 GCs, we examine the $R_{\rm GC}$ values for individual cases. As mentioned before, the data set is not deep enough to conduct a comprehensive analysis of the GCLF of dwarf galaxies itself, for example, to study the GCLF TOM and width as a function of the stellar mass of the host galaxies. Additionally, the estimated colours in this work, while suitable for GC selection using flux-dependent colour cuts, are not ideal for studying the colours of individual GCs given their large uncertainties. Therefore, in this paper, we omit the study of the GCLF and of GC colours.

\subsection{Dividing the sample of dwarf galaxies}

\begin{figure}[htbp!]
\begin{center}
\includegraphics[trim={0 0 0 0},angle=0,width=\linewidth]{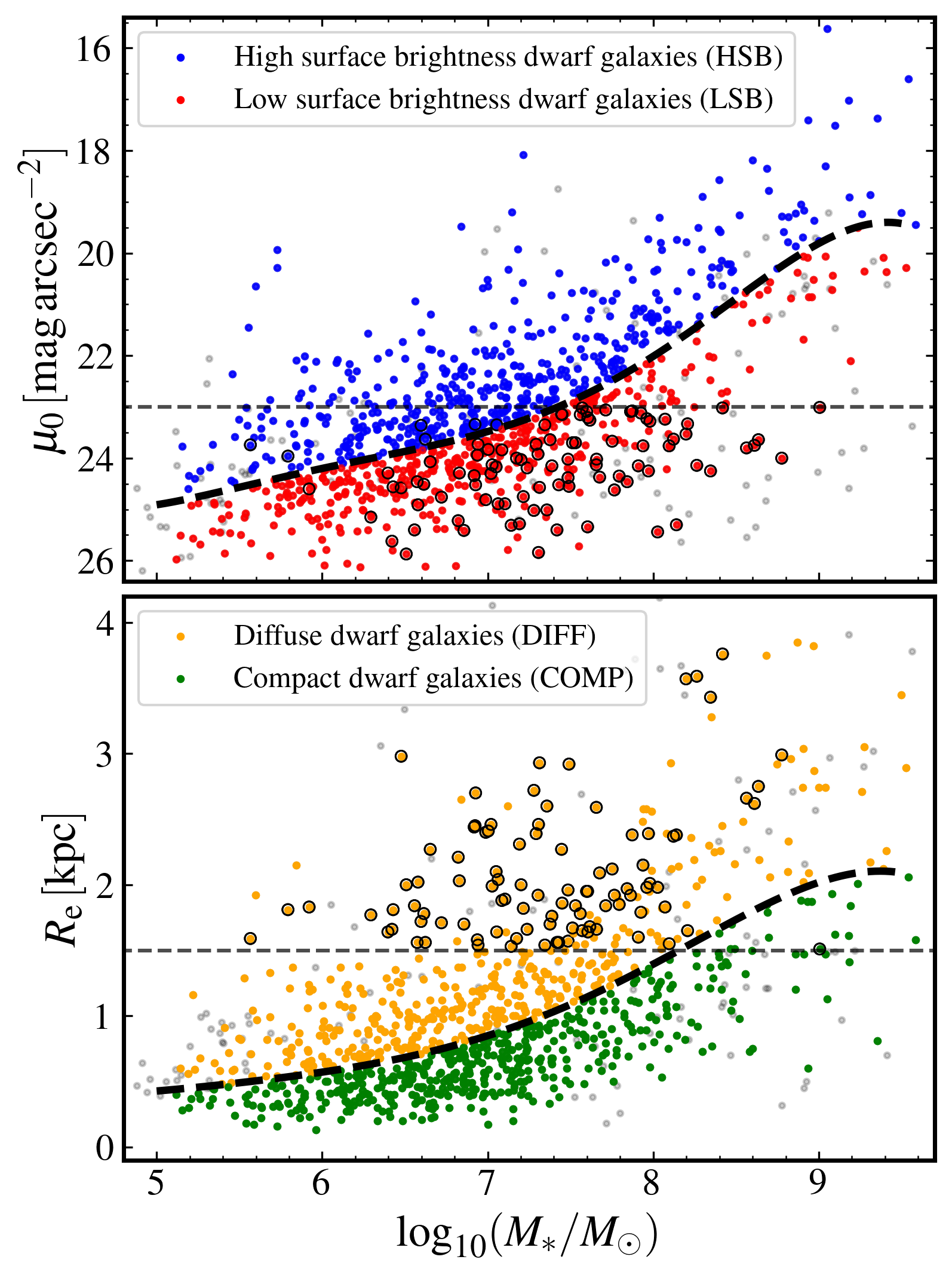}
\end{center}
\caption{Division of the dwarf galaxy sample into high surface brightness dwarf galaxies (HSB, blue points), low surface brightness dwarf galaxies (LSB, red points), compact dwarf galaxies (COMP, green points), and diffuse dwarf galaxies (DIFF, yellow points) based on their central surface brightness ($\mu_0$) and effective radius ($R_{\rm e}$). A distance of 72\,Mpc was assumed to convert angular to physical scale. This division is stellarmass dependent and is indicated with the black dashed curve. Coloured points represent the dwarf galaxies that have been used for the analysis in this paper. Those dwarf galaxies that also satisfy the UDG criteria in $\mu_{0, \IE}$ and $R_{\rm e}$ {($\mu_{0, \IE}$<23\,mag\,arcsec$^{-2}$ and $R_{\rm e}$>1.5\,kpc, shown with dashed horizontal lines)} are indicated with the black circles around the coloured circles. Grey points show those dwarf galaxies that have been removed from the sample after initial assessment (as was mentioned in Sect.\,\ref{sc:Data-dwarf-cat}).}
\label{dwarfs-division}
\end{figure}

Dwarf galaxies at a given stellar mass display a diversity of surface brightnesses and effective radii. Here, we further study trends between the GC populations and the properties of their host galaxy. We do this by dividing our sample of dwarf galaxies into two groups in each of two parameter spaces: the planes of central surface brightness ($\mu_0$) vs. stellar mass ($M_{\ast}$), and of effective radius ($R_{\rm e}$) vs. stellar mass ($M_{\ast}$). Figure\,\ref{dwarfs-division} shows this division. 

Firstly, as shown in the top panel of Fig.\,\ref{dwarfs-division}, presenting $\mu_0$-$M_{\ast}$, we divided the dwarf galaxies into low and high surface brightness sources. This division is made by estimating the median in overlapping bins for a sliding bin with a width of 1\,dex along the mass axis. Note that the dividing line obtained this way is stellar mass dependent. This threshold coincides with that for UDGs (\IE central surface brightness of about 23\,mag\,arcsec$^{-2}$) at a stellar mass of $10^8\,\si{M_{\odot}}$. Secondly, in the lower panel of Fig.\,\ref{dwarfs-division}, showing $R_{\rm e}$-$M_{\ast}$, we make a division between relatively diffuse and compact dwarf galaxies. The terms diffuse and compact correspond respectively to larger and smaller $R_{\rm e}$ in a given stellar mass bin, which relates to the concentration of the light distribution. These divisions are made to explore how $R_{\rm GC}$ and $N_{\rm GC}$ vary with the host galaxies' surface brightness and effective radius at any given stellar mass. From now on in this article, we use the same colour code in all the figures to represent the four categories of objects defined in Fig.\,\ref{dwarfs-division}. We also assign a short label to each category. These four categories (and their labels) are: 

\begin{itemize}

\item \textcolor{blue}{Blue: high surface brightness dwarf galaxies (HSB)}, 
\item \textcolor{red}{Red: low surface brightness dwarf galaxies (LSB)},
\item \textcolor{OliveGreen}{Green: compact dwarf galaxies (COMP)},
\item \textcolor{orange}{Yellow: Diffuse dwarf galaxies (DIFF)}.

\end{itemize}

Figure.\,\ref{dwarfs-division} also shows dwarf galaxies that satisfy the UDG criteria in central surface brightness ($\mu_{0, \IE}$<23\,mag\,arcsec$^{-2}$) and effective-radius ($R_{\rm e}$>1.5\,kpc) with black circles. {These thresholds in $\mu_{0}$ and $R_{\rm e}$ are indicated with a horizontal black dashed line in each panel.} From now on, along with the four categories of dwarf galaxies (LSB/HSB and COMP/DIFF), we examine GC properties of UDGs and non-UDGs, and use a consistent colour-code in the figures throughout the rest of the paper:

\begin{itemize}

\item \textcolor{violet}{Purple: UDGs},
\item \textcolor{gray}{Grey: non-UDGs}

\end{itemize}

\subsection{\label{sc:Results-radial-profile} Radial distributions of GCs}

Considering the small numbers of GCs typically associated with dwarf galaxies, particularly at the lowest galaxy masses and luminosities, studying the GC radial distribution is not in general feasible without stacking samples. With such stacked samples, we are able to study the average GC radial distribution around dwarf galaxies of various categories.

Here, we investigate these stacked radial profiles of GCs around all the dwarf galaxies in the sample across stellar masses. We define eight overlapping galaxy-mass bins, each with a width of 1\,dex in stellar mass, covering the range from  $10^{5.0} \si{M_{\odot}}$ to $10^{9.5} \si{M_{\odot}}$. For dwarf galaxies within such a mass bin, we estimate the galactocentric distances of the GC candidates brighter than the GCLF TOM (considering extinction) and produce a combined (stacked) radial profile. Then we estimate the S\'ersic effective radius of this GC distribution by fitting a S\'ersic function, fixing the S\'ersic index to $n=1$ and allowing for a constant additive background component. The value $n=1$ is motivated by previous works for GCs of dwarf galaxies and UDGs in various environments (\citealp{saifollahi2022,janssens2024,fcc224}), which found values between 0.5 and 2. 

The fitting procedure here is carried out in three steps and in each step, we perform S\'ersic fitting (with a constant background component) with a different aim. The aim of the first fit is to constrain the background component, representing the average GC density in the background, across the frame, at the position of the dwarf galaxies in the sample and in the stellar mass bin. For this step, we consider all the GC candidates within the initial cutout (out to 60\arcsec), using a constant bin width, about 3\arcsec (leading to 20 radial bins in total) for the radial distribution of GC. This approach leads to more data points at larger distances to properly fit and constrain the background. The threshold of 60\arcsec is imposed because of the initial size of cutouts (120\arcsec $\times$ 120\arcsec). For the most massive dwarf galaxies in the sample with $M_{\ast} \sim 10^{9.5} \si{M_{\odot}}$ that are expected to have the most spatially extended GC populations in the sample, based on \citet{lim2020}, we expect $R_{\rm GC} \sim 3$\,kpc. This implies that GCs typically can extend up to about 9\,kpc from their host galaxies, corresponding to about three times $R_{\rm GC}$. The area within this radius is about a fifth of the total area of each cutout. 

Once the GC density of the background is estimated, we perform the second fit and with this fit, we aim to estimate an initial guess for the $R_{\rm GC}$ values which will be used for the third step. For the second fit, we repeat the S\'ersic fitting procedure before, this time for the bins within 10\,kpc (about 30\arcsec) of the galaxy (10 bins in total). This is the radius beyond which we do not expect many GCs associated with dwarf galaxies. This fit is done with the fixed background component calculated from the first fit. 

In the third step, based on the initial $R_{\rm GC}$ of the second fit, we only consider GC candidates within three times the initial $R_{\rm GC}$ (which theoretically encompasses 96\% of the GCs around the host galaxy, \citealp{nacho2001}). To establish the median $R_{\rm GC}$ and its associated uncertainty, for the third fit, we perform the fitting 1000 times using 95\% of the initial number of the GC candidates, randomly selected each time (with replacement), and we adopt the median $R_{\rm GC}$ and the 68\% confidence interval as the final $R_{\rm GC}$ and its uncertainties. During this third (final) fit, we randomly chose the number of radial bins (from 6 to 10) to take into account biases that might arise from our choice for number of radial bins. Note that not all the 1000 iterations would provide a fit and the number of failed fits increases with decreasing the number of the GC candidates used for fitting. For the analysis of the stacked GC distribution, we consider fits with less than 100 failed iterations to be considered. For most cases, the number of failed iterations is zero.

Furthermore, we explore the initial choice of S\'ersic index $n=1$ for GCs of dwarf galaxies in this sample. Repeating the fitting procedure described above with a free S\'ersic index, we find that the most likely $n$ are in the range $n=0.7$--$1.3$ for dwarf galaxies in different stellar mass bins, and on average about 1. This result is consistent with the \citet{janssens2024} where authors find an average $n=1.1$ for six Perseus cluster dwarf galaxies with $N_{\rm GC}>20$. Therefore, $n=1$ seems to be a reasonable assumption for estimating $R_{\rm GC}$ for our sample, consistent with the previous literature. 

\begin{figure}[htbp!]
\begin{center}
\includegraphics[trim={0 0 0 0},angle=0,width=0.99\linewidth]{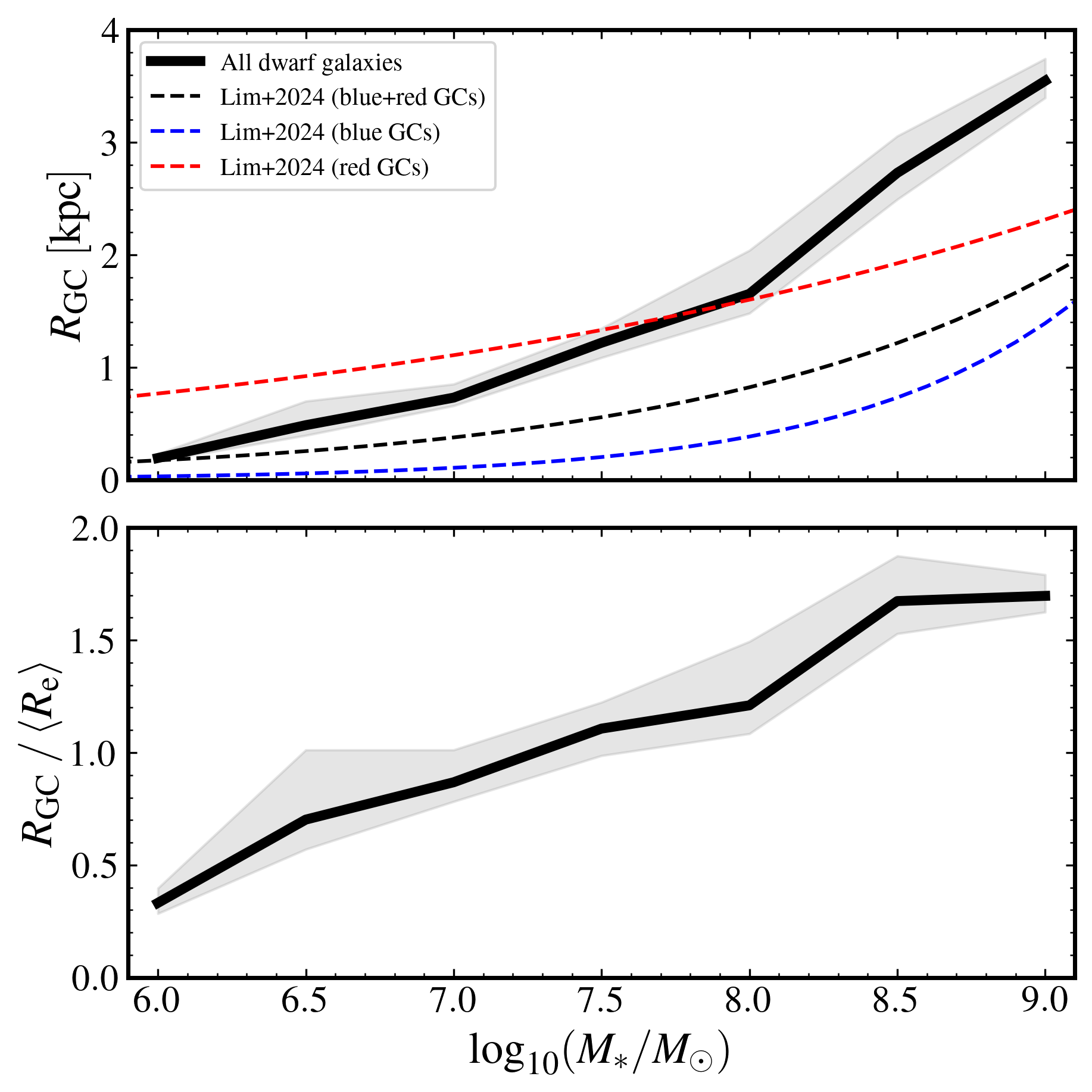}
\end{center}
\caption{GC half-number radius ($R_{\rm GC}$, top panel), and ratio between the GC half-number radius and the average of the host-galaxies effective radius ($R_{\rm GC} / \langle R_{\rm e} \rangle$, bottom panel), as functions of the stellar mass of the host galaxies ($M_{\ast}$) for all dwarf galaxies in the sample. The values are estimated from stacked GC radial profiles that we fit with a S\'ersic function, with a S\'ersic index fixed to $n=1$ (see text). The shaded region around each line represent the 68\% percentiles of the estimated average value. The dashed lines show the trends one would expect from a simple extension of the equations in \citet{lim2024}, that were derived for GCs in massive galaxies, to our lower mass regime (an extrapolation beyond the range examined by these authors). }
\label{dwarfs-stacked-rgc-all}
\end{figure}

\begin{figure}[htbp!]
\begin{center}
\includegraphics[trim={0 0 0 0},angle=0,width=0.99\linewidth]{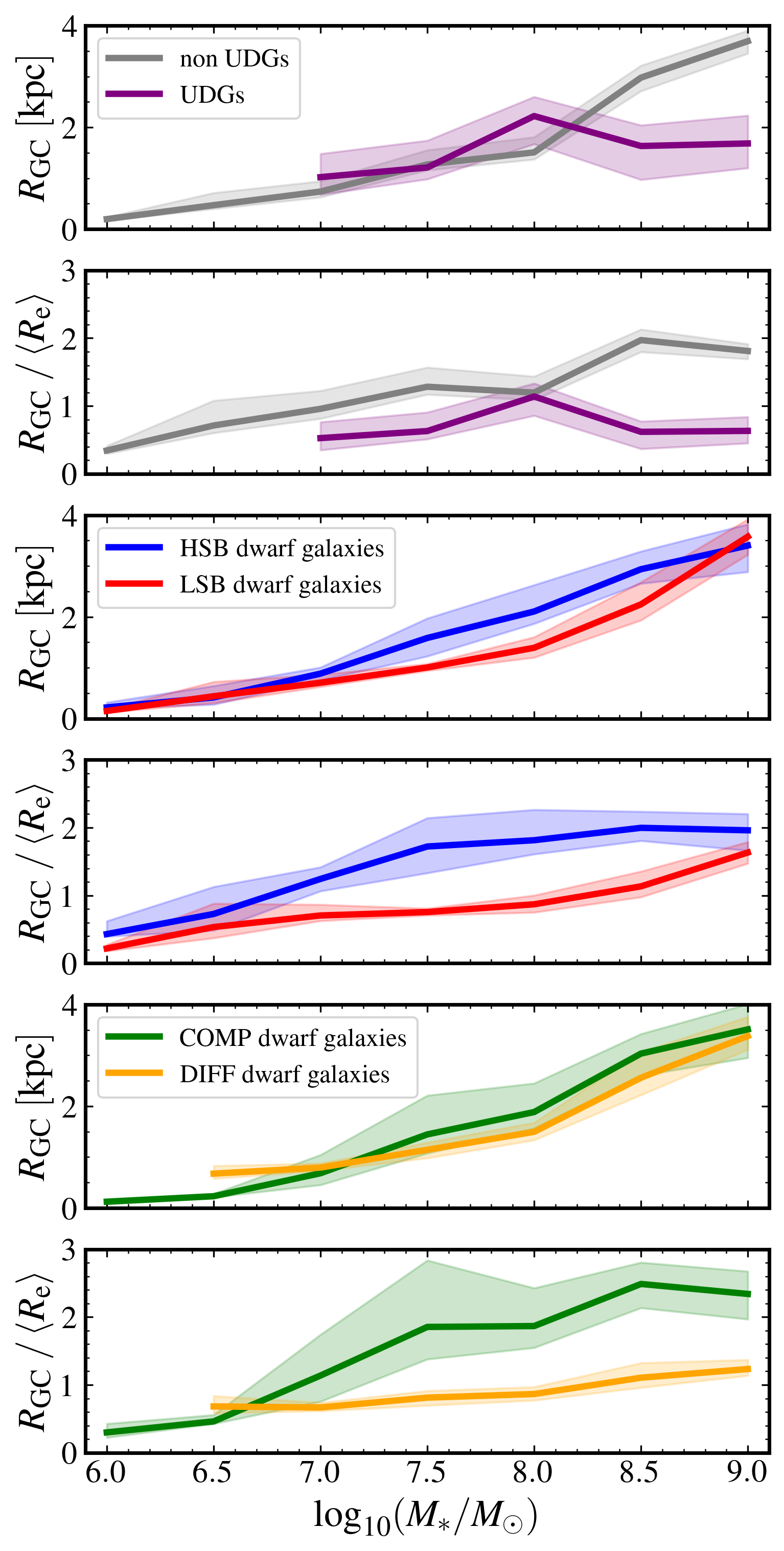}
\end{center}
\caption{Similar to Fig.\,\ref{dwarfs-stacked-rgc-all} for different dwarf galaxy categories. The two top panels represent the estimations for UDGs and non-UDGs (in purple and grey), the two middle panels for LSB and HSB dwarf galaxies (in red and blue), and the bottom two panels for DIFF and COMP dwarf galaxies (in yellow and green).}
\label{dwarfs-stacked-rgc}
\end{figure}

The estimated $R_{\rm GC}$ for galaxies stacked in a mass bin and the associated uncertainties, as well as the ratio between $R_{\rm GC}$ and the host-galaxy $R_{\rm e}$, are shown in Fig.\,\ref{dwarfs-stacked-rgc-all} for all the adopted stellar mass bins. The ratio presented in the lower panel is $R_{\rm GC} / \langle R_{\rm e} \rangle$, where $\langle R_{\rm e} \rangle$ is the average effective radius of the dwarf sample for the given stellar mass bin. The value of $R_{\rm GC} / \langle R_{\rm e} \rangle$ differs from the average of the individual per galaxy ratios $R_{\rm GC} / R_{\rm e}$, which we write as $\langle R_{\rm GC} / R_{\rm e} \rangle$. The latter, cannot be measured because the number of GCs per dwarf galaxy is too small. Appendix\,\ref{app} gives our assessment of the difference of these two quantities, suggesting that $R_{\rm GC} / \langle R_{\rm e} \rangle$ is up to 28\% smaller than $\langle R_{\rm GC} / R_{\rm e} \rangle$. The correction factor depends little on galaxy mass (10\% change across our mass bins), and drops from a 36\% to a 23\% correction in different subsamples because of the smaller dispersion in $R_{\rm e}$ within each of these subsets. On average, the correction factor is 28\%/33\% for UDGs/non-UDGs, 30\%/35\% for LSB/HSB, and 32\%/36\% for DIFF/COMP. For dwarf galaxies more massive than $M_{\ast} > 10^8 \si{M_{\odot}}$, the contrast between correction factors is higher; 23\%/34\% for LSB/HSB, and 23\%/35\% for DIFF/COMP. This difference between the correction factors implies that the difference between $\langle R_{\rm GC} / R_{\rm e} \rangle$ values for different dwarf categories is stronger than the one seen for $R_{\rm GC} / \langle R_{\rm e} \rangle$ in Fig.\,\ref{dwarfs-stacked-rgc}.

In Fig.\,\ref{dwarfs-stacked-rgc-all}, we observe an overall trend that $R_{\rm GC}$ and $R_{\rm GC} / R_{\rm e}$ decrease with decreasing the stellar mass of dwarf galaxies. {Such positive correlations are also seen in more massive 
galaxies} (\citealp{forbes2017,hudson2018,lim2024}). In this figure, we overplotted the {analytical trends} from \citet{lim2024}, extrapolated {into the stellar mass range of} dwarf galaxies. These equations are derived for all GCs, blue GCs, and red GCs of massive galaxies typically more massive than $10^{9} \si{M_{\odot}}$. {In such massive objects, the blue GCs are usually interpreted as originating from accreted dwarfs, while the red GCs may have formed in-situ or in larger accreted objects. In the dwarf galaxies of the Perseus cluster, the observed GCs have blue colours: when we use single stellar population models to convert their typical Euclid colours to optical colours, these lie in the domain of the blue GCs of \citet{lim2024}. However, in dwarf galaxies, blue colours do not imply the GCs were accreted; more likely they were formed in-situ, in the low-metallicity environment that also produced the dwarfs. In this sense, with all the due precautions associated with the exploitation of extrapolations, it may not be too surprising that the GCs of the Perseus dwarfs have system-radii closer to those extrapolated from the red GCs of massive galaxies than from the blue ones (as seen in Fig.\,\ref{dwarfs-stacked-rgc-all}).} 

We repeat the fitting for each subsample of dwarf galaxies (LSB/HSB, COMP/DIFF, UDG/non-UDGs) separately and the results are shown in Fig.\,\ref{dwarfs-stacked-rgc}. As seen in the middle panels, LSB dwarf galaxies (shown in red) tend to have (marginally) smaller $R_{\rm GC}$, implying a more compact GC distribution, while HSB dwarf galaxies (in blue) have larger $R_{\rm GC}$, implying a more extended GC distribution. There is no statistically significant difference between $R_{\rm GC}$ in COMP and DIFF and between UDGs / non-UDGs, except for the most massive bin with $M_{\ast} = 10^9 \si{M_{\odot}}$. Note that the estimated $R_{\rm GC}$ for this bin is based on only 4 UDGs (as presented in Table\,\ref{gc-dist-values-table}). Therefore, overall it seems that $R_{\rm GC}$ is independent of the host galaxy's stellar distribution, with a marginal (or no) dependency on central surface brightness. Considering the $R_{\rm GC} / \langle R_{\rm e} \rangle$, as seen in Fig.\,\ref{dwarfs-stacked-rgc}, the difference between different categories is stronger. This clearly means that the differences are driven by $R_{\rm e}$ and not by $R_{\rm GC}$. On average, for LSB dwarf galaxies, $R_{\rm GC} / \langle R_{\rm e} \rangle < 1$, while for HSB dwarf galaxies, $R_{\rm GC} / \langle R_{\rm e} \rangle = 1.5$. Furthermore, UDGs and DIFF have $R_{\rm GC} / \langle R_{\rm e} \rangle<1$, while non-UDGs and COMP dwarf galaxies have $R_{\rm GC} / \langle R_{\rm e} \rangle=1.5$--$2$. 

The main indication of the analysis of GC distributions for different types of dwarf galaxies is that the GC distribution is not dependent or has weak dependencies on the host galaxy $\mu_0$ and $R_{\rm e}$, while $R_{\rm GC} / \langle R_{\rm e} \rangle$ would differ, leading to a smaller $R_{\rm GC} / \langle R_{\rm e} \rangle$ for LSB, DIFF, and UDGs compared to HSB, COMP, and non-UDGs. We take this into account in the next section for estimating the total GC numbers ($N_{\rm GC}$) of dwarf galaxies. As shown in Appendix\,\ref{app}, the correction factor for converting $R_{\rm GC} / \langle R_{\rm e} \rangle$  to $\langle R_{\rm GC} / R_{\rm e} \rangle$ is different between different dwarf categories; on average, the correction factor on average is 1.28/1.33 for UDGs/non-UDGs, 1.30/1.35 for LSB/HSB, and 1.32/1.36 for DIFF/COMP. For dwarf galaxies more massive than $M_{\ast} > 10^8 \si{M_{\odot}}$, the contrast between correction factors is higher; 1.23/1.34 for LSB/HSB, and 1.23/1.35 for DIFF/COMP. This difference between the correction factors implies that the difference between $\langle R_{\rm GC} / R_{\rm e} \rangle$ values for different dwarf categories is stronger than the one seen for $R_{\rm GC} / \langle R_{\rm e} \rangle$ in Fig.\,\ref{dwarfs-stacked-rgc}.

\begin{figure}[htbp!]
\begin{center}
\includegraphics[trim={0 0 0 0},angle=0,width=0.99\linewidth]{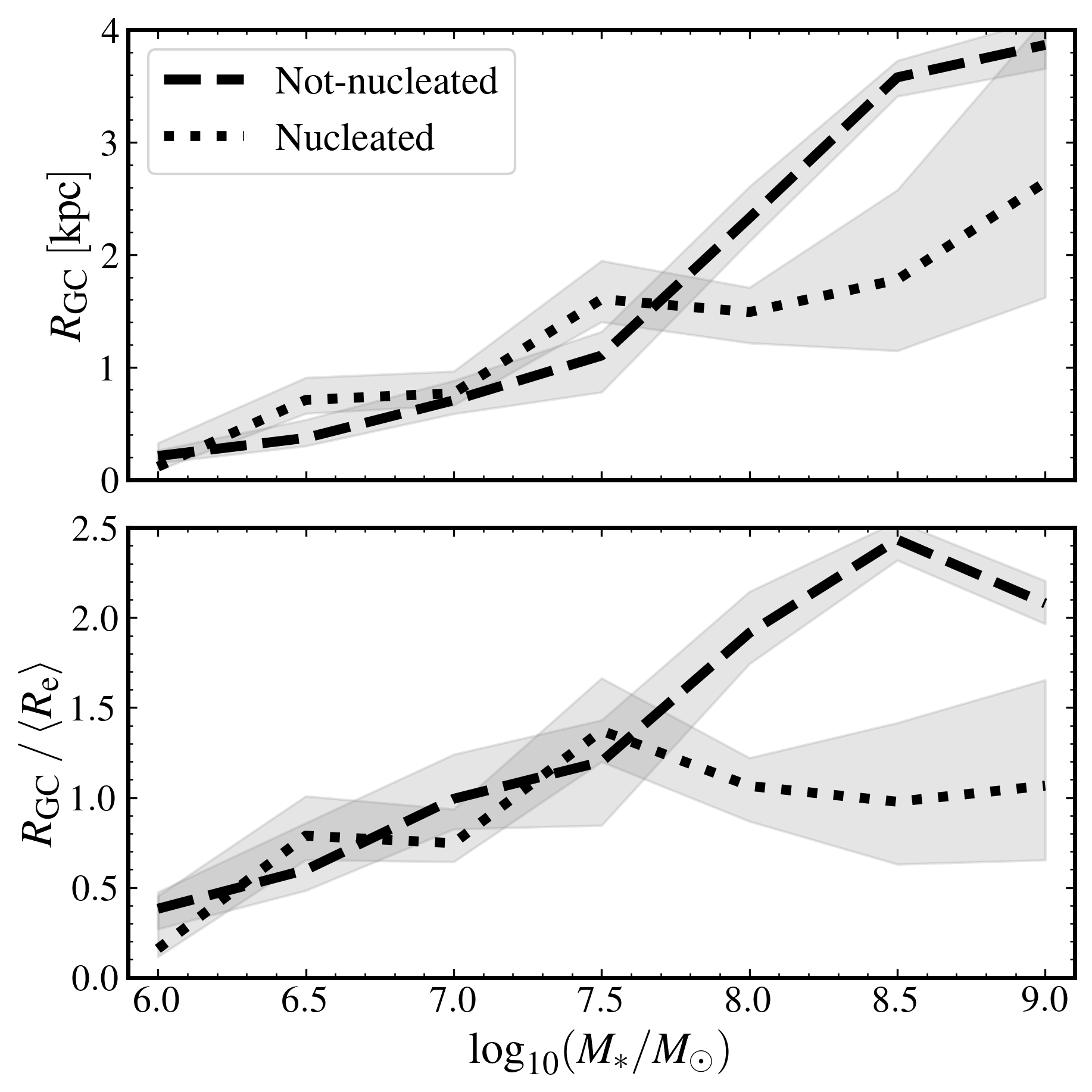}
\end{center}
\caption{Same as Fig.\,\ref{dwarfs-stacked-rgc-all} and Fig.\,\ref{dwarfs-stacked-rgc}, but for nucleated and non-nucleated dwarf galaxies in our sample.}
\label{rgc-nsc}
\end{figure}

Lastly, we repeated our analysis of GC distributions for nucleated and non-nucleated dwarf galaxies. The separation between these two is based on the dwarf catalogue provided by \citet{EROPerseusDGs}. The results are shown in Fig.\,\ref{rgc-nsc}, where it is seen that there is no meaningful difference between $R_{\rm GC}$ and $R_{\rm GC} / \langle R_{\rm e} \rangle$ for nucleated/non-nucleated dwarf galaxies below $M_{\ast} = 10^{8} \si{M_{\odot}}$. Above this stellar mass, nucleated dwarf galaxies show smaller $R_{\rm GC}$ and $R_{\rm GC} / \langle R_{\rm e} \rangle$ compared to non-nucleated dwarf galaxies. The smaller $R_{\rm GC}$ in nucleated dwarf galaxies could be an indication of the stronger dynamical friction of GCs within these dwarf galaxies which has also led to the formation of a nuclear star cluster (NSC, see \citealp{nscreview}) in these dwarf galaxies (\citealp{fahrion2022,roman2023}) for more than 50\% of the dwarf population (\citealp{denbrok14,ruben19}). The results presented in figures in this section ($R_{\rm GC}$ and $R_{\rm GC} / \langle R_{\rm e} \rangle$), and some details such as number of dwarf galaxies and GC candidates in each stellar mass bin are provided in Table\,\ref{gc-dist-values-table}.

\subsection{\label{sc:Results-numbers}GC number counts}

\begin{figure}[htbp!]
\begin{center}
\includegraphics[trim={0 0 0 0},angle=0,width=\linewidth]{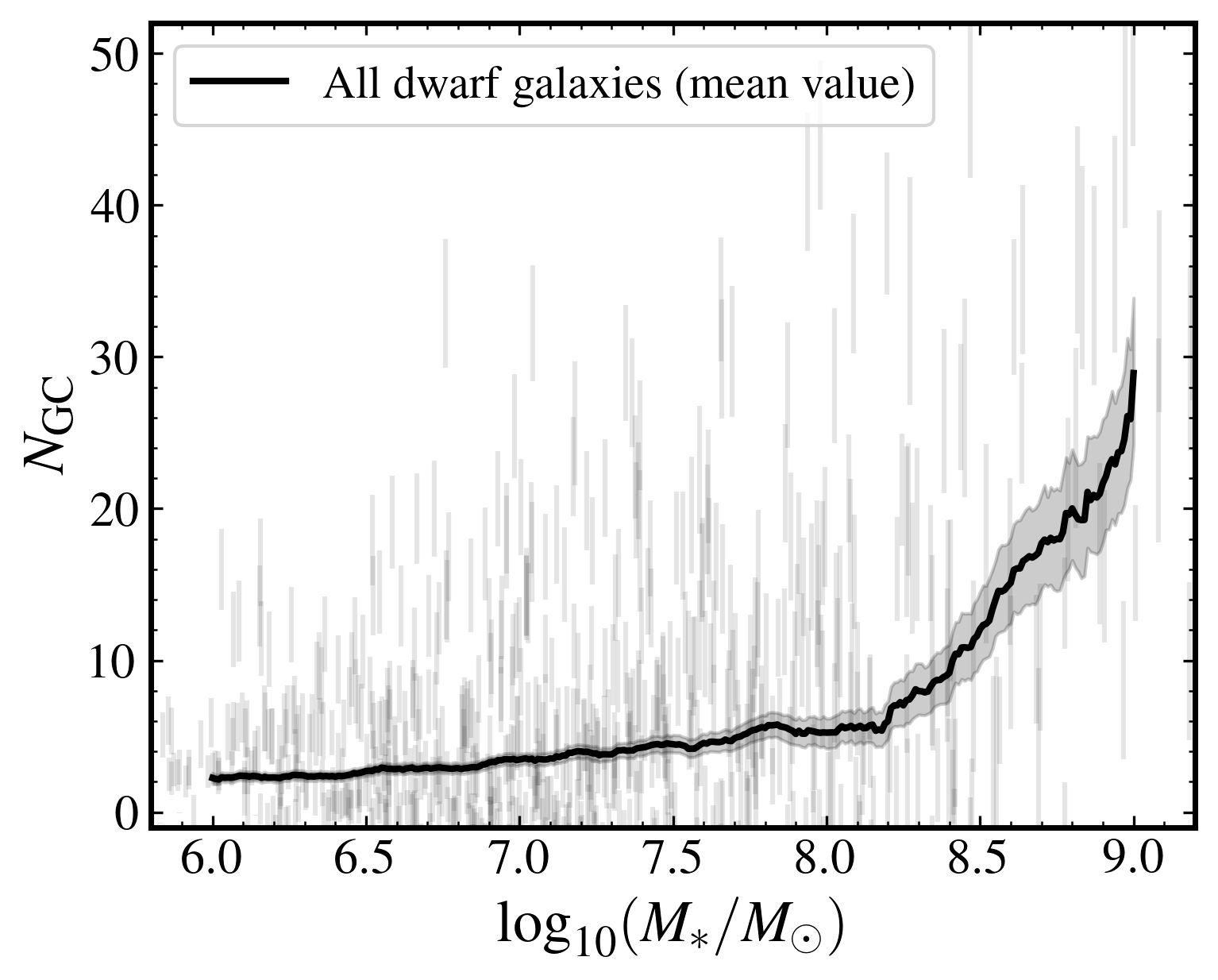}
\end{center}
\caption{Total GC numbers ($N_{\rm GC}$) for all the dwarf galaxies in this work. The vertical (grey) lines represent the $N_{\rm GC}$ and their uncertainties for individual dwarf galaxies. The solid curve and the shaded regions around it represent the average (mean) $N_{\rm GC}$ and 68\% confidence intervals for dwarf galaxies within a moving bin along stellar mass ($x$-axis) with a width of 0.5\,dex. The confidence intervals are based on 1000 bootstrap-repetitions of the estimates using bootstrap sample-sizes of 95\% of the initial sample sizes.}
\label{dwarfs-ngc-all}
\end{figure}

\begin{figure}[htbp!]
\begin{center}
\includegraphics[trim={0 0 0 0},angle=0,width=\linewidth]{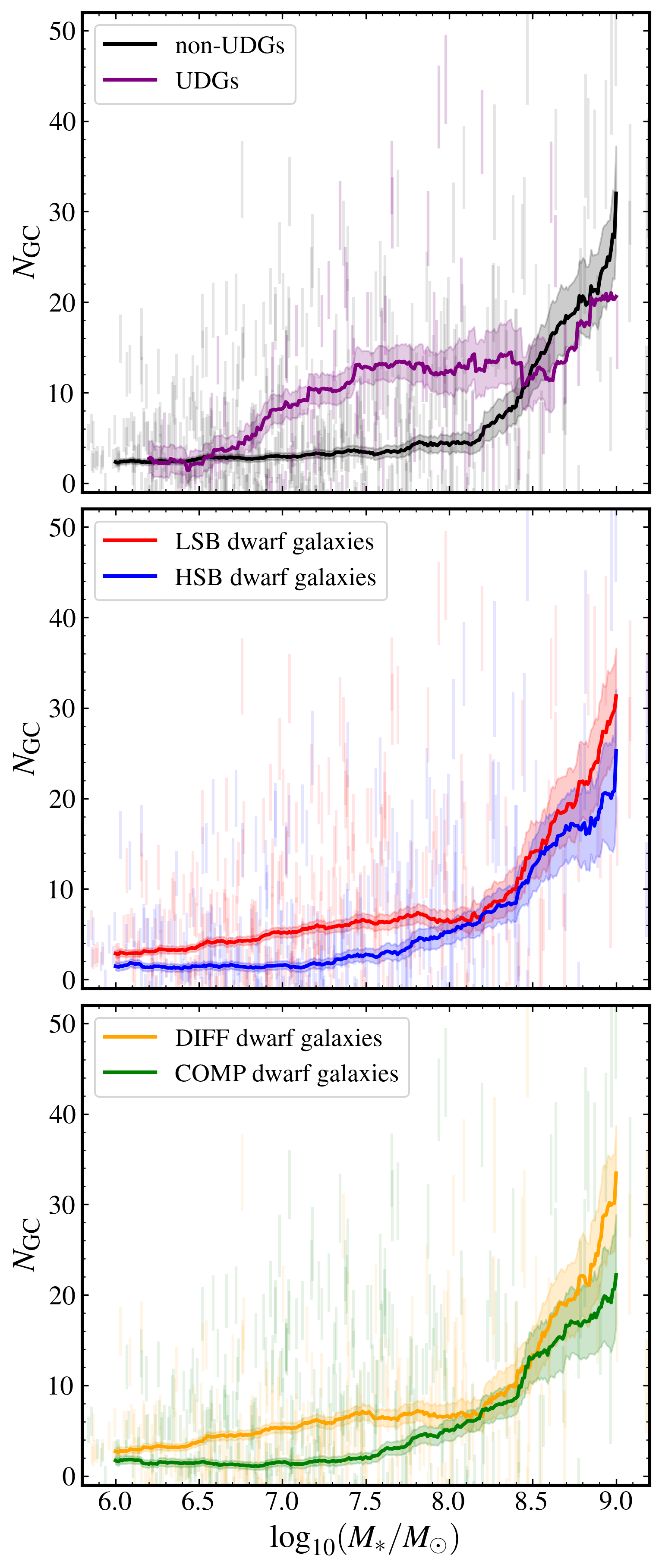}
\end{center}
\caption{Similar to Fig.\,\ref{dwarfs-ngc-all} for UDGs/non-UDGs (top), LSB/HSB dwarf galaxies (middle), and COMP/DIFF dwarf galaxies (bottom). The vertical lines in each panel represent the $N_{\rm GC}$ and its uncertainties for individual galaxies. The solid line and the shaded region around it represent the average (mean) $N_{\rm GC}$ and 68\% confidence intervals for dwarf galaxies within a moving bin along stellar mass ($x$-axis) with a width of 0.5\,dex.}
\label{dwarfs-ngc}
\end{figure}

\begin{figure}[htbp!]
\begin{center}
\includegraphics[trim={0 0 0 0},angle=0,width=\linewidth]{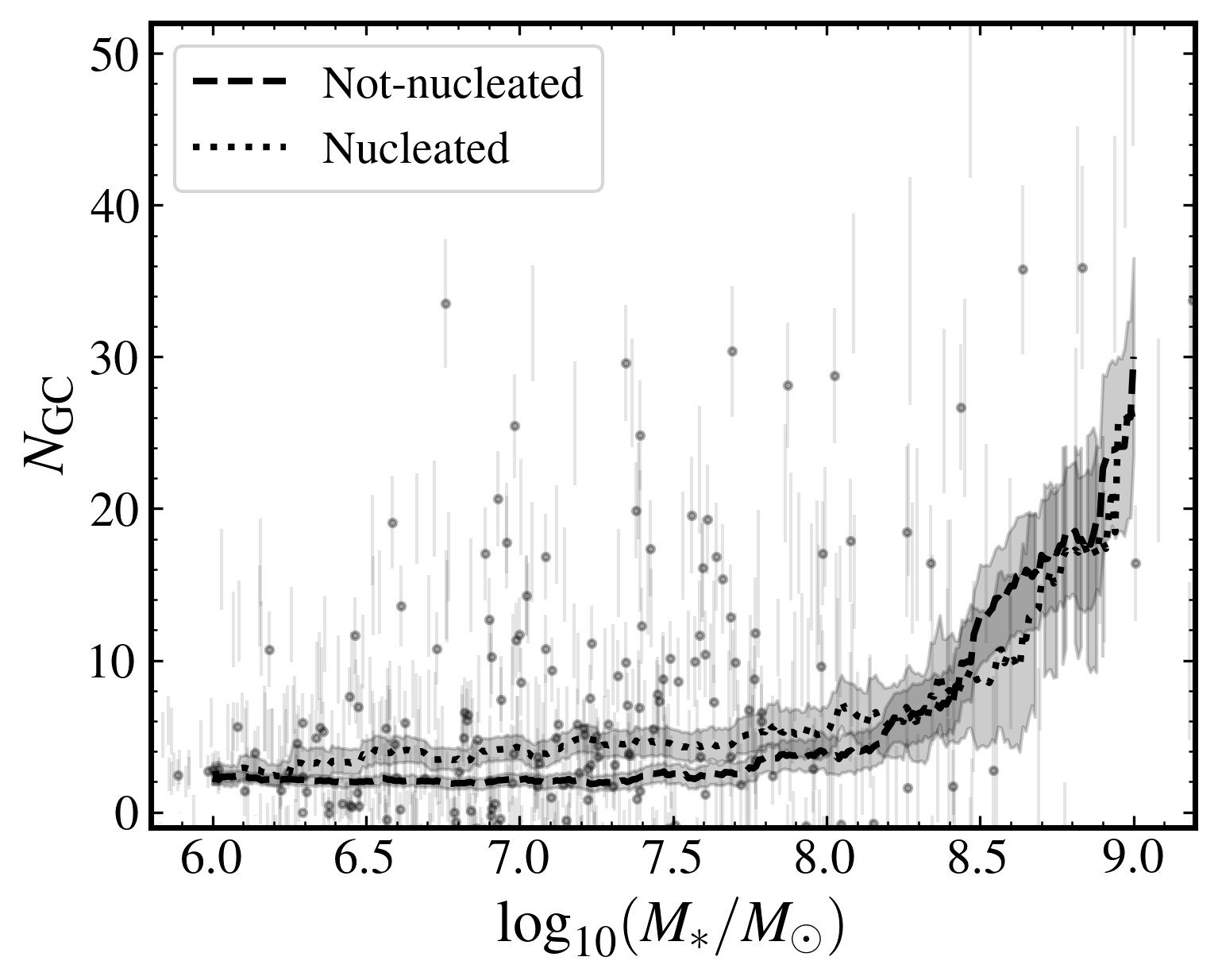}
\end{center}
\caption{Similar to Fig.\,\ref{dwarfs-ngc-all} and Fig.\,\ref{dwarfs-ngc}, for the nucleated and not-nucleated dwarf galaxies in the sample. The vertical lines with and without a central small circle indicate nucleated and not-nucleated dwarf galaxies, respectively.}
\label{dwarfs-ngc-nsc}
\end{figure}

The total GC number counts of dwarf galaxies ($N_{\rm GC}$) have been studied and debated extensively in recent years, with particular attention towards UDGs. The methodology for estimating $N_{\rm GC}$ of dwarf galaxies based on imaging data consists of several steps to take into account: (i) incompleteness in the GC candidate catalogue, (ii) the contamination from GCs that are not associated with the dwarf galaxy (including intracluster GCs, foreground stars and background galaxies). These corrections would ideally require a good understanding of the properties of the GCLF (shape, width, peak magnitude) and the distribution of GCs around their host galaxies. In the case of dwarf galaxies, unless they are very rich in GCs, the GCLF and GC distribution cannot be well constrained for individual galaxies, and therefore $N_{\rm GC}$ estimates require some assumptions on the GCLF and the GC distribution. 

As the initial step, one needs to count the GC candidates within a radius that encloses almost all the GCs. For a S\'ersic distribution, this radius can be estimated using the half-number radius of the GCs ($R_{\rm GC}$) for the S\'ersic index ($n$) of the distribution. For $n=1$, more than 99\% are within $5R_{\rm GC}$, therefore, we do not expect any GCs beyond this radius. However, such a large radius also includes many contaminants, increasing the uncertainties in estimating $N_{\rm GC}$. An alternative approach is to count GCs within a smaller radius which encircles a large fraction of GCs, and later to correct for the (small) missing fraction. For example, choosing a radius of $2R_{\rm GC}$ (as in this work), while including 85\% of the GCs, we reduce the number of contaminants by a factor of 6.25, which reduces the uncertainties (Poisson) by a factor of 2.5. In the end, the estimated $N_{\rm GC}$ will be corrected for the missing 15\% of the GCs. 

Because $R_{\rm GC}$ is typically not well-constrained for individual dwarf galaxies, in many works it is calculated indirectly from $R_{\rm GC}/R_{\rm e}$, by measuring the host galaxies $R_{\rm e}$. Typically a single value of $R_{\rm GC}/R_{\rm e}$ is used, with a value between 0.7 and 1.5. However, in the previous section we showed that this ratio is dependent on the central surface brightness ($\mu_0$) and effective radius ($R_{\rm e}$) of the host galaxy, and therefore using a single value for a wide range of dwarf galaxies could bias the $N_{\rm GC}$ estimate. 

Our analysis also shows that $R_{\rm GC}$, in contrast, is almost independent of properties other than stellar mass, except for a possible weak dependence on the central surface brightness. Here, we investigate the $N_{\rm GC}$ of dwarf galaxies with this new understanding of $R_{\rm GC}$, $R_{\rm GC}/R_{\rm e}$, and their possible dependence (only marginally, based on our data) on $\mu_0$. For each dwarf galaxy, characterised by its stellar mass and central surface brightness, we construct the stacked radial profile of the GC systems of all the dwarf galaxies within 0.5\,dex in stellar mass and 1\,mag in central surface brightness, and estimate $R_{\rm GC}$. We then use $R_{\rm GC}$ (rather than a single $R_{\rm GC}/R_{\rm e}$ value for all dwarf galaxies) to quantify the spatial extent of the GC system of the initial galaxy and to evaluate its $N_{\rm GC}$. This approach should be less biased than the alternatives based on $R_{\rm GC}/R_{\rm e}$.

In practice, we first count the number of GC candidates brighter than the GCLF TOM within 2$R_{\rm GC}$. Here we limit our GC sample to GCs brighter than the GCLF TOM because our detection completeness drops quickly below this magnitude. The value of the GCLF TOM used here is $\IE=26.6$ plus the \IE foreground extinction correction for the line of sight of a given dwarf galaxy. Then, we correct for this number of GC candidates for background contamination by counting GC candidates within the cutout that are farther than 5$R_{\rm GC}$ (where we do not expect any GC for $n=1$), normalised to the ratio of areas within 2$R_{\rm GC}$ and beyond 5$R_{\rm GC}$. Additionally, using the estimated completeness function (Sect.\,\ref{sc:Methods-art}), we correct the GC numbers for incompleteness in detection up to TOM and GCs beyond 2$R_{\rm GC}$ (about 15\% for a S\'ersic profile with $n=1$), and double the resulting number for the GCs fainter than TOM to obtain the total GC number. We also estimate uncertainties in $N_{\rm GC}$, taking into account the Poisson errors of GC candidate numbers within 2$R_{\rm GC}$ and farther than 5$R_{\rm GC}$ in the background. These GC numbers are also compared with the numbers in previous works (\citealp{janssens2024,EROPerseusDGs}) in Appendix\,\ref{app-comparison}.

\begin{figure*}[htbp!]
\begin{center}
\includegraphics[trim={0 0 0 0},angle=0,width=0.19\linewidth]{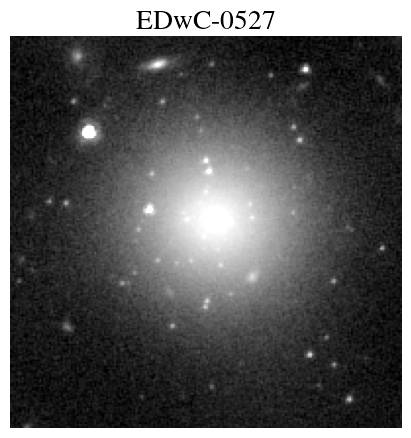}
\includegraphics[trim={0 0 0 0},angle=0,width=0.19\linewidth]{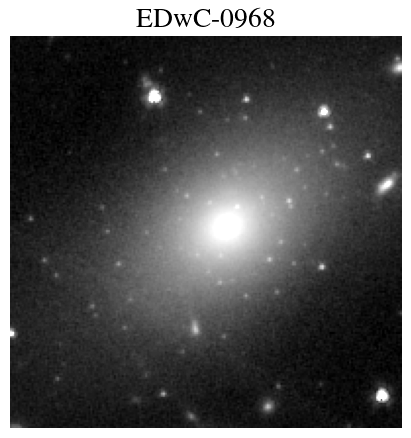}
\includegraphics[trim={0 0 0 0},angle=0,width=0.19\linewidth]{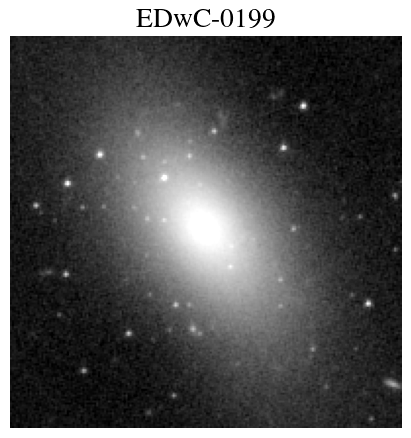}
\includegraphics[trim={0 0 0 0},angle=0,width=0.19\linewidth]{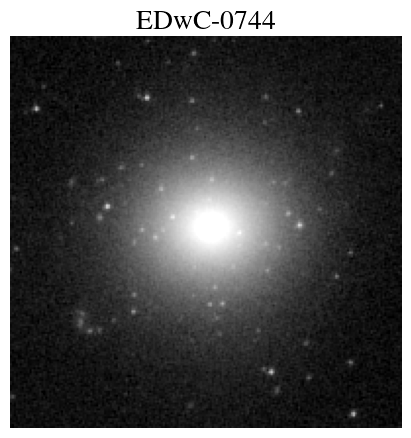}
\includegraphics[trim={0 0 0 0},angle=0,width=0.19\linewidth]{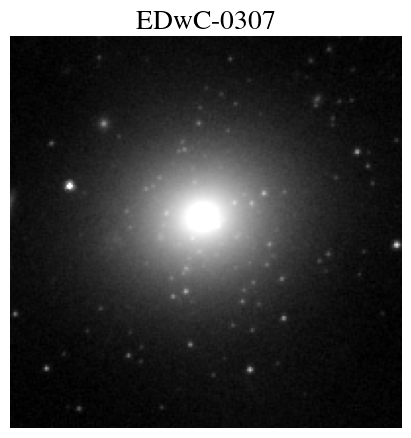}

\includegraphics[trim={0 0 0 0},angle=0,width=0.19\linewidth]{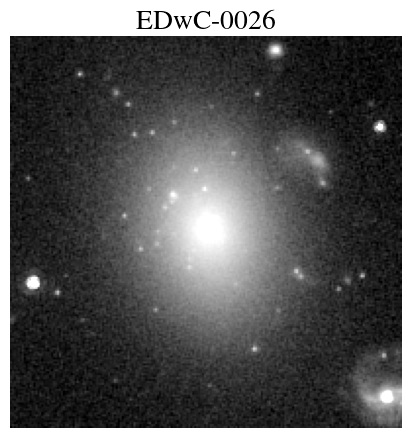}
\includegraphics[trim={0 0 0 0},angle=0,width=0.19\linewidth]{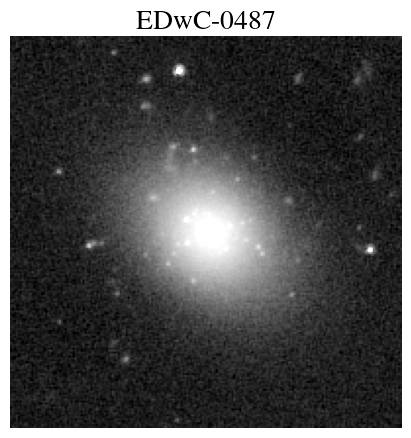}
\includegraphics[trim={0 0 0 0},angle=0,width=0.19\linewidth]{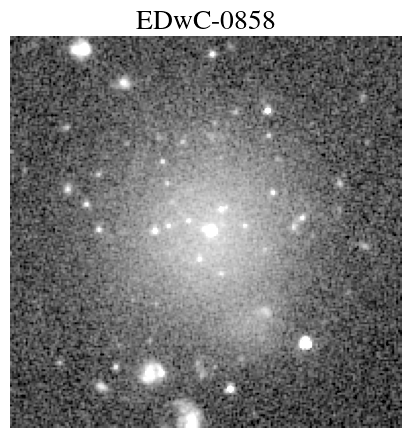}
\includegraphics[trim={0 0 0 0},angle=0,width=0.19\linewidth]{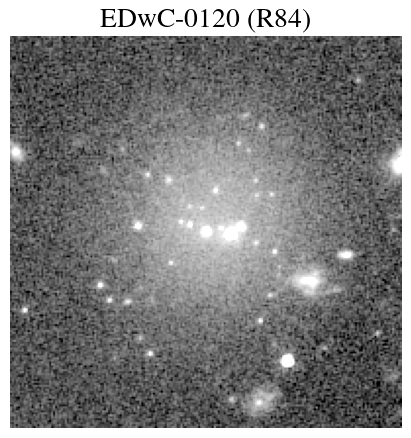}
\includegraphics[trim={0 0 0 0},angle=0,width=0.19\linewidth]{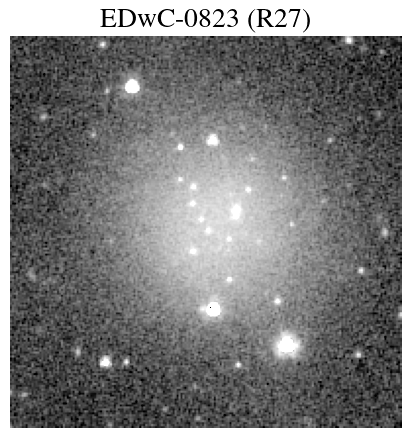}

\includegraphics[trim={0 0 0 0},angle=0,width=0.19\linewidth]{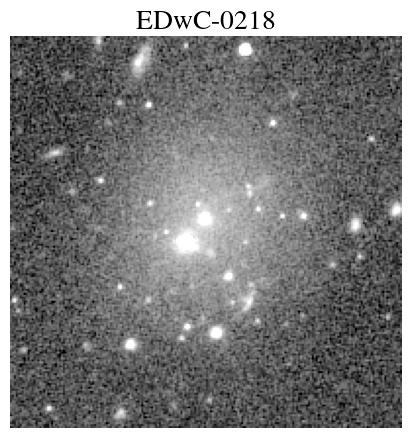}
\includegraphics[trim={0 0 0 0},angle=0,width=0.19\linewidth]{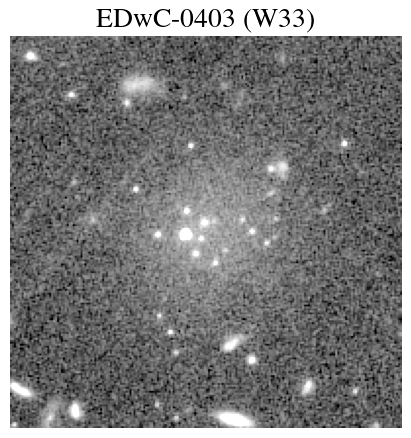}
\includegraphics[trim={0 0 0 0},angle=0,width=0.19\linewidth]{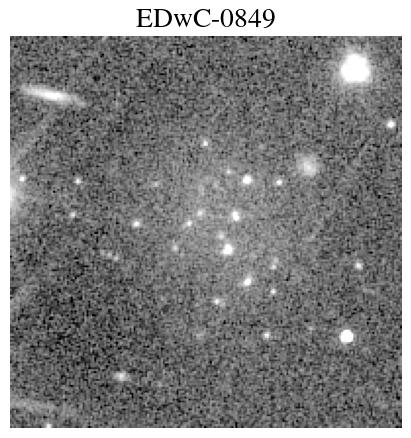}
\includegraphics[trim={0 0 0 0},angle=0,width=0.19\linewidth]{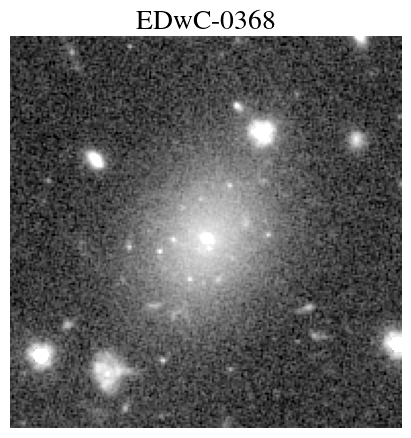}
\includegraphics[trim={0 0 0 0},angle=0,width=0.19\linewidth]{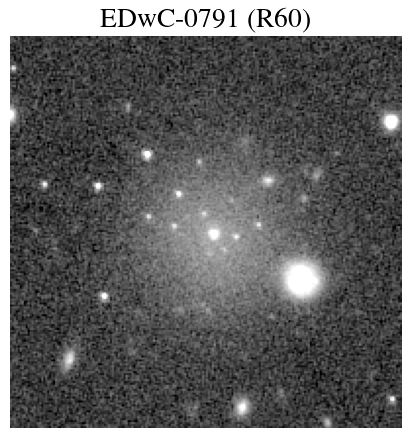}

\end{center}
\caption{A few examples of Perseus cluster dwarf galaxies with more than 10 GCs ($N_{\rm GC}>10$) in $\IE$ data. The majority of such dwarf galaxies have a stellar mass of $M_{\ast} > 10^8 \si{M_{\odot}}$ while they cover a range of central surface brightness ($\mu_0$) and effective radius ($R_{\rm e}$). The \Euclid ID of dwarf galaxies from \citet{EROPerseusDGs} are written on top of each cutouts, as well as their name in \citet{janssens} when available (written in parenthesis). The cutouts have dimensions of $40\arcsec \times 40\arcsec$ (about 14\,kpc $\times$ 14\,kpc).}
\label{dwarfs-and-gcs-vis-gc-rich}
\end{figure*}

We examine the behaviour of $N_{\rm GC}$ for all the dwarf galaxies in Fig.\,\ref{dwarfs-ngc-all}, and present the counts separately for UDGs/non-UDGs, LSB/HSB and COMP/DIFF dwarf galaxies in Fig.\,\ref{dwarfs-ngc}. In each panel, we also show the average (mean) $N_{\rm GC}$  at a given stellar mass (taking into account dwarfs with zero and negative $N_{\rm GC}$). As expected, $N_{\rm GC}$ increases with stellar mass in all cases. In general, we do not observe any hint of the presence of a distinct GC-rich population of dwarf galaxies in the $N_{\rm GC}$--$M_{\ast}$ space. Inspecting Fig.\,\ref{dwarfs-ngc}, we find a higher $N_{\rm GC}$ for LSB, DIFF, and UDGs compared to HSB, COMP, and non-UDGs. The difference between UDGs and non-UDGs is already established (\citealp{lim2018,lim2020}). Here we also find relatively higher $N_{\rm GC}$ when we subdivide the samples into LSB or HSB galaxies, or into DIFF or COMP galaxies. The previously observed GC excess in UDGs arises from the combination of surface brightness and effective radius; neither characteristic alone is sufficient to produce this effect. Assuming that $N_{\rm GC}$ is a proxy for the total (dynamical) mass of galaxies (\citealp{spitler2009,burkert2020,zaritsky2022}), this indicates that at a given stellar mass, LSB, DIFF, and UDG dwarf galaxies are more massive (in total mass) than HSB, COMP, and non-UDG dwarf galaxies. These trends were previously discussed for UDGs; however, our results show that such trends are not unique to UDGs and can be extended to all cluster dwarf galaxies, over a range of stellar masses.

Lastly, we examine the $N_{\rm GC}$ values of nucleated and not-nucleated dwarf galaxies and the result is shown in Fig.\,\ref{dwarfs-ngc-nsc}. At a given stellar mass and for $M_{\ast} > 10^{8} \si{M_{\odot}}$, there is no meaningful difference between the average $N_{\rm GC}$ in nucleated and not-nucleated dwarf galaxies. However, for $M_{\ast} < 10^{8} \si{M_{\odot}}$, nucleated dwarf galaxies show a higher $N_{\rm GC}$ than not-nucleated dwarf galaxies. This observation, while it might look counter-intuitive, is consistent with the previous findings (\citealp{carlsten2022}). This could be the outcome of galaxy evolution in a rich environment, which leads to a more efficient NSC formation as well as a higher $N_{\rm GC}$ for dwarf galaxies at a given stellar mass.

\subsection{\label{sc:Results-gc-rich-poor}GCs of individual dwarf galaxies}

\begin{figure*}[htbp!]
\begin{center}
\includegraphics[trim={0 0 0 0},angle=0,width=0.49\linewidth]{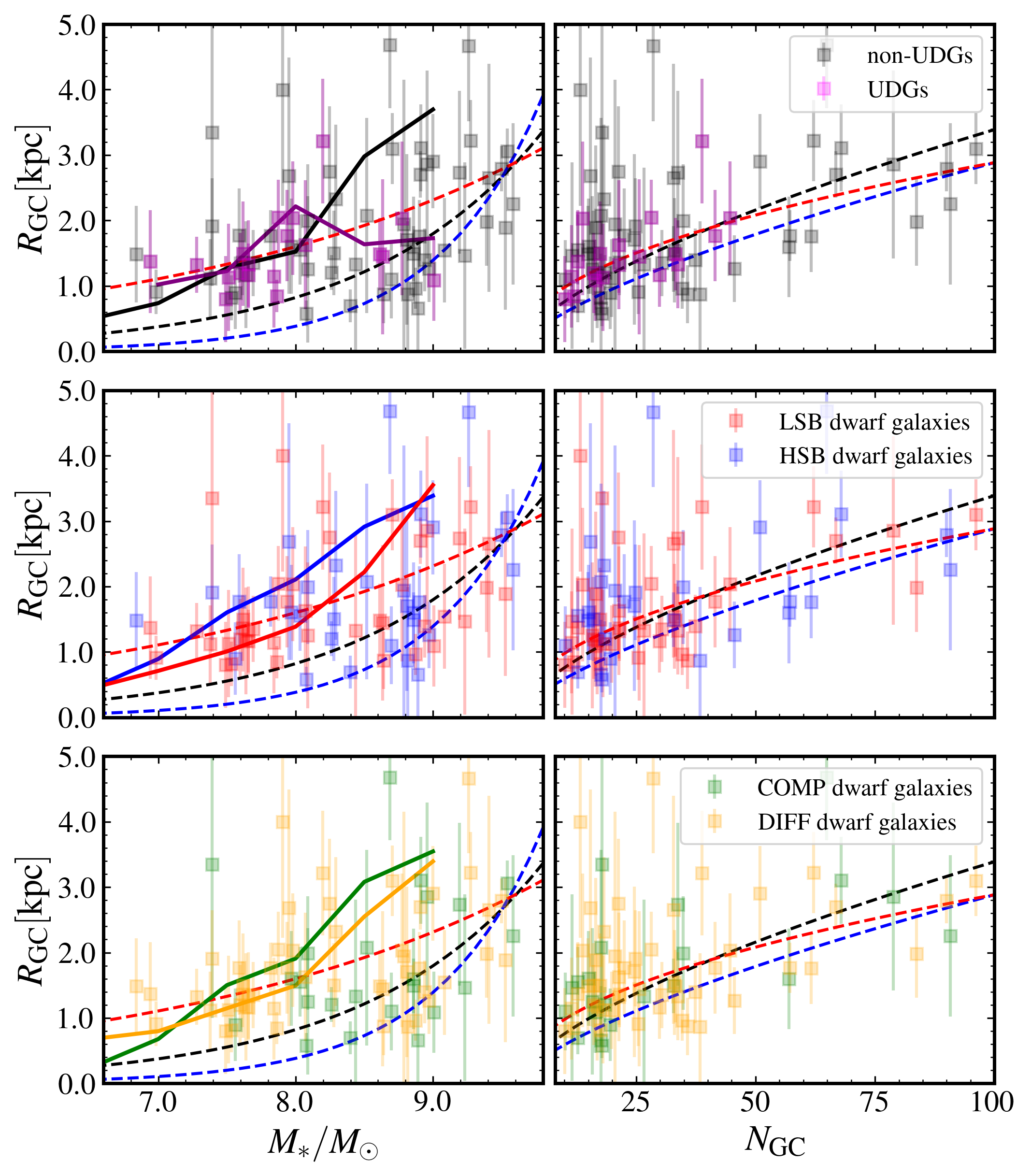}
\includegraphics[trim={0 0 0 0},angle=0,width=0.49\linewidth]{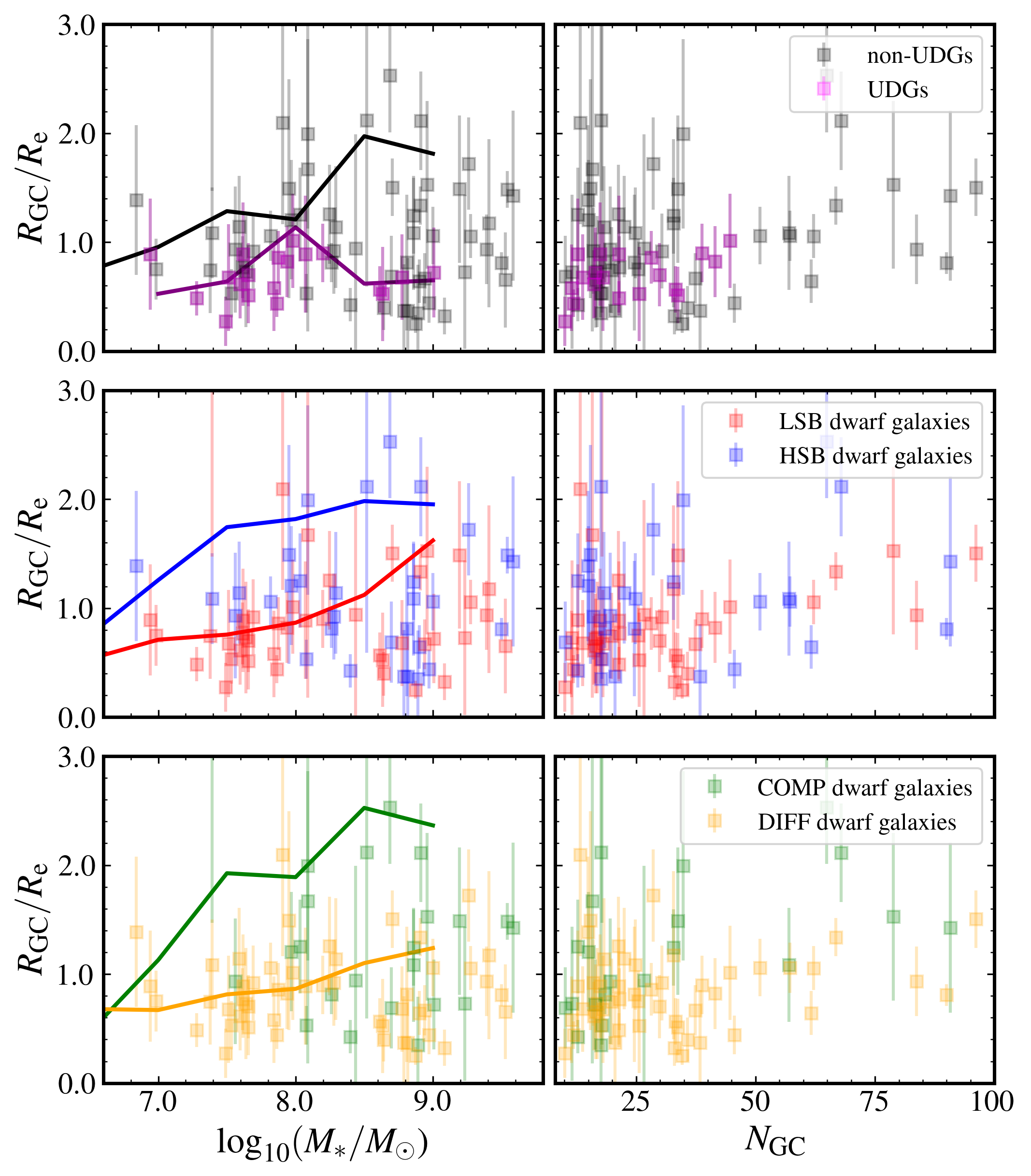}
\end{center}
\caption{Properties of the radial distributions of the GC systems of dwarf galaxies with $N_{\rm GC}>10$. Panels on the left and right show $R_{\rm GC}$ and $R_{\rm GC}/R_{\rm e}$ values for different categories of dwarf galaxies versus their $M_{\ast}$ and $N_{\rm GC}$. The panels on the left also show the extrapolated equations from \citet{lim2024} for all GCs (black dashed-line), blue GCs (blue dashed-line) and red GCs (red dashed-line). We recall that these curves are our extrapolations of analytical expressions derived at higher galaxy masses into the low mass regime. Solid lines show the results from the stacked GC distributions in Sect.\,\ref{sc:Results-radial-profile} for each dwarf galaxy category. {Therefore, these lines are not a fit on the presented data-points, but the curves presented previously in Fig.\,\ref{dwarfs-stacked-rgc}.}}
\label{dwarfs-rgc1}
\end{figure*}

In previous sections, we examined the stacked $R_{\rm GC}$ of dwarf galaxies, mainly because dwarf galaxies do not host many GCs. As a result, the trends discussed correspond to the general population of dwarf galaxies. However, it is likely that some dwarf galaxies, although small in number, do not behave similarly. Here, we examine $R_{\rm GC}$ for dwarf galaxies with more than 10 GCs (164 dwarf galaxies), where we might gain insights into their individual GC properties. Images of some of these dwarf galaxies are shown in Fig.\,\ref{dwarfs-and-gcs-vis-gc-rich}. For these dwarf galaxies, we relaxed the number of failed iterations (for the third fit, as described in Sect.\,\ref{sc:Results-radial-profile}) by adjusting the threshold of the failed iterations to 900 (out of 1000). Additionally, we only consider dwarf galaxies with $R_{\rm GC}$ errors less than 2\,kpc. 

The results of our analysis of individual galaxies are presented in Fig.\,\ref{dwarfs-rgc1} for LSB/HSB, COMP/DIFF, and UDG/non-UDG dwarf galaxies. Out of 164 dwarf galaxies with more than 10 GCs, we could measure $R_{\rm GC}$ for 78 of them, with an overall uncertainty of less than 2\,kpc. {The uncertainties are estimated using bootstrapping, similar to what was done earlier for the analysis of stacked GC profiles in Sect.\,\ref{sc:Results-radial-profile}.} Note that dwarf galaxies with $N_{\rm GC}>10$ are among the most massive dwarf galaxies in our Perseus sample (stellar mass mostly between $10^8\,\si{M_{\odot}}$ and $10^9 \si{M_{\odot}}$), but are only a subset of those most massive dwarfs. Therefore, some caution should be exercised when comparing these figures with Fig.\,\ref{dwarfs-stacked-rgc-all} and Fig.\,\ref{dwarfs-stacked-rgc}.

All panels of Fig.\,\ref{dwarfs-rgc1} display considerable dispersion. Simulations of GC systems drawn randomly from a S\'ersic distribution with $n=1$ lead to distributions of recovered $R_{\rm GC}$ of which the 2.5\% and 97.5\% quantiles are found at 0.55 and 1.67 times the true $R_{\rm GC}$ for $N_{\rm GC}=10$, and at 0.68 and 1.42 times the true $R_{\rm GC}$ for $N_{\rm GC}=25$. The bootstrap estimates {shown by the shaded regions} in the figure are representative of this inevitable uncertainty. We also compare our results for $R_{\rm GC}$ of individual dwarf galaxies with the values provided by \citet{janssens2024} in Fig.\,\ref{dwarfs-comp1} in Appendix\,\ref{app-comparison}. Of the 40 dwarf galaxies common to both studies, five dwarf galaxies have $R_{\rm GC}$ values in both works. Two dwarf galaxies out of five, namely EDwC-0120 and EDwC-0823 (see Fig.\,\ref{dwarfs-and-gcs-vis-gc-rich}), are among the most GC-rich dwarf galaxies in our sample and within the uncertainties, the estimated $R_{\rm GC}$ values are consistent between the two analyses, with $R_{\rm GC} = 2.7^{+1.0}_{-1.0}$\,kpc and $R_{\rm GC} = 1.7^{+0.5}_{-0.7}$\,kpc (and $R_{\rm GC} = 2.4^{+0.4}_{-0.3}$\,kpc and $R_{\rm GC} = 2.2^{+0.4}_{-0.3}$\,kpc in \citealp{janssens2024}). For the other three dwarf galaxies, namely EDwC-1011, EDwC-0403, and EDwC-0791, our values of $R_{\rm GC}$ are, on average, 1.5\,kpc smaller. Examining these galaxies (Fig.\,\ref{dwarfs-and-gcs-vis-gc-rich}), we can see a clear over-density of point sources (most likely GC candidates within the apparent half-light radius of dwarf galaxies DwC-0403 and EDwC-0791). We take this observation as a sign that $R_{\rm GC} / R_{\rm e} < 1.0$ for these dwarf galaxies, as estimated in this work, is justified. {To assess this a bit further, we compared the $R_{\rm GC}$ values with \citet{li2025}, which uses the same {\it HST} data as \citet{janssens2024}, and applies a novel approach to derive the GC properties of dwarf galaxies. We find a better consistency between our estimates of $R_{\rm GC}$ and \citet{li2025} -- see Fig.\,\ref{dwarfs-comp1}.}

We focus on $R_{\rm GC}$ in the left panels of Fig.\,\ref{dwarfs-rgc1}. In the subset of dwarf galaxies with $N_{\rm GC}>1$, we again find a positive correlation between $R_{\rm GC}$ and host-galaxy stellar mass, with no major systematic difference between UDG/LSB/DIFF objects on one hand, and non-UDG/HSB/COMP ones on the other. {This is similar to the trends found earlier using the stacked GC radial profiles (shown with solid coloured curves). These curves, are derived and presented earlier in Sect.\,\ref{sc:Results-radial-profile} and Fig.\,\ref{dwarfs-stacked-rgc}, and they are not a fit to the data-points in Fig.\,\ref{dwarfs-rgc1}.} Overall, be it with respect to host-galaxy stellar mass or to $N_{\rm GC}$, the dwarf galaxies of various types seem to follow the behaviour one obtains by extrapolating the trend seen for the red GC subpopulations of more massive galaxies (red dashed curve). This may imply that GCs of dwarf galaxies, even though they are blue and metal-poor, have formed in-situ similarly to red and metal-rich GCs in massive galaxies. 

The $R_{\rm GC}/R_{\rm e}$ values of individual cases, shown in the right-hand panels of Fig.\,\ref{dwarfs-rgc1}, in some cases seem compatible with the trends we found earlier with stacked GC radial profiles (solid lines). In particular, UDGs have smaller $R_{\rm GC}/R_{\rm e}$ on average, with values close to or smaller than 1, while non-UDGs have a higher $R_{\rm GC}/R_{\rm e}$, about 1.5. Similarly, a tendency for having smaller $R_{\rm GC}/R_{\rm e}$ values is seen for DIFF dwarf galaxies compared to the COMP dwarf galaxies. We do not see any significant difference between the $R_{\rm GC}/R_{\rm e}$ ratios of LSB and HSB galaxies. Lastly, we want to emphasise that regardless of the average behaviour, the GC distribution covers a range of values, with $R_{\rm GC}/R_{\rm e}$ from 0.5 to 2. 

\begin{figure*}[htbp!]
\begin{center}
\includegraphics[trim={0 0 0 0},angle=0,width=\linewidth]{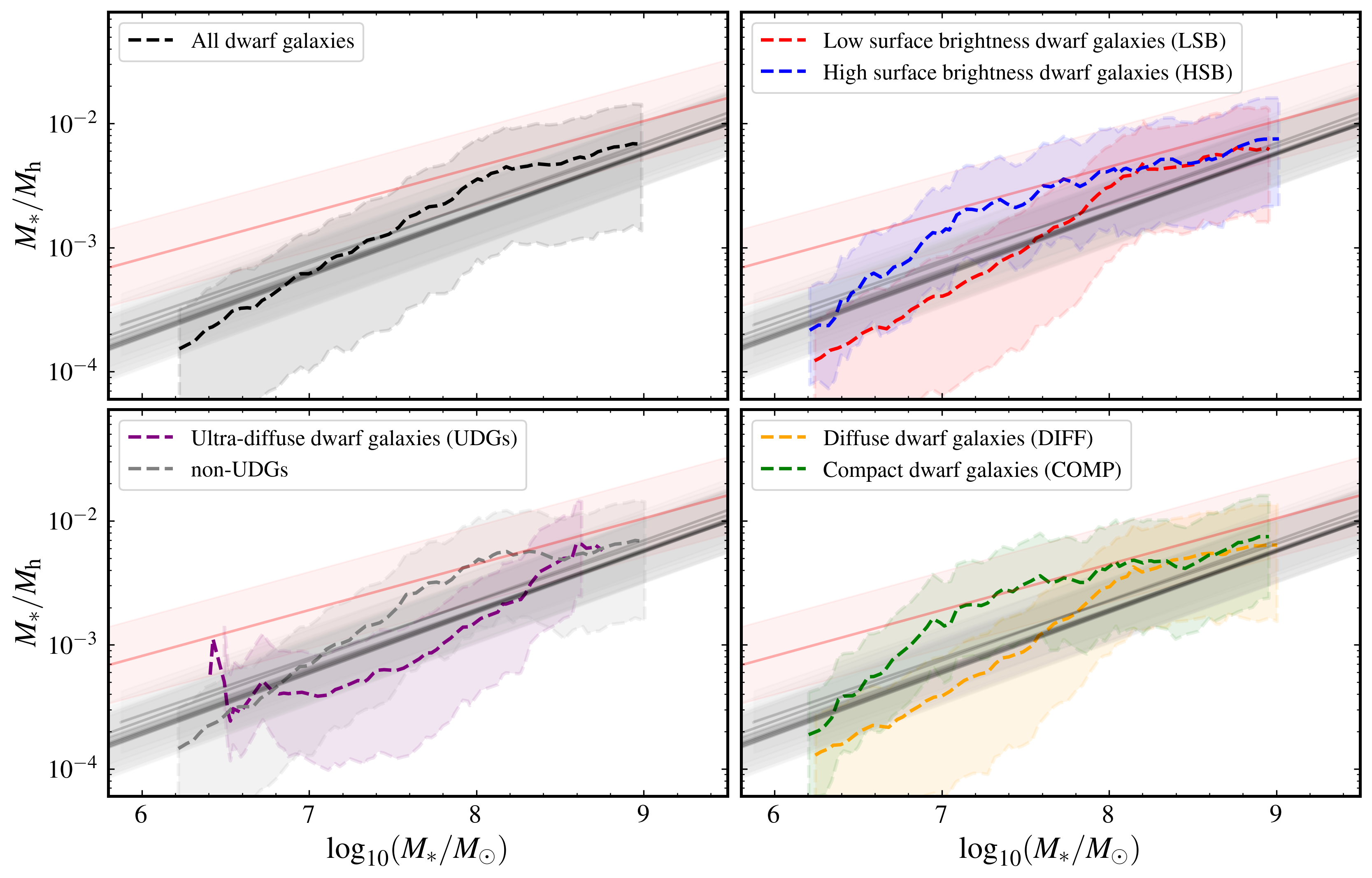}
\end{center}
\caption{Stellar-to-halo mass ratios (SHMR, grey and red solid curves) of dwarf galaxies of various categories at $z=0$ based on their GC numbers, versus models (various techniques). Grey curves represent a compilation of the SHMRs as presented in \citet{behroozi2019}. The shaded area around these SHMRs correspond to 0.25\,dex uncertainty for the SHMR relations. The shown SHMR at these mass regime is an extrapolation of the SHMR obtained for the more massive galaxies. The red solid line and the shaded area around it (in panels on the left) show the SHMR relation for low-mass galaxies (at $z=0$) in \citet{Zaritsky2023} and its uncertainties (0.31\,dex). The shaded area around dashed curves represents the uncertainties in $M_{\rm h}$, converted from uncertainties in $N_{\rm GC}$. {These uncertainties are converted from uncertainties in $N_{\rm GC}$ as estimated in Sec.\,\ref{sc:Results-numbers} (using bootstrapping) and shown in Figures\,\ref{dwarfs-ngc-all} and \ref{dwarfs-ngc}. The shown uncertainties also include to overall uncertainties regarding the $N_{\rm GC}-{M_{h}}$ conversion, which is $\pm (2\times 10^{9} \ N_{\rm GC})$.}}
\label{dwarfs-shmr}
\end{figure*}

\section{\label{sc:discussion}Discussion }

In this work, we inspect several trends that relate properties of the GC systems of dwarf galaxies to properties of these host galaxies (at a given $M_{\ast}$), mainly central surface brightness ($\mu_0$) and effective radius ($R_{\rm e}$). The interpretation of such trends regarding galaxy formation models requires models that take into account galaxy evolution and GC formation and evolution simultaneously; otherwise, interpreting the results is not straightforward. Here, we briefly discuss the immediate implications of the results.

\subsection{Observed trends for GCs of dwarf galaxies}

One of the main findings of this work is that at a given $M_{\ast}$ for dwarf galaxies, the distribution of GCs around dwarf galaxies, as measured through $R_{\rm GC}$, is independent of $R_{\rm e}$ and almost independent of $\mu_0$ (some hints of dependency is found). This is based on the analysis of the stacked radial distribution, which represents the average behaviour of dwarf galaxies. We see the same behaviour for individual dwarf galaxies with enough GCs that we could estimate their $R_{\rm GC}$. Overall, $R_{\rm GC}$ increases with $M_{\ast}$. The increase seems to be more consistent with the extrapolation of the best-fit curve on the red GCs of massive galaxies in \citet{lim2024}. This indicates that GCs of these dwarf galaxies are mostly formed in-situ. This is also the case for $R_{\rm GC}$ versus $N_{\rm GC}$, and dwarf galaxies in this study follow the curve for red-GCs of massive galaxies (relatively, and more than the curve for blue GCs) which supports the in-situ nature of the GCs within dwarf galaxies.

The trends are different for $R_{\rm GC}/R_{\rm e}$, both in the stacked GC profile and in individual cases. We mostly found smaller $R_{\rm GC}/R_{\rm e}$ for diffuse galaxies and lower surface brightness galaxies. This is also the case for UDGs, which are in the overlap zone between low surface brightness and diffuse dwarf galaxies (LSB and DIFF). Our analysis shows that the excess of GC systems of LSB and DIFF dwarf galaxies is mirrored in UDGs, but with a larger amplitude. This suggests that the excess of GC systems of UDGs results from the combination of diffuse and low surface-brightness properties; neither characteristic alone (e.g., DIFF or LSB) is sufficient to produce the quantitative trend seen for UDGs. Thus, these unique aspects of UDGs conspire to regulate their GC systems.

\subsection{Stellar-to-halo mass ratios based on $N_{\rm GC}$}

The total number of GCs of a galaxy has been shown to correlate tightly with galaxy total mass, mostly dark matter, in the high-mass regime and to remain correlated, though with more uncertainties and dispersion, down to total masses of a few $10^9\,\si{M_{\odot}}$ (\citealp{spitler2009,harris2013,burkert2020,jess24,desi-gcs}).
Assuming that this relation between $N_{\rm GC}$ and total mass holds for our observed dwarf galaxies, we may convert the empirical $N_{\rm GC}$ to total mass (about the dark matter halo mass, $M_{\rm h}$), using:

\begin{equation}
    M_{\rm total} / \si{M_{\odot}} = 5.1 \times 10^{9} \ N_{\rm GC} \ .
\label{ngc-eq}
\end{equation}

This conversion is calculated based on the ratio between the total mass of GCs around galaxies and the galaxies total mass, $\eta = M_{\rm GC}/M_{\rm total} = 3.9 \times 10^{-5}$ (estimated value for galaxies in galaxy clusters, \citealp{harris2017}), and assuming an average mass of $2 \times 10^5$ for a GC (\citealp{jordan2007}). With $M_{\rm total}$, we can then obtain the stellar-to-halo mass ratios (SHMR) for the sample. {Different assumptions for values of $\eta$ and the average mass of GCs would lead to a slightly different result than Eq.\,(\ref{ngc-eq}). Our choice for these parameters has led to Eq.\,(\ref{ngc-eq}), which is consistent with the equation $M_{\rm total} / \si{M_{\odot}} = 5.0 \times 10^{9} \ N_{\rm GC}$ provided in \citet{burkert2020}, and slightly different, but by less than a factor of two, from the \citet{zaritsky2022} result $M_{\rm total} / \si{M_{\odot}} = 2.9 \times 10^{9} \ N_{\rm GC}$. The conversation between $N_{\rm GC}$ and $M_{\rm total}$ provided in these works is calibrated directly on the number of GCs, and without any assumption on $\eta$ and on the average mass of GCs. In general, it is important to consider the various uncertainties mentioned above when using a conversation between $N_{\rm GC}$ and $M_{\rm total}$ that results in uncertainties about $\pm (2\times 10^{9} \ N_{\rm GC})$ in $M_{\rm total}$.}

The resulting SHMRs of dwarf galaxies are shown in Fig.\,\ref{dwarfs-shmr}. In each panel, we also show the SHMR fits of various works presented in \citet{behroozi2019}\footnote{see the references therein} at redshift $z=0$, extrapolated to lower masses. Most of these relations are valid for galaxies with $M_{\rm h} > 10^{10} \si{M_{\odot}}$ (or equally $M_{\ast} > 10^{6.5-7} \si{M_{\odot}}$), therefore, interpreting the results for dwarf galaxies that fall below this mass threshold requires caution. These curves are valid for field galaxies, and the halo mass corresponds to the galaxies peak-halo mass, whereas the sample dwarf galaxies of this study represents galaxies in a galaxy cluster environment. In such environments, dwarf galaxies very likely have lost a fraction of their total mass due to tidal interactions with the cluster potential or other galaxies. Here we consider that tidal stripping does not remove any GCs (\citealp{smith2013,smith2015}), and the current GC population represents the GCs at the peak-halo mass (before infall into the cluster). This is a plausible assumption considering that star clusters and GCs are expected to form mostly in the central regions of galaxies, where the conditions for star cluster formation are met and where they are protected by the gravitational potential of the galaxy. Figure\,\ref{dwarfs-shmr} also shows the SHMR presented in \citet{Zaritsky2023}, shown by the red solid curve. \citet{Zaritsky2023} have obtained the SHMR of low-mass galaxies down to $M_{\ast} = 10^5 \si{M_{\odot}}$ for four nearby ($z\sim0$) galaxy clusters (Virgo, Fornax, Hydra, and Coma), using a mass estimator developed based on the photometric properties and scaling relations of galaxies (\citealp{zaritsky2008}). Since this SHMR is based on cluster dwarf galaxies, it is not free from the effect of the environment (e.g. tidal stripping of the halo). This may explain the slightly lower dark matter fraction (higher $M_{\ast}/M_{\rm h}$) for this relation.

As is seen in Fig.\,\ref{dwarfs-shmr}, within the uncertainties, the SHMR obtained for all categories of dwarf galaxies in this work are consistent with the SHMR in the literature. In particular, although UDGs are relatively more dark matter-dominated (lower $M_{\ast}/M_{\rm h}$), they are within the scatter of the SHMR and are not exceptionally over-massive. Examining the SHMR of the different dwarf categories in comparison with \citet{behroozi2019} SHMRs at $z=0$ (grey solid curves), LSB/DIFF/UDG dwarf galaxies have lower $M_{\ast}/M_{\rm h}$ at a given $M_{\ast}$, in the direction to be more compatible with the SHMR at higher redshifts. It is expected that at higher redshifts, the SHMR moves downward, towards lower $M_{\ast}/M_{\rm h}$ values. Even though the uncertainties of SHMRs are large, this may suggest that LSB/DIFF and UDG dwarf galaxies stopped forming stars already at a higher $z$ due to a (potentially environmental) quenching mechanism; however, the exact mechanism is not clear (\citealp{forbes2023}). The early quenching of these dwarf galaxies makes them failed dwarf galaxies (\citealp{buzzo24,anna2024}) that have not been able to convert their mass content into stars as much as other dwarf galaxies of the same mass. Considering the average number of GCs (below 30) and corresponding halo mass (about $10^{11} \si{M_{\odot}}$), these objects are not failed Milky Way galaxies (\citealp{vd15}). In contrast to LSB/DIFF/UDG dwarf galaxies, HSB/COMP dwarf galaxies have higher $M_{\ast}/M_{\rm h}$ at a given $M_{\ast}$ than the average. Although most of the HSB/COMP dwarf galaxies in the Perseus cluster are red and currently quenched, they could be able to maintain their star formation long enough to convert their mass content into stars until very recently.

The scenario above is just one possible scenario that might explain the $M_{\ast}/M_{\rm h}$ of dwarf galaxies shown in Fig.\,\ref{dwarfs-shmr}. In any case, the failure or success of dwarf galaxies in forming stars seems to be independent of their GCs, as $R_{\rm GC}$ behaves similarly in all categories of dwarf galaxies. These scenarios are based on $M_{\rm h}$ from $N_{\rm GC}$, which is uncertain for low-mass and dwarf galaxies. The uncertainty arises from (i) the uncertainties in the estimated $N_{\rm GC}$, and (ii) the uncertainties in the estimated halo mass in dwarf galaxies. The calibration of the $N_{\rm GC}$--$M_{\rm h}$ relation remains one of the main tasks to be done in the coming years. In addition, having $N_{\rm GC}$ for dwarf galaxies in the field (rather than in galaxy clusters), where environmental effects are negligible, would provide further insight into the SHMR of dwarf galaxies.

\subsection{Potential biases when estimating $N_{\rm GC}$ for dwarf galaxies and UDGs}

The dependency of $R_{\rm GC}/R_{\rm e}$ on central surface brightness and effective radius shows that a similar assumption on $R_{\rm GC}/R_{\rm e}$ for all the dwarf galaxies would lead to a biased estimation of $N_{\rm GC}$, resulting in even higher $N_{\rm GC}$ in LSB, DIFF, and UDGs. This is potentially part of the explanation behind the finding by \citet{lim2018} of the very high $N_{\rm GC}$ of UDGs compared to non-UDGs in the Coma cluster, more than any other galaxy cluster/group (\citealp{prole2019,lim2020}). In some previous work, $R_{\rm GC}/R_{\rm e}=1.5$ is adopted to calculate $R_{\rm GC}$ and, subsequently, to count half of the GCs (GC candidates within $R_{\rm GC}$) for UDGs and non-UDGs. The total GC number $N_{\rm GC}$ is then estimated by multiplying this GC half-number count by two. By not taking into account such differences in $R_{\rm GC}/R_{\rm e}$ in UDGs and non-UDGs, such an approach can overestimate $N_{\rm GC}$ in UDGs by about a factor of two. 

Reducing the $N_{\rm GC}$ of UDGs in the Coma cluster does not mean that they have similar $N_{\rm GC}$ as non-UDGs, but it reduces the amplitude of the difference between $N_{\rm GC}$ of UDGs and non-UDGs. The Perseus cluster and Coma cluster are both quite similar environments in richness and mass. The effect of the choice of $R_{\rm GC}/R_{\rm e}$ for estimating $N_{\rm GC}$ is also addressed by \citet{saifollahi2021a,saifollahi2022} for large UDGs (in terms of effective radius) of the Coma cluster, including DF44 and DFX1. Our results can potentially be extended to the majority of UDGs in the Coma cluster. However, note that the values of $R_{\rm GC}$ and $R_{\rm GC}/R_{\rm e}$ cover a wide range of values, and each dwarf galaxy is unique. An average value of $R_{\rm GC}/R_{\rm e}=1$ seems reasonable for UDGs on average (\citealp{matlas-2024}); however, biases in assuming an average value for all galaxies should be taken into account when estimating the total GC numbers.
 
\section{\label{sc:Summary}Summary and Conclusions}

We used the \Euclid ERO data of the Perseus cluster and study the GC number counts and GC radial distribution for a uniquely large and less biased sample {(regarding selection bias)} of dwarf galaxies in a galaxy cluster environment. {In such environments, the majority of dwarf galaxy are early-types and therefore, our sample represents early-type dwarf galaxies. These dwarf galaxies have a stellar mass ranging between $10^5 M_{\odot}$ and $10^{9.5} M_{\odot}$}. The results represent the trends between dwarf galaxies and their GCs in high-density environments such as the Perseus galaxy cluster. We identify trends between GCs and their host dwarf galaxies. For that, we divide dwarf galaxies at a given stellar mass into different categories and studied their properties: low and high surface brightness dwarf galaxies (LSB and HSB), and diffuse and compact dwarf galaxies (DIFF and COMP). We also studied UDGs and non-UDGs, as well as nucleated and not-nucleated dwarf galaxies. Our findings are as follows:

\begin{enumerate}

    \item At a given stellar mass and within the uncertainties, the average $R_{\rm GC}$ of LSB and HSB, and DIFF and COMP, and UDGs and non-UDGs have similar values. This suggests that $R_{\rm GC}$ of dwarf galaxies is independent of dwarf galaxies effective radius ($R_{\rm e}$) and central surface brightness ($\mu_0$). However, our analysis hints at a dependency (not significant based on our data) between $R_{\rm GC}$ and $\mu_0$.\\

    \item At a given stellar mass and within the uncertainties, LSB, DIFF, and UDGs have a smaller $R_{\rm GC}/R_{\rm e}$, about equal to or less than one, compared to HSB, COMP, and non-UDGs with $R_{\rm GC}/R_{\rm e} > 1.5$. Taking into account the previous conclusion, this means that, on average, GCs are distributed over a similar radial extent for all classes of dwarf galaxies examined; however, their distributions are different from the field stars in the host galaxies.\\

    \item For dwarf galaxies at a given stellar mass, LSB, DIFF, and UDGs have higher $N_{\rm GC}$ compared to HSB, COMP, and non-UDGs. Although such behaviour for UDGs was already established in previous work, here we find a similar, albeit smaller, trend for LSB and DIFF galaxies. The excess of GCs in UDGs can be attributed to the unique characteristics of lower surface brightness and a larger effective radius, as shown in the LSB and DIFF sub-samples. UDGs encapsulate both these properties, distinguishing them from other dwarf galaxy populations. \\

    \item At a given stellar mass above $M_{\ast} = 10^8 \si{M_{\odot}}$, nucleated dwarf galaxies have a lower $R_{\rm GC}$ and $R_{\rm GC}/R_{\rm e}$ than not-nucleated dwarf galaxies. Additionally, at a given stellar mass below this threshold, nucleated dwarf galaxies have higher $N_{\rm GC}$ than not-nucleated dwarf galaxies. For more massive dwarf galaxies with $M_{\ast} > 10^8 \si{M_{\odot}}$, no significant difference between $N_{\rm GC}$ in nucleated and not-nucleated dwarf galaxies is seen.\\

    \item Our estimates of dark matter halo mass, based on GC number counts, are entirely consistent with extrapolated trends of the $z=0$ SHMR relation at low masses. This is the case for all the different categories of dwarf galaxies. This implies that, unlike some previous claims, UDGs do not show any exceptional excess in dark matter (being over-massive), and that their SMHR and dark matter content are within the expected ranges.\\

    \item Our results suggest that $R_{\rm GC}/R_{\rm e}$ depends on central surface brightness and effective radius. This has implications for estimations of $N_{\rm GC}$ in dwarf galaxies of different types, as well as in UDGs and non-UDGs, and potentially points to existing biases in the estimated $N_{\rm GC}$ of UDGs in some previous works in the literature.\\

    
\end{enumerate}

The observed trends provide a benchmark to test current/future cosmological simulations that simulate dwarf galaxies and their GCs in galaxy clusters (\citealp{jess23,pfeffer24}). More observations of similar environments and other environments are required to further characterise these trends. In this regard, \Euclid will be fundamental for such studies, as it will observe a large number of dwarf galaxies in various environments (\citealp{Q1-SP001}), including systems within the Fornax-Eridanus supercluster (\citealp{raj}), as well as the Coma cluster and the northern part of the Virgo cluster, and hundreds of nearby dwarf galaxies in isolation. A uniform analysis of such a homogeneous data set promises a better understanding of dwarf galaxies and their GC properties across different environments. {The future data releases of the \Euclid mission also include deep ground-based data (\Euclid-EXT) in the blue optical bands, in the $u$ and $g$ bands, from various surveys (UNIONS / DES / LSST). Additional colour cuts and constraints from these blue bands would improve the selection of the GCs (candidates) and further clean up the GC samples from the remaining contamination in the GC (candidate) samples.}

\begin{acknowledgements}
{We thank the referee for their comments that helped us to improve the quality of the presented work.} We thank Dr. Jonathan Freundlich and Dr. Gauri Sharma for sharing their insights and discussion about stellar-to-halo mass in galaxies. TS acknowledges funding from the CNES postdoctoral fellowship programme. TS, AL acknowledge support from Agence Nationale de la Recherche, France, under project ANR-19-CE31-0022. {TS, AL acknowledge support from the European Union (Widening Participation, ExGal-Twin, GA 101158446).} MC acknowledges support from the ASI-INAF agreement "Scientific Activity for the Euclid Mission" (n.2024-10-HH.0; WP8420) and from the INAF "Astrofisica Fondamentale" GO-grant 2024 (PI M. Cantiello). 
\AckEC
\AckERO
This research has made use of the SIMBAD database (\citealp{simbad}), operated at CDS, Strasbourg, France, the VizieR catalogue access tool (\citealp{vizier}), CDS, Strasbourg, France (DOI : 10.26093/cds/vizier), and the Aladin sky atlas (\citealp{aladin1,aladin2}) developed at CDS, Strasbourg Observatory, France and SAOImageDS9 (\citealp{ds9}). This work has been done using the following software, packages and python libraries: 
\texttt{Topcat} (\citealp{Topcat2005}); 
\texttt{Numpy} (\citealp{numpy}), \texttt{Scipy} (\citealp{scipy}), \texttt{Astropy} (\citealp{astropy}), \texttt{Scikit-learn} (\citealp{scikit-learn}). 
Table\,\ref{gc-dist-values-table} is only available in electronic form at the CDS via anonymous ftp to \url{cdsarc.u-strasbg.fr} (130.79.128.5) or via \url{http://cdsweb.u-strasbg.fr/cgi-bin/qcat?J/A+A/}.
\end{acknowledgements}

\bibliography{mybib}

\begin{appendix}

\section{Tables}

Here, Tables\,\ref{appendix:final-dwarf-catalogue} and \ref{gc-dist-values-table} provide the results of the analysis in this paper.

\begin{table*}[ht]
\centering
\small
\caption{Details and the output of the analysis of the stacked GC distributions.}

\begin{tabular}{lccccccccc}
\hline\hline
\noalign{\vskip 2pt}
Dwarf sample & Parameter & & & & $M_{\ast}$ bin & & & \\
 &  & $10^{6} \si{M_{\odot}}$ & $10^{6.5} \si{M_{\odot}}$ & $10^{7} \si{M_{\odot}}$ & $10^{7.5} \si{M_{\odot}}$ & $10^{8} \si{M_{\odot}}$ & $10^{8.5} \si{M_{\odot}}$ & $10^{9} \si{M_{\odot}}$ \\
\hline
\noalign{\vskip 2pt}
 & $N_{\rm dwarfs}$ & 243 & 377 & 426 & 339 & 211 & 112 & 54 \\
 & $N_{\rm GC, cand, gal}$ & 104 & 777 & 1252 & 1911 & 2229 & 2675 & 2422 \\
\textcolor{black}{\textbf{All dwarfs}} & $N_{\rm GC, cand, back}$ & 37.1 & 550.6 & 890.1 & 1422.7 & 1739.3 & 2090.2 & 1776.4 \\
 & $R_{\rm GC}$ & 0.20 & 0..48 & 0.73 & 1.22 & 1.67 & 2.72 & 3.53 \\
 & $R_{\rm GC} / \langle R_{\rm e} \rangle$ & 0.33 & 0.70 & 0.87 & 1.10 & 1.20 & 1.67 & 1.70 \\
 
\hline

\noalign{\vskip 2pt}
 & $N_{\rm dwarfs}$ & 112 & 187 & 218 & 165 & 108 & 59 & 25 \\
 & $N_{\rm GC, cand, gal}$ &  131 & 235 & 572 & 1805 & 1758 & 1575 & 1351  \\
\textcolor{blue}{HSB} & $N_{\rm GC, cand, back}$ &  90.3 & 158.3 & 499.1 & 1625.3 & 1450.6 & 1214.3 & 1016.3  \\
 & $R_{\rm GC}$ & 0.21 & 0.42 & 0.89 & 1.60 & 2.10 & 2.91 & 3.38  \\
 & $R_{\rm GC} / \langle R_{\rm e} \rangle$ &  0.43 & 0.75 & 1.25 & 1.74 & 1.81 & 1.98 & 1.95 \\
 
\hline

\noalign{\vskip 2pt}
 & $N_{\rm dwarfs}$ & 130 & 188 & 206 & 173 & 102 & 52 & 28  \\
 & $N_{\rm GC, cand, gal}$ &  54 & 421 & 691 & 894 & 699 & 899 & 1159  \\
\textcolor{red}{LSB} & $N_{\rm GC, cand, back}$ & 18.2 & 273.9 & 404.8 & 582.7 & 478.1 & 686.7 & 845.3 \\
 & $R_{\rm GC}$ &  0.14 & 0.43 & 0.71 & 1.00 & 1.38 & 2.22 & 3.55  \\
 & $R_{\rm GC} / \langle R_{\rm e} \rangle$ &  0.21 & 0.53 & 0.71 & 0.75 & 0.86 & 1.12 & 1.62 \\

\hline

\noalign{\vskip 2pt}
 & $N_{\rm dwarfs}$ &  119 & 197 & 220 & 158 & 101 & 56 & 24  \\
 & $N_{\rm GC, cand, gal}$ &   44 & 140 & 721 & 1000 & 1036 & 1680 & 1227  \\
\textcolor{OliveGreen}{COMP} & $N_{\rm GC, cand, back}$ &  8.6 & 72.3 & 620.8 & 850.5 & 819.8 & 1359.5 & 982.5 \\
 & $R_{\rm GC}$ &  0.12 & 0.23 & 0.67 & 1.50 & 1.90 & 3.08 & 3.54  \\
 & $R_{\rm GC} / \langle R_{\rm e} \rangle$ &  0.29 & 0.46 & 1.12 & 1.92 & 1.88 & 2.52 & 2.36 \\

\hline

\noalign{\vskip 2pt}
 & $N_{\rm dwarfs}$ & -- & 178 & 204 & 180 & 109 & 55 & 29  \\
 & $N_{\rm GC, cand, gal}$ &  -- & 455 & 655 & 1053 & 984 & 1297 & 1309  \\
\textcolor{orange}{DIFF} & $N_{\rm GC, cand, back}$ &  -- & 307.8 & 378.1 & 712.8 & 713.6 & 1002.1 & 905.2  \\
 & $R_{\rm GC}$ &  -- & 0.67 & 0.79 & 1.14 & 1.49 & 2.54 & 3.39  \\
 & $R_{\rm GC} / \langle R_{\rm e} \rangle$ &  -- & 0.67 & 0.67 & 0.81 & 0.86 & 1.10 & 1.23\\

\hline

\noalign{\vskip 2pt}
 & $N_{\rm dwarfs}$ &  -- & 47 & 53 & 33 & 13 & 4 & --  \\
 & $N_{\rm GC, cand, gal}$ &  -- & 111 & 439 & 591 & 208 & 67 & --  \\
\textcolor{violet}{UDG} & $N_{\rm GC, cand, back}$ &  -- & 36.7 & 267.9 & 427.7 & 156.6 & 37.6 & --  \\
 & $R_{\rm GC}$ &  -- & 1.02 & 1.22 & 2.21 & 1.63 & 1.72 & --  \\
 & $R_{\rm GC} / \langle R_{\rm e} \rangle$ &  -- & 0.52 & 0.63 & 1.13 & 0.61 & 0.64 & -- \\

\hline

\noalign{\vskip 2pt}
 & $N_{\rm dwarfs}$ &  235 & 353 & 376 & 284 & 176 & 97 & 49  \\
 & $N_{\rm GC, cand, gal}$ &  114 & 797 & 1168 & 1846 & 1652 & 2915 & 2517  \\
\textcolor{gray}{non-UDG} & $N_{\rm GC, cand, back}$ &  47.7 & 586.1 & 883.3 & 1499.4 & 1339.8 & 2382.7 & 1882.1  \\
 & $R_{\rm GC}$ & 0.19 & 0.48 & 0.73 & 1.27 & 1.52 & 2.97 & 3.69  \\
 & $R_{\rm GC} / \langle R_{\rm e} \rangle$ &  0.34 & 0.73 & 0.95 & 1.28 & 1.20 & 1.97 & 1.81 \\

 \hline

\noalign{\vskip 2pt}
 & $N_{\rm dwarfs}$ & 200 & 291 & 291 & 190 & 115 & 71 & 35  \\
 & $N_{\rm GC, cand, gal}$ &  79 & 442 & 708 & 898 & 1925 & 2529 & 1767  \\
\textcolor{black}{Nucleated} & $N_{\rm GC, cand, back}$ &  30.1 & 316.2 & 533.8 & 736.3 & 1662.3 & 2037.6 & 1325.5  \\
 & $R_{\rm GC}$ &  0.20 & 0.36 & 0.69 & 1.09 & 2.32 & 3.56 & 3.87  \\
 & $R_{\rm GC} / \langle R_{\rm e} \rangle$ &  0.38 & 0.58 & 0.98 & 1.19 & 1.91 & 2.42 & 2.08 \\

 \hline

\noalign{\vskip 2pt}
 & $N_{\rm dwarfs}$ &  41 & 76 & 112 & 108 & 56 & 18 & 6 \\
 & $N_{\rm GC, cand, gal}$ &  31 & 207 & 355 & 949 & 478 & 233 & 260  \\
\textcolor{black}{Not-nucleated} & $N_{\rm GC, cand, back}$ &  12.1 & 121.8 & 225.1 & 749.8 & 386.5 & 208.8 & 199.8 

 \\
 & $R_{\rm GC}$ &  0.10 & 0.69 & 0.76 & 1.61 & 1.48 & 1.74 & 2.60 \\
 & $R_{\rm GC} / \langle R_{\rm e} \rangle$ &  0.14 & 0.76 & 0.74 & 1.37 & 1.06 & 0.95 & 1.05 \\

\hline
\end{tabular}
\label{gc-dist-values-table}\\
\vspace{1ex}
\raggedright
{\raggedright {\bf Notes.} The table is divided into nine parts for different categories of dwarf galaxies (column 1). Each part provides details/results for a given category, specified in column 2. These parameters are: number of dwarf galaxies in a given bin ($N_{\rm dwarf}$), total number of GC candidates within 3\,$R_{GC}$ of the dwarf galaxies in the sample ($N_{\rm GC, cand, gal}$), number of background sources beyond 3\,$R_{\rm GC}$ and normalised to the area within 3\,$R_{\rm GC}$ of the dwarf galaxies in the sample ($N_{\rm GC, cand, back}$), the estimated GC half-number radius ($R_{\rm GC}$, in kpc) and the ratio between GCs half-number radius and the average half-light radius of the dwarf sample ($R_{\rm GC} / \langle R_{\rm e} \rangle$).} Columns 3 to 9 present the values corresponding to seven stellar mass bins used for the analysis of the stacked GC distributions, ranging from $10^{5}$ to $10^{9} \si{M_{\odot}}$. The value associated with each column corresponds to the centre of the bin. Stellar mass bins have a width of 1\,dex. 
\end{table*}


\begin{table}[ht!]
\caption{Table of the GC properties of the dwarfs galaxies in this work.}
\label{appendix:final-dwarf-catalogue}

\centering
\begin{tabular}{ccccc}
\hline
\hline
\noalign{\smallskip}
ID  &
$M_{\rm *}/M_{\rm \odot}$  & 
$N_{\rm GC}$ & 
$R_{\rm GC}$  &  
$R_{\rm GC}/R_{\rm e}$ \\[3pt]
(1) & (2) & (3) & (4) & (5) \\
\hline
\noalign{\smallskip}
EDwC-0017 & $ 7.95$ & $ 15.5 \pm  3.5$ & $ {2.68}^{+1.78}_{-1.40}$ & $ {1.49}^{+0.99}_{-0.77} $ \\[3pt]
EDwC-0023 & $ 8.64$ & $ 35.8 \pm  5.6$ & $ {0.87}^{+0.43}_{-0.33}$ & $ {0.40}^{+0.20}_{-0.15} $  \\[3pt]
EDwC-0026 & $ 8.706$ & $ 96.2 \pm  8.1$ & $ {3.11}^{+0.53}_{-0.52}$ & $ {1.51}^{+0.26}_{-0.25} $  \\[3pt]
EDwC-0046 & $ 8.26$ & $ 18.5 \pm  5.6$ & $ {1.21}^{+0.26}_{-0.27}$ & $ {0.82}^{+0.17}_{-0.18} $  \\[3pt]
EDwC-0082 & $ 7.76$ & $ 8.8 \pm  2.2$ & $ {1.04}^{+0.62}_{-0.60}$ & $ {0.54}^{+0.32}_{-0.31} $ \\ [3pt]
EDwC-0086 & $ 7.39$ & $ 24.8 \pm  3.6$ & $ {1.91}^{+0.73}_{-0.72}$ & $ {1.08}^{+0.41}_{-0.41} $  \\ [3pt]
EDwC-0089 & $ 9.00$ & $ 50.9 \pm  7.0$ & $ {2.93}^{+0.63}_{-0.73}$ & $ {1.07}^{+0.23}_{-0.27} $  \\[3pt]
EDwC-0112 & $ 7.84$ & $ 11.5 \pm  2.6$ & $ {1.14}^{+0.79}_{-0.73}$ & $ {0.58}^{+0.40}_{-0.37} $  \\[3pt]
EDwC-0120 & $ 7.94$ & $ 41.6 \pm  4.6$ & $ {1.77}^{+0.59}_{-0.72}$ & $ {0.82}^{+0.27}_{-0.33} $  \\[3pt]
EDwC-0156 & $ 8.82$ & $ 38.4 \pm  6.8$ & $ {0.86}^{+0.98}_{-0.46}$ & $ {0.37}^{+0.42}_{-0.20} $  \\[3pt]
... & ... & ... & ... &  ... \\
\hline 
\end{tabular}
\tablefoot{Column 1 present the \Euclid ID (name) of the dwarf galaxies from \citet{EROPerseusDGs}. Column 2 presents the stellar mass of dwarf galaxies from \citet{EROPerseusOverview}. Columns 3, 4, and 5 present the GC properties of the dwarf galaxies, as estimated in this work. The full table will be available at the Strasbourg astronomical Data Center (CDS) as supplementary material.}
\end{table}

\section{Completeness in the near-infrared bands (\YE, \JE, \HE)}

\label{app-comp-nisp}

\begin{figure}[htbp!]
\begin{center}
\includegraphics[trim={0 -10 0 -10},angle=0,width=0.99\linewidth]{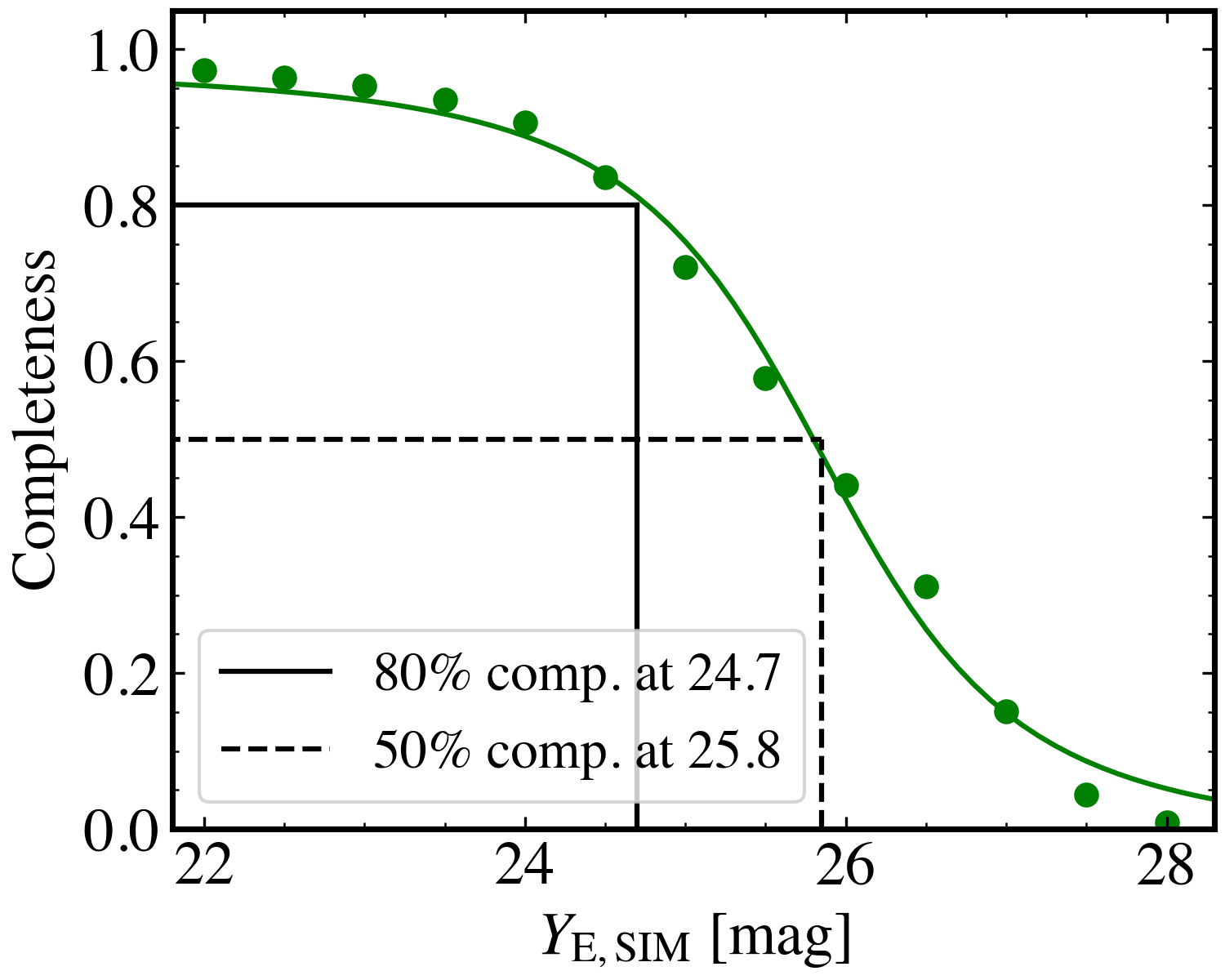}
\includegraphics[trim={0 -10 0 -10},angle=0,width=0.99\linewidth]{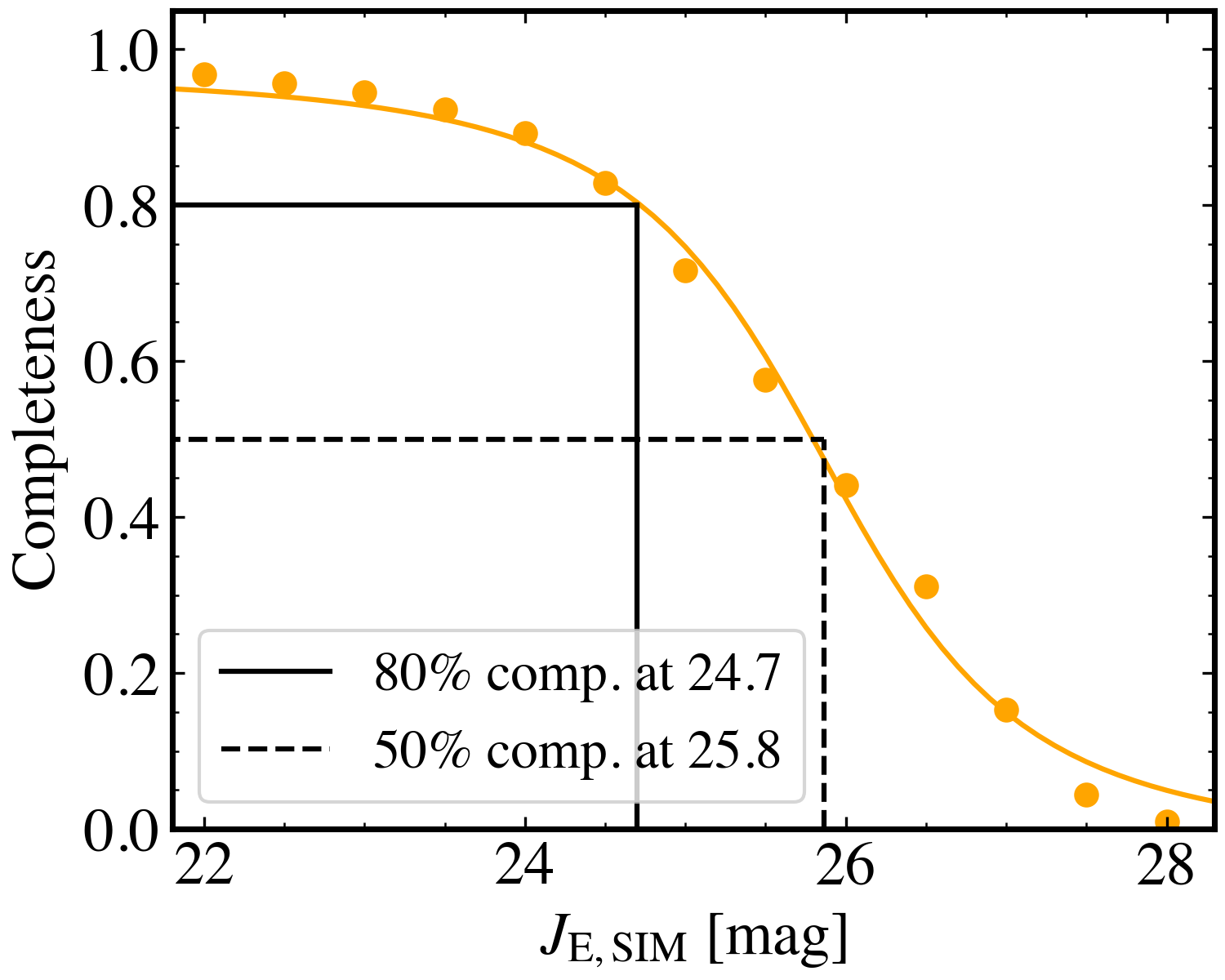}
\includegraphics[trim={0 -10 0 -10},angle=0,width=0.99\linewidth]{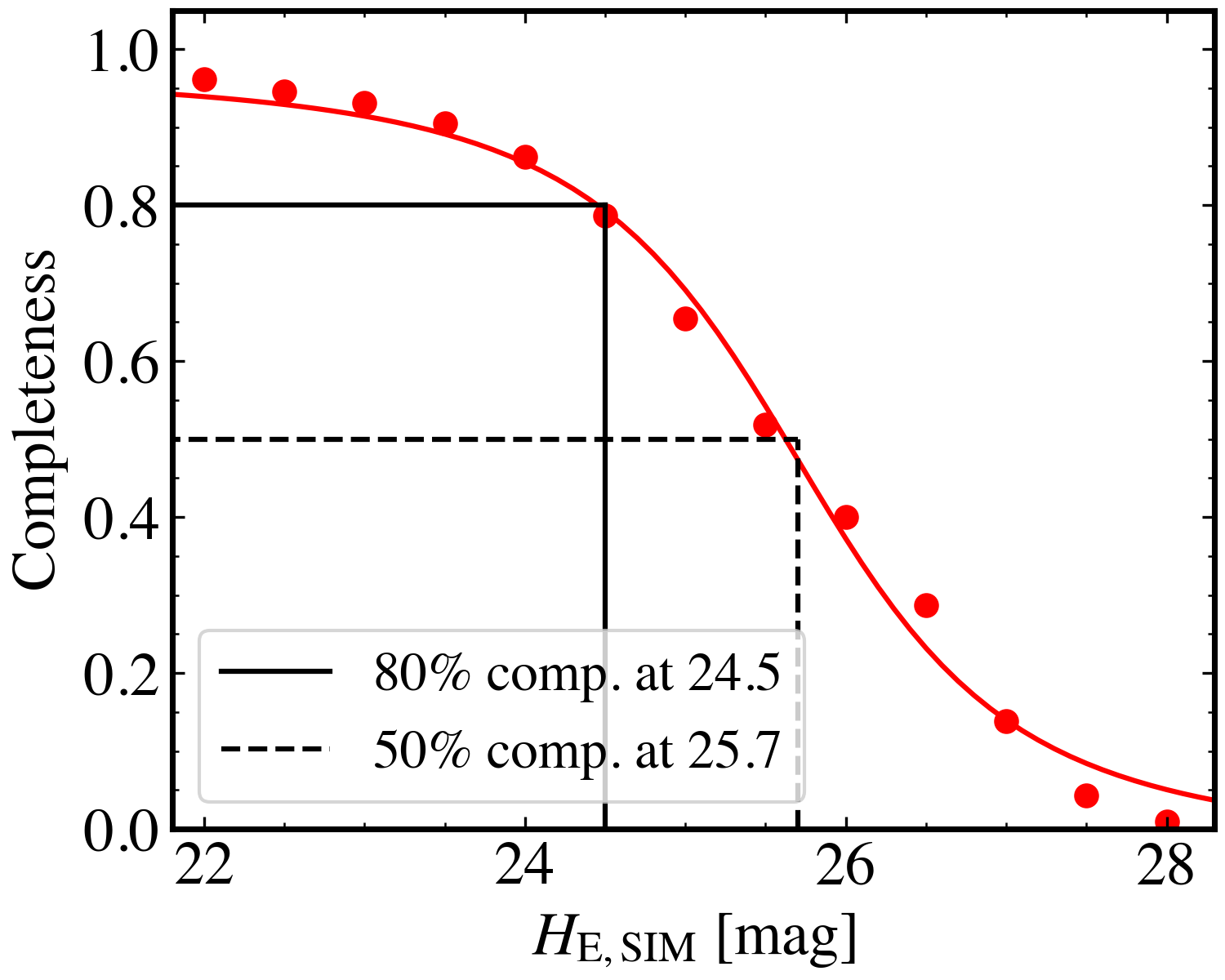}
\end{center}
\caption{Completeness of the performed source detection in \YE, \JE, and \HE based on artificial GC tests. Our assessment shows that the GC detection is 80\% complete at magnitude 24.7, 24.7, and 24.5 in \YE, \JE, and \HE, respectively.}
\label{completeness-nisp}
\end{figure}

Here we provide the results of the completeness assessment for source detection and photometry in the near-infrared bands of the ERO data, in \YE, \JE, \HE. The results are shown in Fig.\,\ref{completeness-nisp}.

\section{Ratio between $\langle R_{\rm GC} / R_{\rm e} \rangle$ and $R_{\rm GC} / \langle R_{\rm e} \rangle$ }

\label{app}

 We assessed the ratio between $\langle R_{\rm GC} / R_{\rm e} \rangle$ and $R_{\rm GC} / \langle R_{\rm e} \rangle$, referred to as $\alpha$, with simulations constructed for each galaxy-mass bin under the following assumptions:
 \begin{itemize}
\item Within a bin, the one-dimensional radial distribution of the GC-population of any galaxy is a sample originating from a single parent distribution that has a median radius ${R^{th}_{\rm GC}}$, a theoretical quantity for which we adopt the empirical value $R_{\text GC}$ shown in Fig.\,\ref{dwarfs-stacked-rgc-all} (i.e. the median radius of the stacked GC-populations of the mass bin). \\
\item Within a bin, the two-dimensional density distributions of the GC-populations around their host galaxy centre all follow a S\'ersic law $\rho_{\rm 2D}(r)$ with the same index $n$ (which we may vary in our experiments). This law can be written as
$$\rho_{\rm 2D}(r)\ \propto \ \exp \left\{ -b_n \left[\left(\frac{r}{r_{\rm e}}\right)^{\frac{1}{n}}  -1 \right]\right\},$$
where $r$ is the distance to the galaxy centre, and $b_n$ is the constant that ensures that $r_{\rm e}$ indeed is the effective radius, which in our context means $r_e=R_{\rm GC}$, since this law is used to describe the radial distribution of GCs.
\end{itemize}

For each galaxy of a given bin, the effective radius $R_{\rm e}$ of the light profile is available from measurements (\citealp{EROPerseusDGs}). 
If the effective radius of the GC-population is the single value ${R^{th}_{\rm GC}} \equiv R_{\rm GC}$, then $\langle R_{\rm GC} / R_{\rm e} \rangle = R_{\rm GC} \, \langle 1 / R_{\rm e} \rangle$, and on average 
$\alpha = \langle 1 / R_{\rm e} \rangle \,
\langle R_{\rm e} \rangle$ (a reference value written $\alpha_0$ hereafter). The values of $\alpha_0$ are between 1.28 and 1.34 for all the mass bins.\\

In practice, the empirical $R_{\rm GC}$ in Fig.\,\ref{dwarfs-stacked-rgc-all} are only estimates of ${R^{th}_{\rm GC}}$ from finite samples. Before we divide the bins into subsets and repeat this experiment (with final numbers that are reported in the main text), we make sure that all the galaxy-mass bins that we consider contain more than 100 GCs after stacking. For S\'ersic distributions with $n\leqslant 2$, 100 GCs provide an estimate of the effective radius (i.e. $R_{\rm GC}$) with a standard deviation smaller than $\pm 15\%$ (checked numerically, and also in agreement with the width of the shaded areas in the top panel of Fig.\,\ref{dwarfs-stacked-rgc-all}). Therefore the values of $\alpha_0$ are representative of the values of $\alpha$ that apply in our bins. {We can simulate the effect of finite samples by assigning to the galaxies of one bin somewhat dispersed effective radii for their GC distributions.}

\section{Comparison with the literature}

\begin{figure}[htbp!]
\begin{center}
\includegraphics[trim={0 0 0 0},angle=0,width=0.99\linewidth]{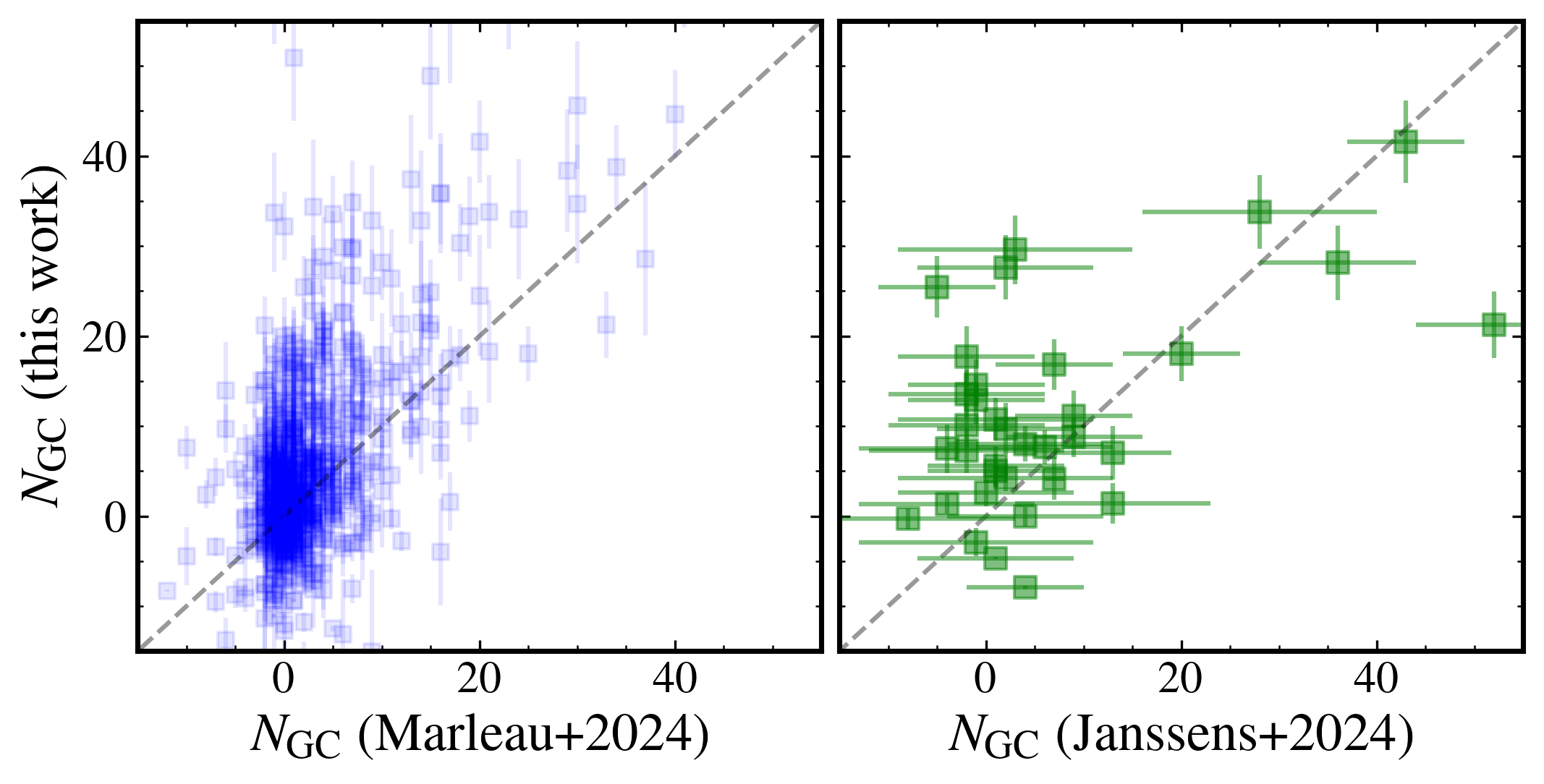}
\includegraphics[trim={0 0 0 0},angle=0,width=0.99\linewidth]{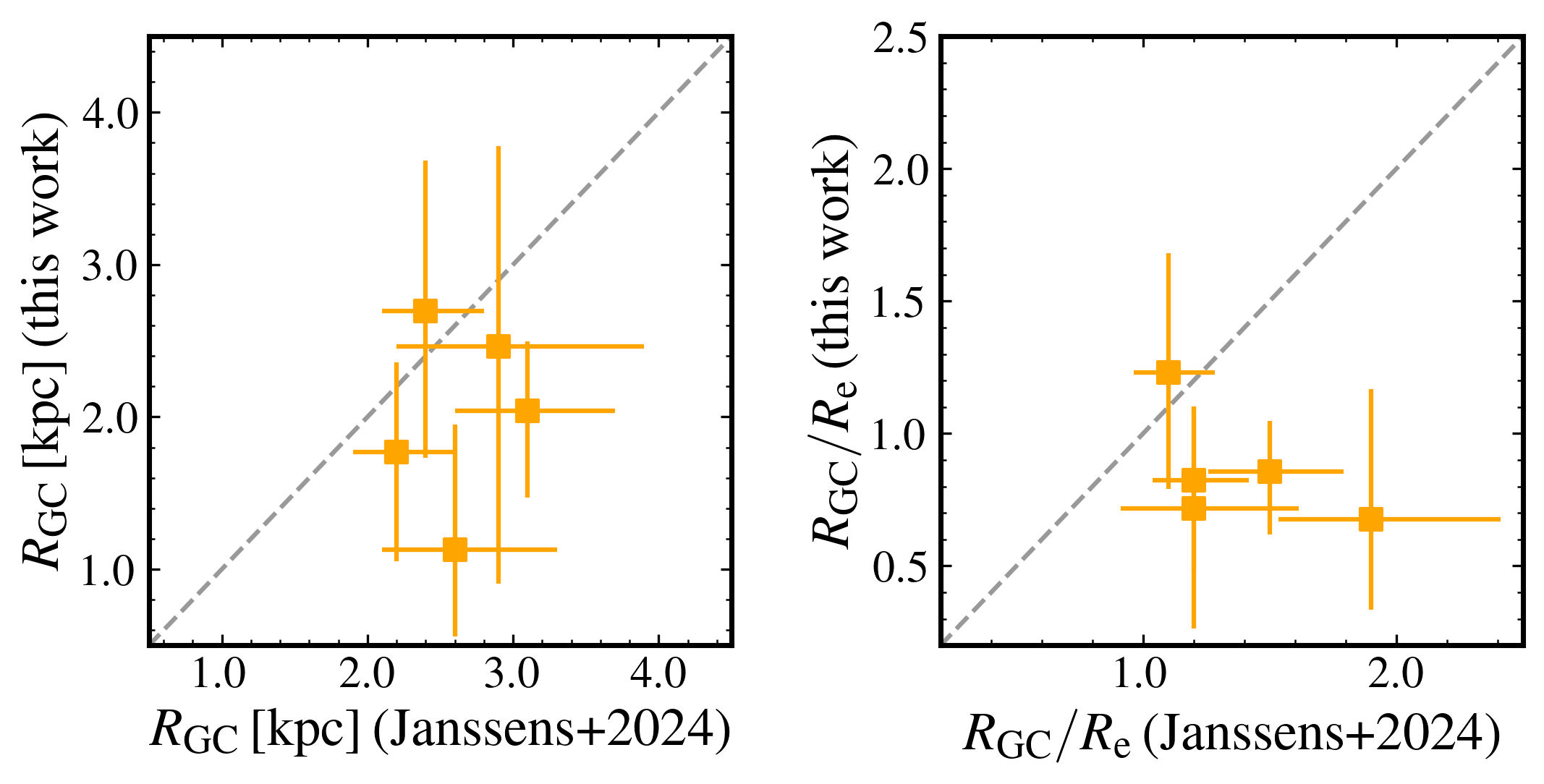}
\includegraphics[trim={0 0 0 0},angle=0,width=0.99\linewidth]{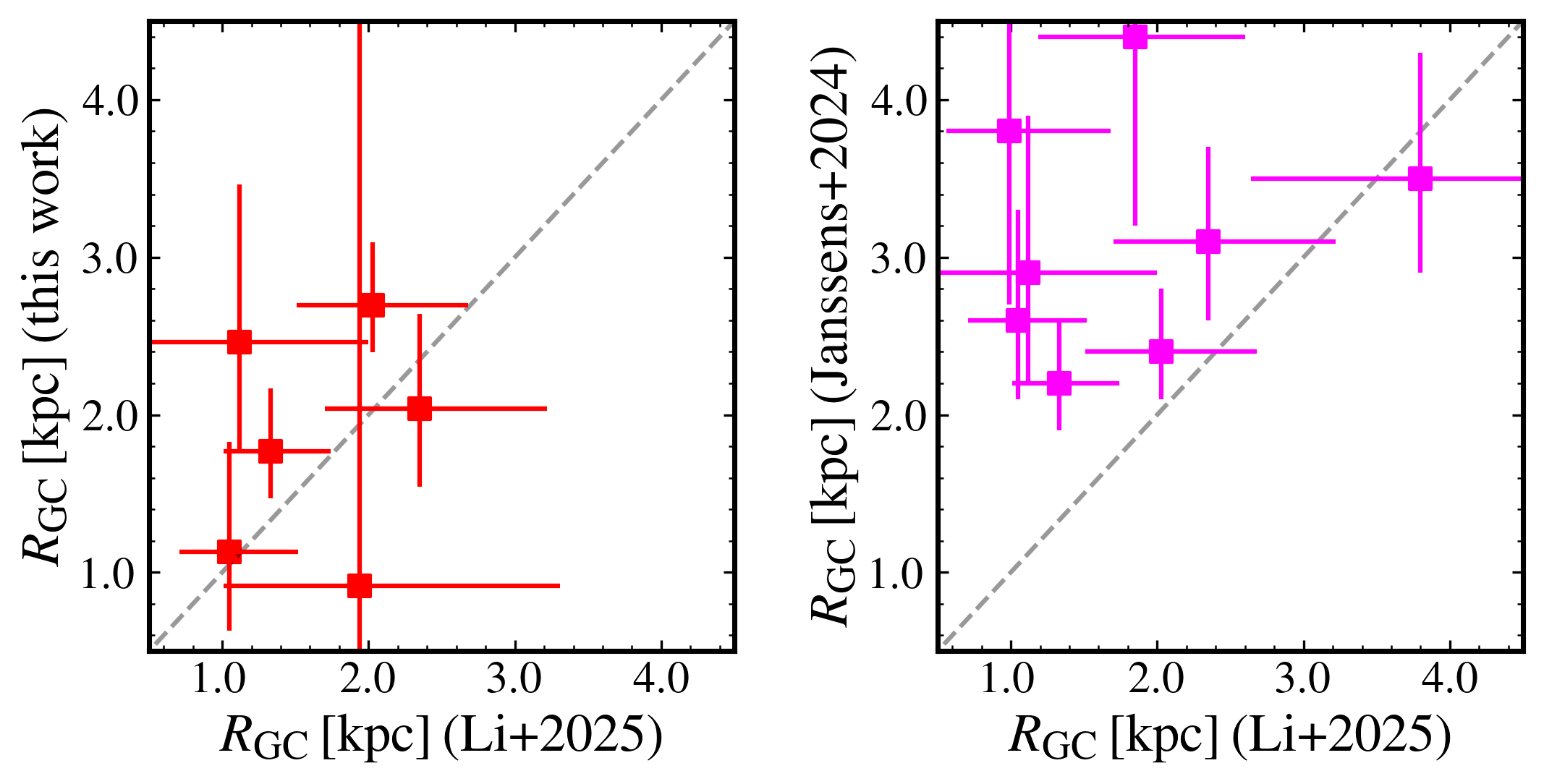}
\end{center}
\caption{Comparison between GC properties in this work and the literature. Top: Number counts estimated in this work versus \citet{EROPerseusDGs} and \citet{janssens2024}. {Middle: $R_{\rm GC}$ and $R_{\rm GC} / R_{\rm e}$ estimated in this work and \citet{janssens2024}. Bottom:  $R_{\rm GC}$ values estimated in \citet{li2025} versus this work, and \citet{janssens2024}.}}
\label{dwarfs-comp1}
\end{figure}

Here we provide a comparison between the GC number counts ($N_{\rm GC}$) and GC distributions ($R_{\rm GC}$ and $R_{\rm GC} / R_{\rm e}$) of dwarf galaxies in this work versus the values published in \citet{EROPerseusDGs} and \citet{janssens2024}. Generally we find a good agreement with their results. Comparing our results with $N_{\rm GC}$ from \citet{EROPerseusDGs}, as is seen in Fig.\,\ref{dwarfs-comp1} on the top, on average, we find a larger $N_{\rm GC}$. In this work we used the same \Euclid ERO data as \citet{EROPerseusDGs} and with a different methodology (e.g., by including colours in the GC selection procedure) for GC selection that leads to a different GC catalogue. Additionally, we adopt different approach for estimating $N_{\rm GC}$, mainly, a fainter GCLF TOM and varying $R_{\rm GC}$ as a function of stellar mass and central surface brightness of dwarf galaxies. Regarding the values $R_{\rm GC}$ and $R_{\rm GC} / R_{\rm e}$ values for five dwarf galaxies, as is seen in Fig.\,\ref{dwarfs-comp1} on the bottom, our analysis leads to, on average, lower values compared to \citet{janssens2024}, however, for the two most GC-rich dwarf galaxies in the sample, measurements from two works are consistent with each other. For other three dwarf galaxies, the difference mainly arises from low number statistics when dealing with GCs in these dwarf galaxies. 

\label{app-comparison}

\end{appendix}

\end{document}